\documentclass[12pt,notitlepage]{article}

\usepackage{amsmath}         
\usepackage{amssymb}         
\usepackage{amsfonts}        
\usepackage{graphicx}        

\usepackage{color}         
\usepackage{slashed}       
\usepackage{framed}        
\usepackage{subcaption}    
\usepackage[mathscr]{euscript} 
\usepackage{cite}          
\usepackage{booktabs}      
\usepackage{youngtab}	    
\usepackage{arydshln} 	    
\usepackage{tocloft}		   
\usepackage{setspace}	   
\usepackage{listings}       
\usepackage{wasysym}
\usepackage{amsthm}
\usepackage{changepage}
\usepackage{enumitem}

\usepackage[margin=2cm]{geometry}   
\graphicspath{{figures/}}	        
\numberwithin{equation}{section}    
\renewcommand{\tilde}{\widetilde}   
\renewcommand{\vec}[1]{\mathbf{#1}} 

\newcommand{\ket}[1]{\left|#1\right\rangle}    
\newcommand{\bra}[1]{\left\langle#1\right|}    

\newcommand{\email}[1]{\href{mailto:#1}{#1}}

\newenvironment{institutions}[1][2em]{\begin{list}{}{\setlength\leftmargin{#1}\setlength\rightmargin{#1}}\item[]}{\end{list}}

\let\oldenumerate\enumerate
\renewcommand{\enumerate}{
  \oldenumerate
  \setlength{\itemsep}{1pt}
  \setlength{\parskip}{0pt}
  \setlength{\parsep}{0pt}
}

\let\olditemize\itemize
\renewcommand{\itemize}{
  \olditemize
  \setlength{\itemsep}{1pt}
  \setlength{\parskip}{0pt}
  \setlength{\parsep}{0pt}
}

\usepackage{bbm}

\usepackage[
	colorlinks=true,
	citecolor=black,
	linkcolor=black,
	urlcolor=blue,
	hypertexnames=false]{hyperref}

\renewcommand{\eqref}[1]{Eq.~(\ref{#1})}

\newcommand{\nlm}{\ensuremath{{n\ell m}}}

\DeclareMathOperator\sinc{sinc}

\newcounter{choices}
\newcommand\newchoice[1]{\refstepcounter{choices}
\begin{adjustwidth}{4em}{0em}
\vspace{1ex}
\begin{enumerate}
\item[\textbf{Choice~\thechoices.}] #1
\end{enumerate}
\vspace{1ex}
\end{adjustwidth}}

\newcounter{radialoptions}

\begin{document}

\begin{center}

    {\LARGE Wavelet-Harmonic Integration Methods}

    \vskip .7cm

    { \bf Benjamin~Lillard} 
    \\ \vspace{0em}
    { 
    \footnotesize
    \email{blillard@uoregon.edu}
    }
	
    \vspace{-.2cm}

    \begin{institutions}[3.5cm]
\begin{center}
    \footnotesize
    {\it 
	    Institute for Fundamental Science and Department of Physics, \\
	    Willamette Hall, University of Oregon, Eugene, OR 97401, U.S.A.
	    }   
\end{center} 
    \end{institutions}

\today
\end{center}


\begin{abstract}

A new integration method drastically improves the efficiency of the dark matter direct detection calculation. In this work I introduce a complete, orthogonal basis of spherical wavelet-harmonic functions, designed for the new vector space integration method. This factorizes the numeric calculation into a ``vector'' that depends only on the astrophysical velocity distribution; a second vector, depending only on the detector form factor; and a scattering matrix defined on the basis functions, which depends on the details of the dark matter (DM) particle model (e.g.~its mass). For common spin-independent DM--Standard Model interactions, this scattering matrix can be evaluated analytically in the wavelet-harmonic basis. This factorization is particularly helpful for the more complicated analyses that have become necessary in recent years, especially those involving anisotropic detector materials or more realistic models of the local DM velocity distribution. With the new method, analyses studying large numbers of detector orientations and DM particle models can be performed more than 10~million times faster.

This paper derives several analytic results for the spherical wavelets, including an extrapolation in the space of wavelet coefficients, and a generalization of the vector space method to a much broader class of linear functional integrals. Both results are highly relevant outside the field of DM direct detection.

\end{abstract}

\tableofcontents

\section{Introduction}

 A direct detection scattering rate prediction requires input from three branches of physics; astrophysics, for the dark matter (DM) velocity distribution; a DM particle physics model, specifying the DM mass and the nature of its interactions with SM particles; and, from condensed matter or physical chemistry, a form factor encoding the SM physics of the detector response to a DM scattering event. 
Each of these items, especially the first two, is subject to uncertainty; propagating these uncertainties into the rate prediction for a given DM particle model may require scanning over ensembles of DM velocity distributions and SM detector response functions. At each point in the parameter space, the rate is given by an integral over the DM velocity $\vec v$, the momentum transfer $\vec q$ to the SM target, and possibly the energy deposited, $E$, with an integrand that does not necessarily have a closed form analytic expression.
The calculation must be repeated for every change to the input functions or parameters. 

For anisotropic detector materials the computational expense is particularly severe, because the DM scattering rate also depends on the orientation of the detector, necessitating scans over  elements of $SO(3)$ on top of everything else.
A rigorous, detailed analysis can be prohibitively expensive. 
To solve this problem, a companion paper~\cite{Lillard:2025ixi} introduces a vector space integration method, where the DM velocity distribution and the detector response functions are approximated by sums of carefully chosen velocity and momentum basis functions.
With some foresight, the original multidimensional numeric integral that defines the DM--SM scattering rate can be integrated analytically for every pair of basis functions, generating a scattering matrix that connects the velocity and momentum vector spaces.

There are two fundamental benefits to this approach.
First, by design, the scattering matrix depends only on the DM model parameters (i.e.~the DM particle mass and the momentum dependence of its interactions with free SM particles), while the vectorized versions of the astrophysical and material form factors can be calculated independently. 
This factorization of the scattering rate greatly simplifies any analysis that includes large sets of velocity distributions, 
whether to account for the time-dependent variations caused by the Earth's motion~\cite{Drukier:1986tm,Freese:1987wu,Collar:1992qc,Hasenbalg:1997hs,Avignone:2008cw,Lee:2015qva},
or to address intrinsic asymmetries in the galactic frame velocity distribution~\cite{Diemand:2008in,Klypin:2010qw,Guedes:2011ux,Hopkins:2017ycn,Kuhlen:2012fz,Necib:2018igl,Riley:2018lbh,Evans:2018bqy}.

Second, the choice to use spherical harmonics as part of the velocity and momentum basis functions makes rotations of the detector (or, equivalently, rotations of the DM sky) almost trivially easy to implement. Because spherical harmonics transform as representations of $SO(3)$, the action of a rotation on a function is given by matrix multiplication acting on its basis functions. 
Consequently, scans over large sets of rotations can be completed quickly, making it easy to optimize daily modulation analyses for anisotropic detectors~\cite{Catena:2015vpa,Hochberg:2016ntt,Budnik:2017sbu,Hochberg:2017wce,Coskuner:2019odd, Geilhufe:2019ndy,Griffin:2020lgd,Coskuner:2021qxo,Sassi:2021umf,Blanco:2021hlm,Blanco:2022pkt,Boyd:2022tcn,Heikinheimo:2022erq,Catena:2023qkj,Catena:2023awl,Dinmohammadi:2023amy}.
While this facility with rotations is irrelevant to isotropic detector targets~\cite{DarkSide:2018ppu,XENON:2018voc,SuperCDMS:2018mne,PICO:2019vsc,DAMIC:2019dcn,Blanco:2019lrf,SENSEI:2020dpa,EDELWEISS:2020fxc,PandaX-II:2021nsg,XENON:2022ltv,LZ:2022ufs,XENON:2023sxq,DAMIC-M:2023gxo}, the factorization of the rate integral can simplify an isotropic analysis by a substantial margin as well.

The drastically streamlined calculation makes it possible to propagate the uncertainties from astrophysics~\cite{Wu:2019nhd,Radick:2020qip,Maity:2020wic,Buckley:2022tjy,Maity:2022enp} and the detector physics~\cite{Knapen:2021run,Knapen:2021bwg,Griffin:2021znd} into the rate prediction, 
or to perform halo-independent analyses of the DM particle model parameters~\cite{Fox:2010bu,Fox:2010bz,Frandsen:2011gi,Gondolo:2012rs,Kavanagh:2013eya,Feldstein:2014gza,Chen:2021qao,Chen:2022xzi}.
A statistically significant signal of dark matter in multiple channels could even be used to measure components of the DM velocity distribution directly~\cite{Peter:2011eu,Peter:2013aha,Lee:2014cpa}.

\medskip

Spherical harmonics and other sets of orthogonal functions have been used to make many physical problems analytically tractable (e.g.~\cite{legendre}), even in the context of the DM velocity distribution~\cite{Lee:2014cpa}. 
What has so far gone unappreciated is the dramatic reduction in complexity that follows from the choice to represent the detector response function and the free DM--SM scattering operator in the same Hilbert spaces. 

This paper investigates the choices that make the vector space method so efficient, with particular attention given to the radial basis functions.
The accuracy and practicality of the vector space integration method is ultimately determined by how quickly the radial basis function expansions converge, and on whether certain intermediate integrals can be completed analytically.
Even mundane decisions, such as which rest frame to use for the velocity distributions, offer unanticipated opportunities to reduce the computation time by {additional} orders of magnitude. 

Section~\ref{sec:method} reviews the basic vector space integration method of ref.~\cite{Lillard:2025ixi}, and derives the analytic simplifications that allow the scattering rate to be written in terms of the partial rate matrix, in Section~\ref{sec:partialrate}.
 In principle, the method could be used with any orthogonal basis of radial functions, but in practice, I find that familiar options such as Fourier or Bessel functions are suboptimal in a few respects.
A better choice is presented in Section~\ref{sec:haar}: the spherical wavelets, which I designed for this application. 
The basis functions are piecewise constant, and simple enough to permit the analytic evaluation of the scattering matrix operator that appears in Section~\ref{sec:method}.

Like the Haar wavelets~\cite{haar} from which they are derived, each higher order spherical wavelet vanishes identically outside of a narrow region. In the large $n$ limit, the width of the $n$th wavelet is proportional to $1/n$. So, the projection of a function onto the wavelet-harmonic basis actually becomes easier, rather than harder, at large $n$, whereas the integrands for the Fourier or Bessel expansions are increasingly oscillatory and numerically challenging.
Section~\ref{sec:power} introduces a test for global convergence of a basis function expansion, and finds that the spherical wavelets also converge substantially faster (as a higher power of $1/n$) than the alternatives.  

Section~\ref{sec:demo} demonstrates the wavelet-harmonic integration method for a toy detector model (an idealized particle in a rectangular box) and a sample velocity distribution, comprised of a galactic DM halo and three large streams of differing widths. The demonstration culminates in a direct detection analysis combining the two models, concluding with a discussion of the computation time for each part of the calculation. 
Finally, the generalization of the wavelet-harmonic integration method is described in Section~\ref{sec:future}.  
Several other useful analytic results are provided in the appendices.

As a bonus, Section~\ref{sec:extrapolation} derives an apparently novel extrapolation procedure for wavelet transformations, where an initial set of $n \leq n_\text{max}$ coefficients is used to predict the values of a much larger set of additional $n > n_\text{max}$ coefficients, using algebraic methods. This is a unique property of wavelet transformations, not shared by other orthogonal bases (e.g.~Fourier, Legendre, Bessel, etc.). 
It relies only on the assumption that the original function is well approximated by its Taylor series within the (increasingly narrow) bases of support of the $n$th wavelets at $n \sim n_\text{max}$.
The wavelet extrapolation method can generate smooth interpolating functions and highly precise inverse wavelet transformations from a relatively small number of initial coefficients.

Section~\ref{sec:future} provides a generalization of the wavelet-harmonic integration method for problems with more than two input functions, alternative scattering operators, and/or input functions of arbitrary $d$-dimensional coordinates $\vec x_i$.
After mentioning several avenues for further development, Section~\ref{sec:implications} lists some physical properties of the partial rate matrix that impact the search for dark matter.

The figures in this paper and the timing information of Section~\ref{sec:demo} were generated using the Python implementation  VSDM (Vector Spaces for Dark Matter), available at
\begin{align*}
\texttt{https://github.com/blillard/vsdm} \, .
\end{align*}
This package can be installed via the Python Package Index (PyPI):
\begin{align*}
\texttt{pip install vsdm}\,. 
\end{align*}

\subsection{Scattering Rate}

The DM--SM scattering rate $R$ for continuum final states is given in the nonrelativistic limit by~\cite{Lewin:1995rx,Essig:2011nj,Essig:2015cda,Trickle:2019nya,Catena:2022fnk}:
\begin{align}
R &=   N_T \frac{\rho_\chi}{m_\chi} \int\! dE\,d^3q \, d^3v \, g_\chi(\vec v)\, f_S^2(\vec q, E)
\,\delta\Big( E + \frac{q^2}{2 m_\chi} - \vec{q} \cdot \vec{v} \Big) \frac{\bar\sigma_0 F_\text{DM}^2(\vec q, \vec v) }{4\pi \mu_\chi^2 } ,
\label{rate:continuum}
\end{align}
where $N_T$ is the number of SM targets (e.g.~particles, molecules, or unit cells of a crystal) in the detector; $\rho_\chi$ is the local DM mass density; $m_\chi$ is the DM mass; $\mu_\chi = m_\chi m_\text{SM} / (m_\chi + m_\text{SM})$ is the reduced mass of the DM--SM particle system, where $m_\text{SM}$ is the mass of the Standard Model particle being scattered;  and $E$ and $\vec{q}$ are the energy and momentum transferred from the DM to the SM target.
Here $f_S^2(\vec q, E)$ is a detector response function, with units of inverse energy.
$R$ is the total rate for the detector target, i.e.~the number of scattering events in the detector per unit of time.
The lab-frame DM velocity distribution $g_\chi(\vec v)$ is normalized as 
\begin{align}
\int\! d^3  v \, g_\chi(\vec v) \equiv 1 ,
\end{align} 
and $\bar\sigma_0 F_\text{DM}^2( q)$ is the free DM--SM particle scattering cross section, defined at some reference momentum transfer $q_r$:
\begin{align}
\bar\sigma_0 = \sigma( q=  q_r),
&&
F_\text{DM}(  q_r) \equiv 1.
\end{align}
The differential rate $dR/dE$ is given by:
\begin{align}
\left.\frac{dR}{dE} \right|_E &= N_T \frac{\rho_\chi}{m_\chi} \int\! d^3q \, d^3v \, g_\chi(\vec v) \,f_{S}^2(\vec q, E) 
\,\delta\Big( E + \frac{q^2}{2 m_\chi} - \vec{q} \cdot \vec{v} \Big) \frac{\bar\sigma_0 F_\text{DM}^2(\vec q, \vec v) }{4\pi \mu_\chi^2 } .
\end{align}
This is primarily relevant for experiments that can measure the energy of the outgoing SM state.

In this paper I focus on spin-independent scattering, where $F_\text{DM}(\vec q, \vec v) = F_\text{DM}(q)$ is an isotropic function of the transferred momentum, $q \equiv |\vec q|$. Spin-dependent interactions (e.g.~from a vector boson mediator) can generate additional velocity dependence~\cite{Fitzpatrick:2012ix,Catena:2015vpa}, but even in this case $F_\text{DM}$ is spherically symmetric unless the DM or the detector has polarized spins: i.e.~$F_\text{DM}(\vec q, \vec v) = F_\text{DM}(q, v, \vec q \cdot \vec v)$.

The typical convention for DM--electron scattering defines $\bar\sigma_0$ at $q_0 = \alpha m_e$, the inverse Bohr radius. For nuclear scattering the common choices are $q_0 = m_\chi v_0$, for a $v_0 \approx 220\text{--}240\,\text{km/s}$ chosen to match the assumed virial velocity $v_\sigma$ of the DM distribution; or, if the limit is well defined, $q_0 \rightarrow 0$.
For finite mediator mass $m_\phi$,\footnote{This interaction corresponds to the $\mathcal O_1$ Lagrangians of~\cite{Fitzpatrick:2012ix,Catena:2022fnk}, or the $c_1$ interactions of~\cite{Trickle:2020oki}.
In the notation of~\cite{Boyd:2022tcn}, $\mathcal F(q) = F_\text{DM}(q)$, while in~\cite{Trickle:2019nya} the same quantity is written as $\mathcal F_\text{med}(q)$. 
Ref.~\cite{Catena:2022fnk} represents $F_\text{DM}(q, v)$ as $\mathcal R_l(q, v)$, while the material response function $f_S^2(\vec q, E)$ maps onto $\mathcal W_l(\vec q, E)$.
Note that refs.~\cite{Trickle:2019nya,Hochberg:2021pkt,Boyd:2022tcn} use the letter $R$ for the rate per unit mass, $R/M_T$, where $M_T$ is the mass of the detector.
}
\begin{align}
F_\text{DM}(q) &= \frac{q_0^2 + m_\phi^2 }{q^2 + m_\phi^2} .
\end{align}
In the limiting cases of heavy or light mediator masses, $m_\phi \gg q$ and $m_\phi \ll q$, $F_\text{DM}(q) \rightarrow 1$ or $(q_0/q)^2$, respectively.

\paragraph{Scattering On Free Particles:}

If the SM targets are free particles, e.g.~isolated atomic nuclei, 
then the response function $f_S^2(q, E)$ is especially simple.
Labeling the final states by their lab-frame recoil energy $E_R$,
\begin{align}
E_R &\equiv \frac{q^2}{2 m_A },
\end{align}
with $m_\text{SM} = m_A$ the mass of the atomic nucleus, 
the total energy deposited during inelastic nuclear scattering is
\begin{align}
E &= \Delta E +  E_R = \Delta E + \frac{q^2}{2 m_A },
\label{nuclear:ER}
\end{align}
with $\Delta E$ the energy of the excited (nuclear) state. In the special case of elastic scattering, $\Delta E = 0$.  

In both scenarios, $f_S^2(\vec q, E)$ is simply a $\delta$ function that enforces \eqref{nuclear:ER}:
\begin{align}
f_S^2(\vec q, E) &= \delta\!\left( \frac{q^2}{2 m_A} + \Delta E - E \right) ,
\label{fS2:nuclear}
\\
\left. \frac{dR}{dE_R} \right|_{E_R} &= \frac{N_T \rho_\chi m_A}{2 m_\chi \mu_\chi^2} \int\frac{d^3 v}{v} g_\chi(\vec v) \Theta( v - v_\text{min}) \, \bar\sigma_0 F_\text{DM}^2(E_R) ,
\end{align}
where $v_\text{min} = E/q + q/(2 m_\chi)$, $E = \Delta E + E_R$, and $E_R = q^2/( 2 m_A)$. 
This $v_\text{min}$ appears in many other contexts: here, it reduces to
\begin{align}
v_\text{min}(E_R) &= \frac{\sqrt{ 2 m_A E_R}}{2 \mu_\chi} + \frac{ \Delta E}{\sqrt{ 2 m_A E_R}}.
\end{align}
See e.g.~\cite{Lewin:1995rx,Fitzpatrick:2012ix,10.1093/ptep/ptaa104} for a review.

\paragraph{Scattering On Interacting States:}

If the initial and/or final SM states are not momentum eigenstates, then \eqref{fS2:nuclear} no longer applies. Instead, the momentum form factor $f_{S}^2(\vec q, E)$ is related to the overlap between the initial and final state wavefunctions, given some transfer of momentum $\vec{q}$. In terms of the dynamic structure factor $S(\vec q, \omega)$ of ref.~\cite{Trickle:2019nya},
\begin{align}
f_S^2(\vec q, \omega) &\equiv   \frac{V_\text{cell}}{2\pi}  S(\vec q, \omega) 
=  \frac{ V_\text{cell}}{V_T} \sum_f |\langle f | \mathcal O_{\vec q} | 0 \rangle |^2 \,  \delta(\omega - \Delta E_f  ) .
\label{fs2:structurefactor}
\end{align}
 The operator $\mathcal O_\vec{q}$ applies the momentum transfer from the DM to the final state $\ket{f}$, and $\Delta E_f$ is the difference in energies between the $\ket{f}$ and $\ket{0}$ states. 
 $V_T$ is the total volume of the detector target, which arises in~\cite{Trickle:2019nya} from the normalization of the continuum wavefunctions $\ket{f}$. 
 For position space wavefunctions $\ket{0} = \phi_0(\vec r)$, $\mathcal O_\vec{q} = e^{i \vec q \cdot \vec r}$.
Note that \eqref{fs2:structurefactor} can be applied to DM--$e^-$ or DM--nuclear scattering, e.g.~for DM-induced phonon production.

In the dark photon formalism of ref.~\cite{Hochberg:2021pkt,Boyd:2022tcn}, the DM--$e^-$ scattering rate can be expressed
in terms of the dielectric loss function of the electron, $\epsilon(\vec q, \omega)$:
\begin{align}
f_S^2(\vec q, \omega) &= \frac{V_\text{cell}  q^2 }{4\pi^2} \frac{1}{\alpha} \text{Im}\left( \frac{-1}{\epsilon(\vec q, \omega) } \right) .
\end{align}
In these expressions, $V_\text{cell}$ is the volume of the crystal cell or a single particle target, so that $V_T/V_\text{cell} = N_T$ is the number of SM scattering targets in the detector. Alternatively, for a fluid of mass density $\rho_T$, $V_\text{cell} = m_\text{cell} / \rho_T$, where $m_\text{cell}$ is the mass of a single scattering target, $ M_T / m_\text{cell} = N_T$. 

These normalization choices for $f_S^2$ ensure that all of the material properties, including the detector target density, are contained within $f_S^2$. 
It is an intensive quantity (i.e.~it is independent of the detector size), with units of inverse energy, representing a scattering probability within one unit cell (or a single SM particle) of the target.
While the normalization of $f_S^2$ depends on the definition of the unit cell, the combination $N_T f_S^2$ is independent of this choice.

\paragraph{Discrete Final States:}

For discrete final state energies $\Delta E_i$, the rate for inducing a $\ket{0} \rightarrow \ket{s}$ excitation in the material is given by:
\begin{align}
R_s = N_T \frac{\rho_\chi}{m_\chi} \int\! d^3q \, d^3v \, g_\chi(\vec v) \,f_{s}^2(\vec q) 
\,\delta\Big( \Delta E_s + \frac{q^2}{2 m_\chi} - \vec{q} \cdot \vec{v} \Big) \frac{\bar\sigma_0 F_\text{DM}^2(\vec q, \vec v) }{4\pi \mu_\chi^2 } ,
\label{rate:discrete}
\end{align}
where $\Delta E_s = E_s - E_0$ is the difference in energy between the initial and final SM states,
and where 
\begin{align}
f_{S}^{2}(\vec q, E) &= f_s^2(\vec q) \, \delta(E - \Delta E_s).
\label{compare:fSfs}
\end{align}
This $f_s^2(\vec q)$ form factor is dimensionless.

For spin-independent DM scattering off of single-particle SM states $0\rightarrow s$, the momentum form factor is given simply in terms of the wavefunctions of the ground state $\ket{0}$ and excited state $\ket{s}$:\footnote{In ref.~\cite{Catena:2022fnk}, some spin-dependent interactions couple to a vectorial form factor $\vec f_s(\vec q)$, where $\mathcal O_{\vec q} = e^{i \vec q \cdot \vec r} \nabla_{\vec r}$ acts on the position-space ground state wavefunction in \eqref{def:fgs}. The \eqref{def:fgs2} definition of $f_s^2(\vec q)$ corresponds to $\mathcal B_1$ in ref.~\cite{Catena:2022fnk}, while $\mathcal B_{2,3,4}(\vec q)$ include scalar products of $\vec q \cdot \vec f$ or $\vec f \cdot \vec f^\star$.
}
\begin{align}
f_{s}(\vec q) & \equiv  \left\langle s \Big| \hat{\mathcal O}_{\vec q} \Big| 0 \right\rangle
=  \int \! d^3 r \, \psi_s^\star(\vec r) \, e^{i \vec q \cdot \vec r } \, \psi_0(\vec r) 
=  \int \! d^3 k \, \tilde\psi_s^\star(\vec k + \vec q) \tilde\psi_0(\vec k)  ,
\label{def:fgs}
\\
f_s^2(\vec q) & \equiv \left| f_{s}(\vec q)  \right|^2,
\label{def:fgs2}
\end{align}
for position-space or momentum-space wavefunctions $\psi$ and $\tilde\psi$, with normalization $\int d^3r\, \psi^\star \psi = \int d^3 q \, \tilde\psi^\star \tilde \psi = 1$.
The total rate for one observable may include contributions from multiple final states, in which case:
\begin{align}
R_\text{tot} = \sum_s R_s.
\end{align}
This is common in scintillators (a.k.a.\ fluorescent dyes)~\cite{Blanco:2019lrf,Blanco:2021hlm}.

\subsection*{Detector Rotations}

Thanks to the $\vec q \cdot \vec v$ term in the energy-conserving $\delta$ function, the continuous and discrete versions of the scattering rate both depend on the orientation of the detector with respect to the DM velocity distribution. If the detector is rotated by some $\mathcal R \in SO(3)$, the new scattering rate can be calculated by applying the rotation operator to the momentum form factor,  $f_S^2(\vec q, E) \rightarrow \mathcal R \cdot f_S^2(\vec q, E) = f_S^2(\mathcal R^{-1} \vec q, E)$, or to the DM velocity distribution, $g_\chi(\vec v) \rightarrow \mathcal R^{-1} \cdot g_\chi(\vec v) =  g_\chi(\mathcal R  \vec v)$. This can be represented compactly as
\begin{align}
R(\mathcal R) = R\big(g_\chi f_S^2 \longrightarrow g_\chi \cdot \mathcal R \cdot f_S^2 \big),
\label{rotation:rate}
\end{align}
for $g_\chi$ and $f_S^2$ defined in some initial orientations, where the rotation operator acting to the left acts as the inverse, $g_\chi \cdot \mathcal R = \mathcal R^{-1} \cdot g_\chi$.

\subsection{Review of the Standard Approach} \label{sec:standard}

Evaluating \eqref{rate:continuum} or \eqref{rate:discrete} requires a six- or seven-dimensional integral, which must be repeated for every dark matter mass $m_\chi$ and form factor $F_\text{DM}(q)$, and for every orientation of the detector $\mathcal R$. 
If an analytic halo model for $g_\chi$ is available, however, the integrand can be simplified by performing the velocity integral first.
The usual derivations assume either than the material form factor is isotropic ($f_S^2(\vec q, E) = f_S^2(q, E)$) or that the DM velocity distribution is isotropic in the galactic rest frame. 
When neither condition is satisfied, the integrated forms of the velocity distribution cannot be used to simplify the scattering calculation.

\medskip

If the detector material has an isotropic response, $f_S^2(\vec q, E) = f_S^2(q, E)$, then the velocity integral can be separated from the rest of the problem by performing an angular average over $\vec q$.
Defining
\begin{align}
\eta(q, E) &\equiv  2 q \int\! \frac{d \Omega_q d^3 v}{4\pi}  g_\chi(\vec v) \, \delta\!\left( E + \frac{q^2}{2m_\chi} - \vec q \cdot \vec v \right)
\label{eta:def}
\end{align}
and completing the $d\Omega_q$ angular average, this $\eta(q, E)$ becomes: 
\begin{align}
\eta(q, E, m_\chi) &= \int\! \frac{d^3 v}{v} \, g_\chi(\vec v) \,\Theta\!\left( v - v_\text{min}(q, E, m_\chi) \right),
\label{eta:altdef}
\\
v_\text{min}(q, E, m_\chi) &\equiv \frac{E}{q} + \frac{q}{2 m_\chi}.
\label{vmin}
\end{align}
This $v_\text{min}$ has a physical meaning: it is the smallest speed for which a DM particle with mass $m_\chi$ is still able to deposit energy $E$ in the detector through elastic scattering. (For inelastic DM, e.g.~\cite{Bramante:2016rdh}, this expression is modified to account for the energy required to excite the DM final state.)
\eqref{eta:altdef} is often taken as the starting point for rate calculations with isotropic detectors, with \eqref{rate:continuum} simplified to
\begin{align}
R &= \frac{N_T \rho_\chi \bar\sigma_0}{ 8\pi m_\chi \mu_{\chi}^2 } \int\! dE \frac{4\pi q^2 dq}{q} \eta(v_\text{min})\,  F_\text{DM}^2(q)\, f_S^2( q, E) ,
\end{align}
and \eqref{eta:altdef} defining $v_\text{min}$ and $\eta$. 

If the material response $f_S^2(\vec q, E)$ is not isotropic, then it is not appropriate to factorize the $d\Omega_q$ angular average in \eqref{eta:def}.
Instead, let us define a related function of 3d momentum,  $\eta(\vec q, E)$:
\begin{align}
\eta(\vec q, E) &\equiv 2q  \int d^3 v \, g_\chi(\vec v) \, \delta\!\left(E + \frac{q^2}{2 m_\chi} - \vec q \cdot \vec v \right),
\label{eta:vecdef}
\\
R(\mathcal R) &= \frac{N_T \rho_\chi \bar\sigma_0}{8\pi m_\chi \mu_\chi^2} \int\! dE \frac{d^3 q}{q}  F_\text{DM}^2(q)\,  \eta(\vec q, E) \cdot \mathcal R \cdot f_S^2(\vec q, E) .
\label{rate:gq}
\end{align}
Note that the angular average of this $\eta(\vec q, E)$ reproduces the more familiar $\eta(v_\text{min})$ of \eqref{eta:def}, 
and that both versions of $\eta$ have units of inverse speed.\footnote{For comparison with the $g(\vec q, E)$ of ref.~\cite{Trickle:2019nya}, note that $g(\vec q, E) = (\pi/q) \, \eta(\vec q, E)$.}
The only loss of generality between this expression and \eqref{rate:continuum} is the assumption that $F_\text{DM}$ is velocity independent. This is a valid assumption for spin-independent DM--SM couplings. For spin-dependent scattering with unpolarized DM and SM spins, the isotropic $F_\text{DM}(q, v)$ can be expanded in powers of $v$, 
\begin{align}
F_\text{DM}^2(q, v) = \sum_{\beta} c_\beta v^\beta F_{\beta}^{2}(q) , 
\end{align}
and the factor of $v^\beta$ can be moved into an associated integrated velocity distribution $\eta_j$, 
\begin{align}
\eta^{(\beta)} (\vec q, E) &\equiv 2q  \int d^3 v \, v^\beta \, g_\chi(\vec v) \, \delta\!\left(E + \frac{q^2}{2 m_\chi} - \vec q \cdot \vec v \right).
\end{align}

\medskip 

If a model for the DM velocity distribution is isotropic in some rest frame, then $\eta(\vec q, E)$ can be simplified via an angular average over the rest frame velocities. 
With $\vec v_E$ the lab velocity in the rest frame of the galactic halo, 
\begin{align}
g_\chi^\text{gal}(v) = g_\chi^\text{lab}(\vec v - \vec v_E),
\label{eq:gXiso}
\end{align}
$\eta(\vec q, E)$ can be expressed as a 1d integral in the rest frame:
\begin{align}
\eta(\vec q, E, \vec v_E, m_\chi) &= \int_{v_-^2(\vec q, E, \vec v_E, m_\chi)}^\infty \! 2\pi \, dv^2\, g_\chi^\text{gal}(\sqrt{v^2} )  ,
\label{etaveciso}
\\
v_-(\vec q, E, \vec v_E, m_\chi) &\equiv \frac{E}{q} + \frac{q}{2 m_\chi} + \frac{\vec q \cdot \vec v_E}{q} .
\label{vminus}
\end{align}
If $g_\chi^\text{gal}(v)$ has an analytic expression, it may be possible to derive a closed form expression for $\eta(\vec q, E)$. 

Note that unlike $g_\chi(\vec v)$, the integrated velocity distributions $\eta(q, E)$ and $\eta(\vec q, E)$ depend on the DM mass parameter, $m_\chi$; so, if \eqref{eta:altdef} or \eqref{etaveciso} must be evaluated numerically, then the calculation must be repeated for every $m_\chi$.

\paragraph{Standard Halo Model (SHM):}

The typical  simplified model of $g_\chi$ is a Maxwell--Boltzmann distribution truncated sharply at the galactic escape velocity $v_\text{esc}$,
\begin{align}
g_\chi(\vec v) &\approx \frac{1}{N_0} \exp\left( - \frac{|\vec v + \vec v_E |^2 }{v_\sigma^2 } \right) \Theta(v_\text{esc} - |\vec v + \vec v_E| ),
\label{gX:MB}
\end{align}
where $N_0$ enforces the normalization $\int d^3 v \, g_\chi \equiv 1$, and 
with dispersion velocity $v_\sigma \approx 220\text{--}240\, \text{km/s}$ and $v_\text{esc} \approx 544\text{--}600\,\text{km/s}$~\cite{Lewin:1995rx,Baxter:2021pqo}.
The normalization condition $\int d^3 v \, g_\chi \equiv 1$ sets the value of $N_0$ in terms of the other parameters.

From \eqref{eq:gXiso},  $\eta(\vec q, E)$ becomes:
\begin{align}
\eta(\vec q, E, \vec v_E, m_\chi) &= \frac{2\pi v_\sigma^2 }{ N_0} \left( e^{- v_-^2/v_\sigma^2 } - e^{- v_\text{esc}^2/v_\sigma^2 } \right) \Theta(v_\text{esc} - v_-(\vec q, E, \vec v_E)),
\label{gq:MB}
\end{align}
where $v_-(\vec q, E, \vec v_E, m_\chi)$ is defined in \eqref{vminus}.

\paragraph{Rotations:}

As noted below \eqref{rotation:rate}, the rotation operator $\mathcal R$ can be applied to the integrated velocity distribution $\eta$, rather than to the detector form factor $f_S^2$:
\begin{align}
R(\mathcal R) &= \frac{N_T \rho_\chi \bar\sigma_0}{8\pi m_\chi \mu_\chi^2} \int\! dE \frac{d^3 q}{q}  F_\text{DM}^2(q)\,  f_S^2(\vec q, E)\cdot \mathcal R^{-1} \cdot   \eta(\vec q, E, \vec v_E),
\label{rate:SHM}
\\
\mathcal R^{-1} \cdot   \eta(\vec q, E, \vec v_E ) &= \eta(\vec q, E,  \mathcal R \vec v_E ) .
\end{align}
Because of the azimuthal symmetry of the  lab frame SHM velocity distribution $g_\chi(\vec v)$ with respect to rotations about the $\vec v_E$ axis, a 3d scan over $\mathcal R \in SO(3)$ can be reduced to a simpler 2d scan over the sphere, $\mathcal R \vec v_E \in S^2$.

In less idealized models of the velocity distribution, it may still be possible to write down a closed form version of $\eta(\vec q, E)$: e.g.~if $g_\chi(\vec v)$ is a sum of components that would be isotropic in their own individual rest frames. An example of this type appears in Section~\ref{sec:demo}. For more complicated $g_\chi(\vec v)$, e.g.~derived from galactic-scale simulations or inferred from astronomical data, the integrated velocity distribution $\eta$ would need to be evaluated numerically for every value of the DM mass $m_\chi$.

\subsection{A Need for Factorization}

In a generic anisotropic analysis, \eqref{rate:gq} must be evaluated for every detector orientation $\mathcal R \in SO(3)$ as well as for every value of the parameters $m_\chi$, $\vec v_E$, and for each version of $F_\text{DM}^2(q, m_\phi)$. 
If both $\eta(\vec q, E)$ and $f_S^2(\vec q, E)$ have closed-form analytic expressions, it might still be possible to integrate \eqref{rate:gq} analytically for every orientation $\mathcal R$. In this best case scenario, $R(m_\chi, F_\text{DM}, \mathcal R, g_\chi, f_S^2)$ has a closed form solution, and no further numerical improvement is needed. 
Only in the simplest cases (e.g.~nuclear scattering in isotropic noble gases) is this generally possible. 

More often, $f_S^2(\vec q, E)$ is obtained numerically, and \eqref{rate:gq} requires repeated numerical integration over every combination of variables, parameters, and orientations, which is a very time-consuming proposition.
Such a calculation can also be very memory-intensive: if an accurate discretization of $f_{S}^2(\vec q, E)$ requires $10^2$--$10^3$ values along each axis, then each 3d or 4d grid would include, respectively, $10^6$--$10^9$ or  $10^8$--$10^{12}$ points.

Compounding the difficulty of the analysis, it may be necessary to include multiple lab-frame velocity distributions: e.g.~to account for the $\pm 15 \, \text{km/s}$ change in the Earth velocity over the course of the year, or to account for unvirialized components of $g_\chi(\vec v)$ not included in the SHM~\cite{Diemand:2008in,Klypin:2010qw,Guedes:2011ux,Hopkins:2017ycn,Kuhlen:2012fz,Necib:2018igl,Riley:2018lbh}.
The impact of such components on the scattering rate can be substantial, in some cases larger than the expected annual variation~\cite{Wu:2019nhd,Radick:2020qip,Maity:2020wic,Buckley:2022tjy,Maity:2022enp}.

To simultaneously test multiple ensembles of $g_\chi^\text{gal}$ or $f_S^2$ distributions, to quantify the modeling uncertainties or to compare different physical cases, 
the integral \eqref{rate:gq} must be repeated
\begin{align}
N_\text{integrals} =   N_{\mathcal R} \times N_\text{DM} \times N_{g_\chi} \times N_{f_S}
\label{counting:old}
\end{align}
many times, where 
$N_{\mathcal R}$ is the number of detector orientations;
$N_\text{DM}$ is the number of DM particle models, i.e.~the number of $m_\chi$ values times the number of $F_\text{DM}$ form factors to test;
$N_{g_\chi}$ is the number of lab-frame DM velocity distributions $g_\chi$, i.e.~the number of values of the Earth velocity $\vec v_E(t)$ times the number of distinct galactic-frame DM distributions;
and $N_{f_S}$ is the number of detector form factors $f_S^2$ to be tested.

For $g_\chi$ models with no analytic description, we must also evaluate $\eta(\vec q, E, \vec v_E, m_\chi)$ for every relevant combination of variables and parameters. This can be a substantial addition to the computational difficulty: \eqref{eta:vecdef} must be evaluated
\begin{align}
\Delta N_\text{integrals} = N_\text{DM} \times N_{g_\chi} \times n_\text{grid points}
\end{align}
many times, where $n_\text{grid points}$ counts the number of $(\vec q, E)$ points at which $\eta(\vec q, E)$ is to be evaluated.
Depending on the size of $n_\text{grid points}$, it may be faster to evaluate $R$ directly from \eqref{rate:continuum} in this case, skipping the intermediate step that defines $\eta$.

\medskip

The method presented in ref.~\cite{Lillard:2025ixi} saves us from having to integrate anything so many times. Rather than evaluating $\eta(\vec q, E)$ and $f_S^2(\vec q, E)$ on a grid of points, the new method projects $f_S^2(\vec q, E)$ and the original velocity distribution $g_\chi(\vec v)$ onto bases of orthogonal functions.
The $dE\, d^3 q \, d^3v$ integral of \eqref{rate:continuum} is replaced by integrals over the basis functions, producing a matrix that acts on the $\vec v$ and $(\vec q, E)$ vector spaces.
To evaluate the scattering rate, one simply multiplies this matrix by the vectorized versions of $g_\chi$ and $f_S^2$,
an action that must be repeated
\begin{align}
N_\text{matrix products}^{\text{(new)}} =  N_\text{integrals}^\text{(old)}  
\label{vectorsarethenewblack}
\end{align}
many times, following \eqref{counting:old}. Matrix multiplication is \emph{extremely} fast compared to multidimensional numeric integration, especially if the matrices in question are not particularly large. The reduction in computation time can be extreme, particularly if the basis functions utilize spherical harmonics.

Sections~\ref{sec:method} and~\ref{sec:haar} outline the technical details of the method and introduce the piecewise-constant orthogonal ``spherical Haar wavelet'' radial basis functions, for which the remaining $dE \, d^3 q \, d^3v$ integrals can be completed analytically. 
Now, the only integrals that may need to be performed numerically are the projections of $g_\chi$ and $f_S^2$ onto the vector spaces spanned by the Section~\ref{sec:haar} basis functions.
Supposing accurate reconstructions for $g_\chi$ and $f_S^2$ require $N_v$ and $N_q$ coefficients, respectively, the number of integrals to be completed is
\begin{align}
N_\text{integrals}^\text{(new)} = N_v \times N_{g_\chi} +  N_q \times N_{f_S}.
\end{align}
Section~\ref{sec:demo} provides a detailed study of the convergence evaluation time for each part of the calculation, for a simple particle-in-a-box model for $f_S^2(\vec q)$ and a complicated $g_\chi(\vec v)$ distribution, finding that $N_{v,q} \sim 10^3$ provides sufficiently good accuracy in these nontrivial examples.

\medskip 

By replacing the direct evaluation of  \eqref{rate:continuum} or \eqref{rate:SHM} with a much faster rate calculation, the vector space method overcomes one impediment to studies of anisotropic detector materials and/or velocity distributions. 
These analyses have a pressing need for efficiency: the study of daily modulation in crystalline trans-stilbene in ref.~\cite{Blanco:2021hlm}, for example, could only test $N_{\mathcal R} = 48$ detector orientations, enough to provide hourly snapshots of $R(t)$ starting from two different initial orientations.
Even when using the simple SHM velocity distribution \eqref{gq:MB} with a single value of $\vec v_E$, it was too computationally expensive to scan over the much larger $N_{\mathcal R}$ that would have been needed in order to optimize the crystal orientation.

\begin{figure}
\centering
\includegraphics[width=0.98\textwidth]{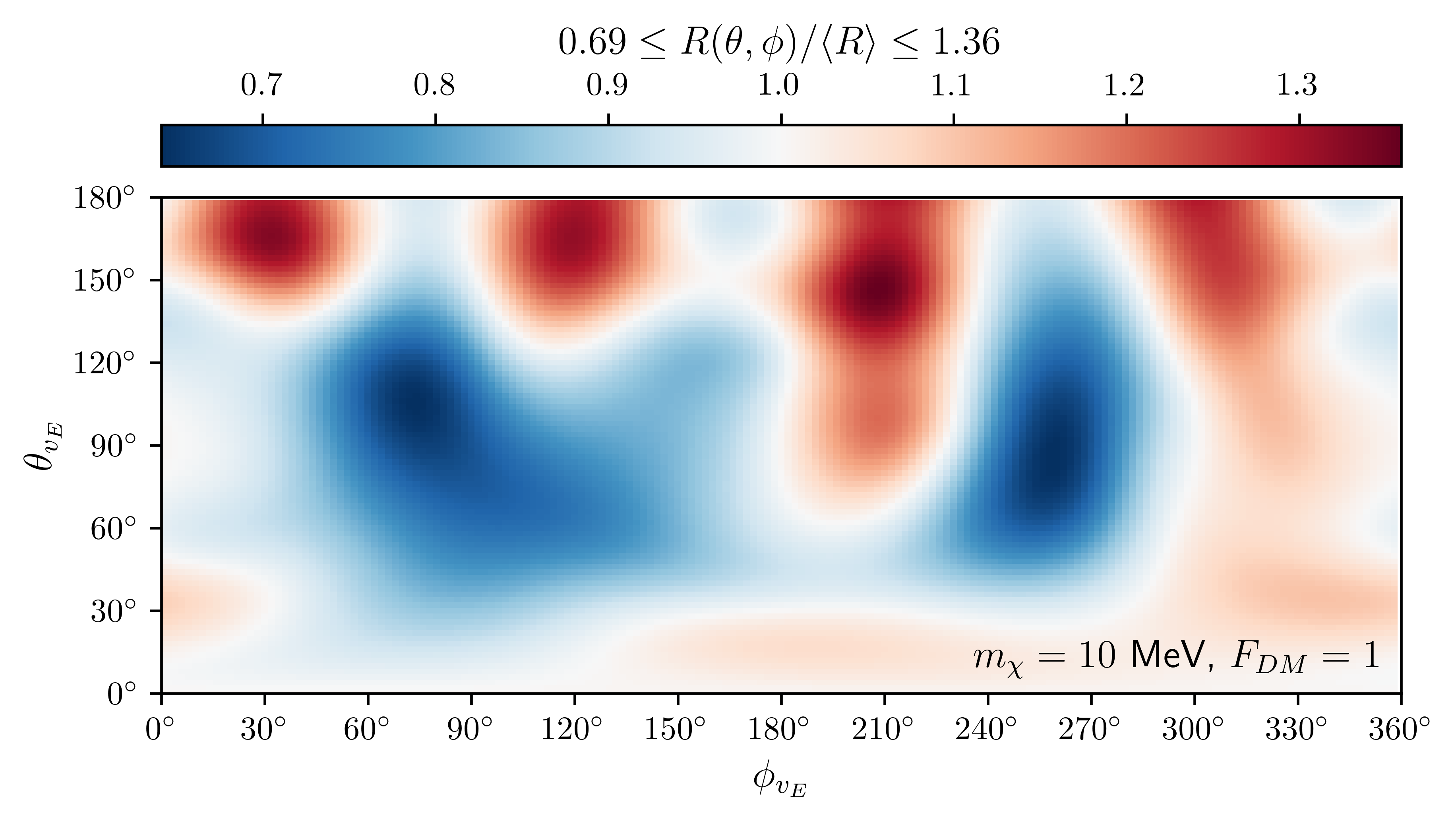}
\caption{Example of a calculation that would have been extremely difficult without the vector space integration method. Here the scattering rate $R(\mathcal R)$ is shown as a function of detector orientation $\mathcal R(\theta, \phi)$, for a toy detector model and velocity distribution presented in Section~\ref{sec:demo}, with a $m_\chi = 10$\,MeV, $F_\text{DM} = 1$ DM candidate. In the coordinate frame of the detector, the $\theta = \phi = 0$ orientation aligns the Earth velocity $\vec v_E$ with the $\hat z$ direction, and $\mathcal R(\theta, \phi)$ rotates the detector about an axis in the $xy$ plane until $\vec v_E$ points towards the spherical coordinate $(\theta, \phi)$ in the detector frame.}
\label{fig:ratedemo}
\end{figure}

The method presented in this paper makes it essentially trivial to scan over large numbers (e.g. $N_{\mathcal R} \gg 10^4$)  of detector orientations, completely solving the problem encountered in ref.~\cite{Blanco:2021hlm}.
Figure~\ref{fig:ratedemo} shows the kind of plot that must be made in order to find the optimal orientations of an anisotropic detector. 
In the simplest analysis with the SHM velocity distribution, the azimuthal symmetry of $g_\chi(\vec v)$ makes the space of orientations effectively two-dimensional: for each value of $|\vec v_E|$, the scattering rate is a function of the two angular coordinates of $\vec v_E$ in the coordinate frame of the detector.  
For asymmetric $g_\chi(\vec v_E)$, one must also specify the rotation angle $\beta$ about the $\hat v_E$ axis. Figure~\ref{fig:ratedemo} shows $R(\mathcal R)$ for a 2d slice of $SO(3)$ at fixed $\beta$, for some toy examples of $g_\chi(\vec v)$ and $f_s^2(\vec q)$ presented in Section~\ref{sec:demo}.
Even assembling this two-dimensional picture required evaluating $R(\mathcal R)$ on a grid of 16200 orientations $\mathcal R$, a task that would be essentially impossible without the method presented in this paper. 

\medskip 

The as-yet-unmeasured DM velocity distribution is subject to substantial uncertainty as well as annual variation in $|\vec v_E|$, which (unlike the uncertainty in the local DM density $\rho_\chi$) cannot be accommodated by simply rescaling the total rate. 
Studies of nuclear~\cite{Wu:2019nhd} and electronic~\cite{Radick:2020qip} DM scattering find that variations in the escape velocity and the velocity dispersion can alter the event rate by factors of $\mathcal O(100\%)$, or even by orders of magnitude for lighter DM candidates with masses close to the kinematic threshold of the detector.

Despite the potentially large effects, it is not currently a standard practice to include any estimate of these systematic uncertainties in projections for direct detection scattering rates.
With the factorized rate calculation presented here, it is quite easy to scan over large numbers of DM velocity profiles, to vary the SHM parameters or to account for likely deviations from the SHM (e.g.~\cite{Evans:2018bqy}). 
Similarly, the vectorized rate calculation makes it much simpler to conduct annual modulation analyses, once every version of the lab-frame $g_\chi(\vec v, \vec v_E)$ has been projected onto the basis of orthogonal velocity functions. 

\medskip

Finally, important insights into the experimental design can be obtained directly from the representation of $g_\chi$, $f_S^2$ and $F_\text{DM}^2$ in spherical harmonics.
As discussed in ref.~\cite{Lillard:2025ixi} and Section~\ref{sec:kinematrix}, if the DM--SM free particle cross section is spherically symmetric, $F_\text{DM}(\vec q, \vec v) = F_\text{DM}(q, v, \vec q \cdot \vec v)$, then the DM--SM scattering does not mix different harmonic modes---so, the $\ell = 1$ harmonics of $g_\chi(\vec v)$ couple only to the matching $\ell = 1$ harmonics of the detector form factor, for example.
Consequently, target materials with central-inversion-symmetric $f_S^2(- \vec q )  = f_S^2( \vec q )$ will be insensitive to the leading-order $\ell = 1$ anisotropies in the lab frame $g_\chi(\vec v)$, because these detector form factors have support only at even $\ell$. 
This point has not been previously appreciated. The $\mathcal O(20\%)$ daily modulation in trans-stilbene~\cite{Blanco:2021hlm}, for example, is due to the $\ell = 2, 4, \ldots$ harmonic modes of the DM velocity distribution, leaving open the possibility that a less symmetric target material might have an even larger modulation amplitude.

\section{Orthogonal Basis Functions} \label{sec:method}

The fundamental idea behind the vector space integration method is to represent $g_\chi$ and $f_S^2$ as sums of basis functions, 
\begin{align}
\phi_{n \ell m}(\vec v) &\equiv r_n(v) Y_{\ell m}(\hat v) ,
&
\varphi_{j n \ell m}(E, \vec q) &\equiv \tilde r_{jn}(E, q) Y_{\ell m}(\hat q) ,
\label{eq:basisfunctions}
\end{align}
where $Y_{\ell m}$ are the real-valued spherical harmonics, and $r_n$ and $\tilde r_{jn}$ are some sets of orthogonal ``radial'' functions, of one or two dimensions respectively (where $v \equiv |\vec v|$,  $\hat v \equiv \vec v/ v$, and $\vec q = q \hat q$). 
In this language, transitions to discrete final states would be represented by 1d radial basis functions $r_n(q)$, i.e.~$R_{jn}(E, q) = r_n(q) \,\delta(E - \Delta E_j)$.
These basis functions are chosen according to two criteria.
First, they should provide relatively compact representations of the functions $g_\chi(\vec v)$ and $f_S^2(\vec q, E)$, so that the sum
\begin{align}
f(\vec u) &\simeq \sum_{n\ell m} c_{n\ell m} \phi_{n\ell m}(\vec u) 
\end{align}
reproduces the original function $f(\vec u)$ to the required accuracy after including a finite number of terms $N_\text{coeffs}$ in the sum. 
Second, the basis functions $\phi$ should be simple enough that \eqref{rate:continuum} can be integrated analytically, after the substitutions $g_\chi(\vec v) \rightarrow \phi_{nlm}(\vec v)$ and $f_S^2(\vec q, E) \rightarrow \varphi_{jnlm}(\vec q)$. 
The penultimate goal is to calculate the ``partial rate matrix'' $K^{(\ell)}_{m m'}$ of ref.~\cite{Lillard:2025ixi} for every pair of spherical harmonic modes $(\ell m)$ and $(\ell m')$, 
for every DM model, form factor, and velocity distribution in the analysis. Ultimately, the scattering rate for any detector orientation $\mathcal R$  is given by a product of $K^{(\ell)}$ with the $(2 \ell + 1) \times (2 \ell + 1)$ matrix representation of the rotation operator $\mathcal R \in SO(3)$.

Real spherical harmonics are defined in terms of the complex $Y_\ell^m$ as follows:
\begin{align}  
Y_{\ell m}(\Omega) &\equiv
\left\{ \begin{array}{l c c}
\sqrt{2} (-1)^m \text{Im}\, Y_{\ell}^{|m|}(\Omega)  
&& \text{for } m < 0,
\\[\medskipamount]
Y_\ell^0(\Omega)  && \text{for } m = 0,
\\[\medskipamount]
\sqrt{2} (-1)^m \text{Re}\,Y_\ell^m (\Omega) 
&& \text{for } m > 0.
\end{array}  
\right.
\label{Ylm:realdef}
\end{align}
Spherical harmonics of fixed $\ell$ (with $- \ell \leq m \leq \ell$) transform as the $(2\ell+1)$ dimensional representations of $SO(3)$. So, for an $f_{S}^2$ or $g_\chi$ expanded in this basis, the rotation operator $\mathcal R$ of \eqref{rotation:rate} can be written as a matrix acting only on the $m$ indices of $\phi_{n\ell m}$ or $\psi_{jn\ell m}$.

Any set of basis functions would permit the factorization of the rate calculation into $\{g_\chi\} + \{f_S^2\} + \{F_\text{DM}^2, m_\chi\}$, but 
the choice to use spherical harmonics greatly simplifies the action of rotations on the basis functions.
Section~\ref{sec:haar} will show that the piecewise-constant ``spherical Haar wavelets'' are an ideal choice for $r_n(u)$: they are sufficiently simple for the integrals, and they converge in a well-defined way in the large $n$ limit for arbitrary functions $f(\vec u)$. 
This section focuses on the simplifications that follow from the choice to use spherical harmonics as part of the basis functions, so until Section~\ref{sec:haar} the form of $r_n$ and $\tilde r_{jn}$ will be left unspecified.

\medskip

For conciseness, I use a bra/ket notation to represent vectors and inner products on the vector spaces: 
\begin{align}
g_\chi(\vec v) &\equiv \ket{g_\chi}=  \sum_{n \ell m} \langle  \phi_{n \ell m} | g_\chi  \rangle \ket{ \phi_{n \ell m} }  ,
\\
f_{S}^2(\vec q, E)  &\equiv \ket{f_S^2} =  \sum_{j n \ell m} \langle  \varphi_{j n \ell m} |  f_{S}^2 \rangle \ket{ \varphi_{j n \ell m} } ,
\label{def:projection}
\end{align}
where $\ket{\phi}$ and $\ket{\varphi}$ are the basis functions defined in \eqref{eq:basisfunctions}. 
When the context is clear, I may abbreviate either type of basis vector by its indices, i.e.~$\ket{\phi_{n \ell m } } \rightarrow \ket{n \ell m}$, $\ket{\varphi_{jn\ell m}} \rightarrow \ket{jn\ell m}$. In other contexts, especially when summing over complete sets of basis vectors, I suppress the $(n \ell m)$ index in $\phi$ or $\varphi$ instead, e.g.~by writing $g_\chi = \sum_{\phi} \langle g_\chi | \phi \rangle \bra{\phi}$.
For real-valued functions the inner products are symmetric, $\langle f | g \rangle = \langle g | f \rangle$, and I draw no distinction between ``bra'' and ``ket'' vectors. 

The inner products in \eqref{def:projection} and the normalizations of the basis functions are defined as
\begin{align}
\langle g_\chi | \phi \rangle &= \int \frac{d^3 v}{v_0^3} \, \phi(\vec v) g_\chi(\vec v) ,
&
\langle\phi' |\phi \rangle &\equiv \delta_{\phi', \phi} ,
\\
\langle \varphi | f_S^2 \rangle &= \int \frac{dE}{E_0} \frac{d^3 q}{q_0^3} \, \varphi(E, \vec q) f_S^2(E, \vec q) ,
&
\langle\varphi' | \varphi \rangle &\equiv \delta_{\varphi', \varphi} .
\label{eq:inner}
\end{align}
Here, $v_0$, $q_0$, and $E_0$ are respectively an arbitrary reference velocity, momentum, and energy, introduced here to make $\phi$ and $\varphi$ dimensionless. With this convention, a function $f$ and its vector representation $\ket{f}$ have the same units; for $g_\chi$ and $f_S^2$ the units are $(\text{velocity})^{-3}$ and $(\text{energy})^{-1}$, respectively. 
Combining \eqref{eq:basisfunctions} with \eqref{eq:inner}, the  normalizations of the radial functions $r_n(v)$ and $R_{jn}(E, q)$ should satisfy:
\begin{align}
\int_0^\infty \! \frac{v^2 dv}{v_0^3}\, r_m(v) r_n(v) = \delta_{mn},
&&
\int_0^\infty \! \frac{dE}{E_0} \frac{q^2 dq}{q_0^3}\, \tilde r_{j n}(E, q) \tilde r_{j'n'}(E, q) = \delta_{j'j} \delta_{n'n}.
\label{eq:radnorm}
\end{align}
Lastly, the basis vectors $\phi$ and $\varphi$ have completeness relations:
\begin{align}
v_0^3 \,\delta^{(3)}(\vec v - \vec v') &= \sum_{\phi} \ket{\phi} \bra{\phi} = \sum_{n = 0}^\infty \sum_{\ell = 0}^\infty \sum_{m = - \ell }^\ell \phi_{n \ell m}(\vec v) \phi_{n \ell m}(\vec v'),
\\
E_0 q_0^3  \, \delta(E - E')\, \delta^{(3)}(\vec q - \vec q') &= \sum_{\varphi} \ket{\varphi} \bra{\varphi} = \sum_{j = 0}^\infty \sum_{n = 0}^\infty \sum_{\ell = 0}^\infty \sum_{m = - \ell }^\ell \varphi_{jn \ell m}(E, \vec q) \varphi_{j n \ell m}(E', \vec q').
\end{align}

\subsection{Kinematic Scattering Matrix} \label{sec:kinematrix}

After expanding $g_\chi \rightarrow \ket{g_\chi}$ and $f_S^2 \rightarrow \ket{f_S^2}$ in the relevant bases, the scattering rate can be written from \eqref{rate:continuum} as 
\begin{align}
R &=   N_T\rho_\chi \bar\sigma_0 E_0 q_0^3 v_0^3 
\sum_{\phi, \varphi}
\langle g_\chi | \phi \rangle 
\left\langle \phi  \left| \frac{F_\text{DM}^2(\vec q)}{4\pi \mu_\chi^2 m_\chi} \, \delta\Big( E + \frac{q^2}{2 m_\chi} - \vec{q} \cdot (\vec{v} - \vec{v}_\text{lab}) \Big)  \right| \varphi \right\rangle
\langle \varphi | f_S^2 \rangle,
\label{rate:vectors}
\end{align}
where I have inserted the completeness relations to factorize $g_\chi$ and $f_S^2$ from the kinematic $\delta$ function,
and where $\vec{v}_\text{lab}$ is the velocity of the laboratory in whatever Galilean reference frame $g_\chi(\vec v)$ is defined in.
The matrix element $\langle \phi | \hat{\mathcal O } | \varphi \rangle$ is defined for an operator $\hat{\mathcal O}$ by
\begin{align}
\left\langle \phi  \left|  \hat{\mathcal O}  \right| \varphi \right\rangle
&\equiv \int\! \frac{d^3 v}{v_0^3} \frac{dE}{E_0} \frac{d^3 q}{q_0^3} \phi(\vec v) \, \mathcal O(\vec v, \vec q, E) \, \varphi(\vec q, E) .
\label{def:mathcalO}
\end{align}
This integral replaces \eqref{rate:continuum}. Because it involves the basis functions, rather than $g_\chi$ and $f_S^2$, the integral may be completed analytically if $\phi$ and $\varphi$ are sufficiently simple. 

In the $(\phi, \varphi)$ basis, the operator $\hat{\mathcal O} = F_\text{DM}^2 \delta(\ldots)$ is a scattering matrix that acts on the vectors $\bra{g_\chi}$ and $\ket{f_S^2}$. 
For the specific case of nonrelativistic DM scattering, I collect all of the $m_\chi$ dependent terms into this scattering matrix, along with enough factors of $v_0$ and $q_0$ to make the resulting $\mathcal M$ dimensionless:
\begin{align}
\mathcal M_{\phi}^{\varphi} &\equiv \left\langle \phi  \left| \frac{(q_0^4 / v_0^2) F_\text{DM}^2(q) }{4\pi \mu_\chi^2 m_\chi} \delta\Big( E + \frac{q^2}{2 m_\chi} - \vec{q} \cdot \vec{v} + \vec{q} \cdot \vec{v}_\text{lab} \Big) \right| \varphi \right\rangle,
\label{def:mathcalM}
\end{align}
so the scattering rate is expressed succinctly as a function of the detector orientation  $\mathcal R \in SO(3)$:
\begin{align}
R(\mathcal R) &=  \left( N_T \rho_\chi \bar\sigma_0 \frac{v_0^2}{q_0}\right)  \sum_{\phi, \varphi} \langle v_0^3 g_\chi | \phi \rangle \cdot \mathcal M_\phi^\varphi \cdot \mathcal R \cdot \langle \varphi | E_0 f_S^2 \rangle.
\end{align}
The prefactor in this expression has units of inverse time, while every object in the double sum is dimensionless. 
To convert this rate into an expected number of events, one simply multiplies $R$ by the exposure time $T_\text{exp}$. From this I define a dimensionless exposure factor $k_0$,
\begin{align}
k_0 &\equiv N_T T_\text{exp} \bar\sigma_0 \rho_\chi \frac{v_0^2}{q_0}= \frac{ M_T}{m^{(\text{mol})}_\text{cell} } T_\text{exp} \mathcal N_A \bar\sigma_0 \rho_\chi \frac{v_0^2}{q_0} ,
\label{def:k0}
\end{align}
which is proportional to $ T_\text{exp}$, and to the number of individual SM targets $N_T$, or equivalently the ratio of the total detector mass $M_T$ to the molar mass of the unit cell, $m_\text{cell}^\text{(mol)}$. 
For future reference, the numeric value of $k_0$ is given by:
\begin{align}
k_0 &= 3288.95 \times \left(   \frac{M_T T_\text{exp} }{1\, \text{kg-yr} } \frac{1\, \text{g} }{m_\text{cell}^\text{(mol)} } \frac{\bar\sigma_0}{10^{-40}\,\text{cm}^2 } \frac{\rho_\chi}{0.4\,\text{GeV/cm}^3} \right)  \left(\frac{v_0 }{220\, \text{km/s} }\right)^2 \left( \frac{\alpha m_e c}{q_0 } \right).
\end{align}

\medskip

A subclass of the operators $\mathcal O(\vec v, \vec q, E)$ depend only on rotational invariants: $q$, $v$, $E$, and $\vec q \cdot \vec v$. These spherically symmetric operators take an especially simple form in vector spaces spanned by spherical harmonics.
Specifically, the angular integrals of \eqref{def:mathcalO} can be completed using the orthogonality of the spherical harmonics, with the result that the matrix $\langle n \ell m | \mathcal O | jn' \ell' m' \rangle \propto \delta_\ell^{\ell'} \delta_{m}^{m'}$ is diagonal in the angular indices. 
In the case of \eqref{def:mathcalM} with $\vec{v}_\text{lab} \equiv 0$ and isotropic $F_\text{DM}^2(q, v)$, 
\begin{align}
\mathcal M_{n \ell m}^{j n' \ell' m'}  &= \delta_\ell^{\ell'} \delta_{m}^{m'} \mathcal I^{(\ell)}_{n, j n'},
\label{eq:angdiag}
\end{align}
where $\mathcal I^{(\ell)}$ is the \emph{kinematic scattering matrix}, 
\begin{align}
\mathcal I^{(\ell)}_{n, j n'} &\equiv \frac{q_0^3 / v_0^3 }{2 m_\chi \mu_\chi^2 }\int_0^\infty \! \frac{dE}{E_0} \frac{qdq}{q_0^2} \, \tilde r_{j n'}(E, q) \int_{v_\text{min}(q, E) }^\infty \! \frac{vdv}{v_0^2} P_\ell\! \left( \frac{v_\text{min}(q) }{v} \right) r_n(v) \, F_\text{DM}^2(q, v)  ,
\label{def:mathcalI}
\end{align}
written here as a function of $v_\text{min}(q)$, the standard combination of $E$, $q$ and $m_\chi$ defined in \eqref{vmin}.
The derivation of \eqref{def:mathcalI} uses completeness relations involving the Legendre polynomials ($P_\ell$) and spherical harmonics to write the $\delta$ function as an infinite sum:
\begin{align}
\delta(a - \hat{q} \cdot \hat{v} ) &= 2\pi \sum_{\lambda = 0}^\infty \sum_{\mu = - \lambda}^{\lambda}  P_\lambda(a) \, Y_{\lambda \mu} (\hat{q}) Y_{\lambda \mu} (\hat{v} ) ,
\label{completeness:legendre}
\end{align}
for $-1 \leq a \leq 1$, where in this case $a \equiv v_\text{min} / v$. The $\vec q$ and $\vec v$ angular integrals are subsequently completed using the orthogonality of the spherical harmonics. Here the lower bound on the remaining $dv$ integral comes from the condition $|a| \leq 1$ in \eqref{completeness:legendre}.

So, a large number of entries in the tensor $\mathcal M_{n\ell m}^{j n' \ell' m'}$ are trivially zero, and those that remain are given by a 3d (rather than 7d) integral, which depends only on the DM model details, $m_\chi$ and $F_\text{DM}$, and the choice of basis functions. If $r_n$ and $\tilde{r}_{jn'}$ are piecewise-constant, and if $F_\text{DM}(q, v)$ is a power series in $v^m q^n$, then \eqref{def:mathcalI} can be evaluated analytically.

This drastic reduction in complexity follows directly from the choice to express $\phi(\vec v)$ in the lab frame ($v_\text{lab} \equiv 0$), which makes the operator $\mathcal O$ spherically symmetric, rather than merely azimuthally symmetric (with respect to rotations about the $\vec v_E$ axis). 
In retrospect, the $\mathcal M \rightarrow \mathcal I$ simplification can be understood as an application of the Wigner--Eckart theorem to the spherically symmetric scattering operator $\mathcal O$.
Wigner--Eckart provides a systematic way to simplify the scattering calculation for anisotropic $F_\text{DM}(\vec q, \vec v)$ in systems in which the SM and/or DM systems are not spin-averaged. 
Combining the spin-dependent operators of refs.~\cite{Fitzpatrick:2012ix,Catena:2015vpa} with a spin-polarized SM target 
would introduce slightly off-diagonal couplings $\delta^{\ell'}_{\ell \pm 1}$ into \eqref{eq:angdiag}, for example. 
The derivation of \eqref{eq:angdiag} under these more generic conditions is left to future work.

For brevity, the companion paper~\cite{Lillard:2025ixi} defines a version of $I^{(\ell)} \propto \mathcal I^{(\ell)}$ without the factors of $q_0$ and $v_0$.
The relationship between $I^{(\ell)}$  and the dimensionless $\mathcal I^{(\ell)}$ is: 
\begin{align}
\mathcal I^{(\ell)}= \frac{q_0}{E_0 v_0^5} I^{(\ell)} .  
\end{align}

\subsection{The Partial Rate Matrix} \label{sec:partialrate}
The complex spherical harmonics $Y_{\ell}^m$ of fixed $\ell$ transform as $2\ell + 1$ dimensional representations of $SO(3)$, with the group action given by:
\begin{align}
\mathcal R \cdot Y_\ell^m(\hat u) = Y_{\ell}^m(\mathcal R^{-1} \cdot \hat u) = \sum_{m' = -\ell}^\ell D^{(\ell)}_{m'm}(\mathcal R) Y_{\ell}^{m'}(\hat u),
\end{align}
or $ \ket{\mathcal R \cdot \phi} = D^{(\ell)}(\mathcal R) \cdot \ket{\phi} $,
where $D^{(\ell)}_{m'm} = \langle \ell m' | \mathcal R | \ell m \rangle$ is the Wigner $D$ matrix.
For the present analysis with real spherical harmonics $\ket{\ell m}$, I define an analogous Wigner $G$ matrix:
\begin{align}
G^{(\ell)}_{m' m} &\equiv \langle \ell m' | \mathcal R | \ell m \rangle.
\label{Gell:def}
\end{align}
Although \eqref{Gell:def} represents the matrix coefficients as integrals over the angular coordinates, its solutions are known polynomials of trigonometric functions, so no integration is actually necessary. Explicit expressions for $G^{(\ell)}$ in terms of the more familiar $D^{(\ell)}$ are provided in Appendix~\ref{appx:realspherical}. 

Both $\mathcal M$ and $G^{(\ell)}$ are diagonal in $\ell$, so the scattering rate can be written explicitly as
\begin{align}
R(\mathcal R) &= \frac{k_0}{T_\text{exp}}  \sum_{\ell= 0}^\infty \sum_{m, m' = -\ell}^\ell \sum_{n, j, n' = 0}^\infty   \langle v_0^3 g_\chi | n \ell m \rangle \cdot  \mathcal I^{(\ell)}_{n , jn'}  G^{(\ell)}_{m m'}(\mathcal R)   \langle j n' \ell m' | E_0 f_S^2 \rangle.
\label{rate:sumR}
\end{align}
Here we reference $k_0$ of \eqref{def:k0} for a compact representation of the numeric factors. 

Each object $\langle g_\chi | \varphi \rangle$, $\mathcal I^{(\ell)}$, $G^{(\ell)}$, and $\langle \varphi | f_S^2 \rangle$ is calculated independently, and the rate is given by the tensor product above.
As an intermediate step, the sum over the radial modes can be completed for each $(\ell ,m, m')$, to assemble the \emph{partial rate matrices}, $\mathcal K^{(\ell)}$:
\begin{align}
\mathcal K^{(\ell)}_{m m'}(g_\chi, f_S^2, m_\chi, F_\text{DM} )  &\equiv \sum_{n=0}^\infty \sum_{ j, n' = 0}^\infty   \langle v_0^3 g_\chi | n \ell m \rangle \cdot  \mathcal I^{(\ell)}_{n , jn'}(m_\chi, F_\text{DM} )  \cdot   \langle j n' \ell m' | E_0 f_S^2 \rangle,
\label{def:mathcalK}
\\
R(g_\chi, f_S^2, m_\chi, F_\text{DM} , \mathcal R) &= \frac{k_0}{T_\text{exp}} \sum_\ell \sum_{m, m'}  \mathcal K^{(\ell)}_{m m'} G^{(\ell)}_{m m'}(\mathcal R)  
= \frac{k_0}{T_\text{exp}} \sum_\ell \text{Tr}\!\left( G^{(\ell)} [\mathcal K^{(\ell)}]^T \right) .
\label{rate:concise}
\end{align}
The infinite sums over radial modes $n$, $j$ and $n'$ are terminated at some $n_\text{max}$, $j_\text{max}$, and $n'_\text{max}$ once the value of $\mathcal K^{(\ell)}_{m m'}$ converges. 
The contribution to $R$ from each $\ell$ mode is given by the ``partial rate'' $\tilde R^{(\ell)}(\mathcal R)$, 
\begin{align}
R(\mathcal R) &= \frac{k_0}{T_\text{exp}} \sum_\ell \tilde R^{(\ell)}(\mathcal R),
\label{def:partialRate}
\\
\tilde R^{(\ell)}(\mathcal R) &\equiv  \text{Tr}\!\left( G^{(\ell)}(\mathcal R) \cdot [\mathcal K^{(\ell)}]^T \right) 
= \sum_{m, m'} G^{(\ell)}_{m m'}(\mathcal R)  \cdot \mathcal K^{(\ell)}_{m m'} .
\label{def:RKG}
\end{align}
The partial rate matrices $\mathcal K^{(\ell)}$ compress the information from $g_\chi$, $f_S^2$ and $(m_\chi, F_\text{DM})$ into the experimentally-accessible observables $R(\mathcal R)$, making them closely related to the Fisher information matrix~\cite{Fisher:1922saa}.
With sufficiently precise measurements of $R(\mathcal R_i)$ for multiple orientations $\mathcal R_i$, an experiment can constrain or measure the coefficients of $\mathcal K^{(\ell)}$. 
By inverting \eqref{def:mathcalK} to isolate $g_\chi$, these measurements can be recast as a constraint or a detection of some components of $\ket{g_\chi}$. 

This $\mathcal K^{(\ell)}$ is more precisely referred to as the \emph{dimensionless} partial rate matrix. Compared to  the partial rate matrix $K^{(\ell)}$ defined in ref.~\cite{Lillard:2025ixi}, it differs by factors of $v_0$ and $q_0$:  
\begin{align}
\mathcal K^{(\ell)}_{m m'} &\equiv \frac{q_0}{v_0^2} K^{(\ell)}_{m m'},  
&
R(\mathcal R) &= N_T \bar\sigma_0 \rho_\chi \sum_{\ell m m'} K^{(\ell)}_{m m'} G^{(\ell)}_{m m'}(\mathcal R) . 
\end{align}

\paragraph{Radial Continuum Limit:} 

In the $n, j, n' \rightarrow \infty$ limit, the partial rate matrix has an alternative definition in terms of the following continuum radial functions at fixed $(\ell, m)$: 
\begin{align}
g_{\ell m }(v) &\equiv \langle \ell m | g_\chi \rangle , 
&
g_\chi(\vec v) &= \sum_{\ell m} g_{\ell m}(v) \, Y_{\ell m}(\hat v) , 
\\
f^2_{\ell m}(q, E) &\equiv \langle \ell m | f_S^2 \rangle, 
&
f_S^2(\vec q, E) &= \sum_{\ell m} f^2_{\ell m}(q, E) \, Y_{\ell m}(\hat q) .
\end{align} 
Applying these substitutions in \eqref{rate:continuum} produces an alternative derivation of $K^{\ell}$, without invoking the radial basis function expansion:
\begin{align}
\mathcal  K^{(\ell)}_{m m'}(g_\chi, f_S^2, m_\chi, F_\text{DM} )  &\equiv \int_0^\infty \! \frac{q\,dq}{q_0^2} dE\, f^2_{\ell m'}(q, E)  \int_{v_\text{min}(q, E, m_\chi)}^\infty \! \frac{v \, dv}{v_0^2} P_\ell\!\left( \frac{v_\text{min} }{v} \right) g_{\ell m}(v)\, \frac{q_0^3 F_\text{DM}^2(q, v) }{2 m_\chi \mu_\chi^2} .
\label{alt:mathcalK}
 \end{align}
In special cases where $\langle g_\chi | \ell m \rangle$ and $\langle f_S^2 | \ell m \rangle$ have analytic expressions, or in analyses with relatively few versions of $g_\chi$ and $f_S^2$, it may be fastest to evaluate $\mathcal K^{(\ell)}$ directly via integration of \eqref{alt:mathcalK}.

\subsection{Summary} \label{sec:genSummary}

Beyond the initial decision to vectorize the scattering rate with basis functions, I have made three important choices so far:
\newchoice{ \label{choice:spherical}
Use real spherical harmonics to describe the angular parts of the velocity and momentum basis functions, following Eqs.~(\ref{eq:basisfunctions}--\ref{eq:radnorm}).
}
\newchoice{ \label{choice:labframe}
Express $\langle g_\chi | \phi \rangle$ and $\mathcal M_\phi^\varphi$   with lab frame  ($\vec v_\text{lab} \equiv 0$) basis vectors $\ket{\phi}$.
}
\newchoice{  \label{choice:FDM}
Collect all $m_\chi$-dependent terms inside the kinematic matrix $\mathcal M$,
following \eqref{def:mathcalM}.
}
The first two choices led to the block diagonalization of $\mathcal M \rightarrow \mathcal I$ in the angular coordinates,
while the third choice allows the complete factorization of the DM model parameters ($m_\chi$, $F_\text{DM}$) from the astrophysics ($\ket{g_\chi}$) and SM material properties ($\ket{f_S^2}$).

The next choice to be made is as consequential: what should be used for the radial basis functions?
Familiar examples include continuous orthogonal functions on $0 \leq u < \infty$, such as the Laguerre or Hermite polynomials; or functions on finite intervals $0 \leq u \leq u_\text{max}$, e.g.~sinusoidal or Bessel functions. 
The  radial functions could be localized at a series of points $u_i$, as in the Whittaker cardinal series of $\sinc$ functions,
or even piecewise-defined within concentric spherical shells, e.g.~to impose different asymptotic behaviors in the $u\rightarrow 0$ and $u \rightarrow \infty$ limits. 
 A good basis should converge quickly enough that the infinite sums over $j,n,\ell$ can be terminated at some finite $j_\text{max}$, $n_\text{max}$ and $\ell_\text{max}$ with an acceptable loss of precision. If the asymptotic behavior in some limit is known, we may impose that behavior on the basis functions; but, we should avoid tailoring the basis functions to any single specific $g_\chi$ or $f_S^2$ examples. 
Simplicity is the final criterion. Ideally, it should be possible to integrate \eqref{def:mathcalI} analytically. 

With these criteria in mind, I construct the ``spherical Haar wavelets'' in Section~\ref{sec:haar}.
After developing some analytic properties of the spherical wavelets, it becomes clear that the spherical Haar wavelets are the obvious choice for this application.

\section{Spherical Haar Wavelets} \label{sec:haar}

A generic wavelet transformation uses basis functions (wavelets) that are related to each other by translation and scaling operations. Haar wavelets~\cite{haar} are a particularly simple example.
For functions on the interval $[0,1]$, all of the Haar wavelets are derived from two ``scaling functions'' $H_{-1}$ and $H_{0,0}$:
\begin{align}
H_{-1}(x) &\equiv 1,
&
H_{0,0}(x) &\equiv \left\{ \begin{array}{c c c}
	+1 && 0 \leq x < 1/2 \\
	-1 && 1/2 < x \leq 1. \end{array} \right.
\end{align}
with $H_i(x) = 0$ for all $x < 0$, $x > 1$. 
Every other wavelet can be expressed as a rescaling and translation of $H_{0,0}(x)$.
Following~\cite{Lillard:2019exp}, the higher-order wavelets $H_{\lambda, \mu}$ are
\begin{align}
H_{1,0}(x) &= \sqrt{2} \, H_{0,0}(2x),
&
H_{2,0}(x) &= 2 \, H_{0,0}(4x),
&
H_{2,2}(x) &= 2 \, H_{0,0}(4x-2),
\label{eq:haardef}
\\
H_{1,1}(x) &= \sqrt{2} \, H_{0,0}(2x-1),
&
H_{2,1}(x) &= 2 \, H_{0,0}(4x-1),
&
H_{2,3}(x) &= 2 \, H_{0,0}(4x-3),
\end{align}
etc. Each function $H_{\lambda, \mu}$ has a base of support of length $2^{-\lambda}$ on which $h_{\lambda, \mu}(x) \neq 0$. The index $\mu$ indicates the position of the wavelet within the interval $[0, 1]$.
while $\lambda$ labels which generation the wavelet belongs to.

\subsection{Definition}

\begin{figure}
\centering
\includegraphics[height=0.48\textwidth]{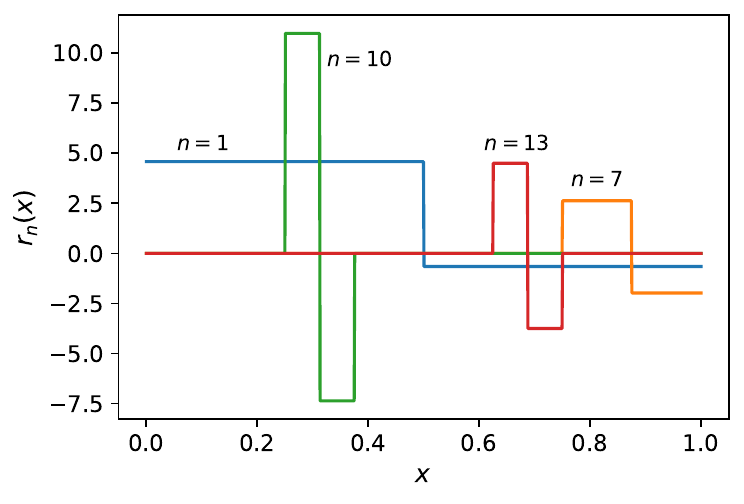}
\caption{Spherical wavelets, for $n=1, 7, 10, 13$. For example, $n=10$ and $n=13$ are of the same generation (equal $\lambda$, different $\mu$); they have equal widths but different heights, so that spherical shells of $|r_n(u)|^2$ occupy identical volumes.}
\label{fig:wavelets}
\end{figure}

To describe the 3d functions of velocity and momentum I define a set of ``spherical Haar wavelets'' $h_{\lambda \mu}(u)$, 
\begin{align}
\int_0^1 \! u^2 du \, h_{\lambda \mu}(u) h_{\lambda' \mu'}(u) &= \delta_{\lambda \lambda'} \delta_{\mu \mu'} ,
\end{align} 
with the explicit form for the $\lambda = 0, 1, 2, \ldots$ wavelets given by:
\begin{align}
{h}_{\lambda \mu} (x) = 
\left\{
\begin{array}{r c l}
+A_{\lambda \mu} && 2^{-\lambda} \mu \leq x <  2^{-\lambda} (\mu + \frac{1}{2} ) ,
\\[4pt]
-B_{\lambda \mu} &&  2^{-\lambda} (\mu + \frac{1}{2} )  < x \leq 2^{-\lambda} (\mu+ 1) ,
\\[4pt]
0 && \text{otherwise} ,
\end{array}
\right.
\label{def:sphericalHaar}
\end{align}
where
\begin{align}
x_1 = 2^{-\lambda} \mu,
&&
x_2 = 2^{-\lambda} (\mu + \tfrac{1}{2} ), 
&&
x_3 = 2^{-\lambda} (\mu + 1) ,
\label{def:x123}
\end{align}
and 
\begin{align}
A_{\lambda \mu} &= \sqrt{ \frac{3}{x_3^3 - x_1^3} \frac{x_3^3 - x_2^3 }{x_2^3 - x_1^3} }
=  \sqrt{ \frac{3  \cdot 8^{\lambda} }{3 \mu^2 + 3 \mu + 1 } \frac{12 \mu^2 + 18 \mu + 7}{12 \mu^2 + 6 \mu + 1}  }
\label{haar:A}
\\
B_{\lambda \mu} &= \sqrt{ \frac{3}{x_3^3 - x_1^3} \frac{x_2^3 - x_1^3}{x_3^3 - x_2^3 } } 
=  \left( \frac{12 \mu^2 + 6 \mu + 1}{12 \mu^2 + 18 \mu + 7}  \right) A_{\lambda \mu} .
\label{haar:AB}
\end{align}
To match the $\ket{\phi} = r_n Y_{\ell m}$ notation, the $(\lambda, \mu)$ indices are mapped onto a single integer $n = 1, 2, 3, \ldots$ via
\begin{align}
n = 2^\lambda + \mu,
\label{nLambdaMu}
\end{align}
for $\lambda = 0, 1, 2, \ldots$ and $0 \leq \mu \leq 2^{\lambda} - 1$. 
Finally, to complete the basis, I must include a constant $h_0(x)$ function analogous to $H_{-1}$, 
\begin{align}
h_{n=0}(0 \leq x \leq 1) &= \sqrt{3} .
\end{align}
A few of these basis functions are shown in Figure~\ref{fig:wavelets}. For small $n$ (e.g.~$n=1$) the magnitudes of $A$ and $B$ are quite different, but for $n$ farther away from the origin (e.g.~$n=7$ or $n=13$) the difference is less noticeable. Wavelets belonging to the same generation $\lambda$ have identical widths, but different heights. 

Continuing to enumerate the decisions about the basis functions:
\newchoice{  \label{choice:haar}
Use spherical wavelets for the velocity and momentum basis functions, and regular Haar wavelets for functions of energy:
\begin{align}
r_n(v) &= h_n(v/v_\text{max}), 
\\
\tilde{r}_{j,n}(E, q) &= h_{n}(q/ q_\text{max}) \cdot H_j(E/E_\text{max})  .
\end{align}
For discrete final state energies, or for evaluating $dR/dE$ at fixed energy $E$, I use momentum basis functions $\tilde{r}_{j,n} = h_n(q/q_\text{max}) \cdot \delta(E - \Delta E_j)$ instead.
}
For the energy-dependent Haar basis functions $H_j(E)$, I take $j = 2^\lambda + \mu$ for $\lambda = 0, 1, 2, \ldots$, and map $j=0$ to the constant $H_{-1}$ basis function. 
Each basis functions vanishes if $v > v_\text{max}$, $E > E_\text{max}$, or $q > q_\text{max}$.
Finite cutoffs for $v$, $q$ and $E$ are entirely appropriate for nonrelativistic scattering, but this basis can in principle be supplemented by additional $u > u_\text{max}$ basis functions, even extending to $u \rightarrow \infty$. Any additional basis functions should vanish on $0 \leq u \leq u_\text{max}$, to ensure their orthogonality with respect to the primary set of wavelet basis functions described here.
Further discussion of this possibility is postponed to Section~\ref{sec:polycap}, as these hybrid $[0, u_\text{max}] + [u_\text{max}, \infty)$ basis functions are not needed for nonrelativistic direct detection.

\paragraph{Consequences of Locality:}

Before moving on, I will highlight two unique numerical benefits for wavelet-type basis functions. 
First: at larger $n$, the integrals $\langle \phi | f \rangle$ actually get easier to perform. 
Suppose a Monte-Carlo integration method, with some fixed precision goal, evaluates the $\langle n | f \rangle$ integrands for $n=0$ and $n=1$ with $N_\text{evals}$ many points. 
Because the higher generation wavelets occupy progressively smaller fractions of the total volume, the $N_\text{evals}$ can be reduced by the same ratio to obtain $\langle n | f \rangle$ at the same level of absolute precision.
Within the $\lambda$th generation, none of the $h_{\lambda \mu}$ basis functions overlap: so, each new generation of wavelet coefficients can be evaluated with a \emph{total} of $N_\text{evals}$ integrand evaluations.
Each new generation doubles the number of coefficients in the expansion, $n_\text{coeffs} = 2^\lambda$; so, the total amount of integration work scales logarithmically with $n_\text{coeffs}$,
\begin{align}
N_\text{evals}^\text{(total)} \sim N_\text{evals}^{(\lambda = 0)} \log_2 n_\text{coeffs} . 
\label{eq:logNeval}
\end{align}
This compares {extremely} favorably to the \emph{increasing} amount of effort needed to numerically integrate the large $n$ coefficients in a Fourier or Bessel series.
These $\phi_n$ are highly oscillatory, with $n$ nodes, while occupying the full integration volume, so it may require $n \cdot N_\text{evals}^{(0)}$ integrand evaluations to accurately calculate a single $\langle \sin(n\pi x) | f(x) \rangle$ coefficient. In this case, the total $N_\text{evals}^\text{(total)}$ scales {quadratically}, rather than logarithmically, with $n_\text{coeffs}$. 
(Using specialized methods for Fourier-type integrals, e.g.~\cite{levin1982},  $N_\text{evals}^\text{(total)}$ can be improved to scale approximately linearly with $n_\text{coeffs}$, though at the cost of some intermediate calculations.)

A second, highly beneficial aspect of wavelets is presented in Section~\ref{sec:extrapolation}. Under the assumption that a function $f(x)$ is well described by its Taylor series in the neighborhood of $x \pm \Delta x$ for sufficiently small $\Delta x$, an initial set of wavelets ($n \leq n_0$, for $n_0 \sim 1/\Delta x$) can be used to predict the values of a much larger set of $n \gg n_0$ coefficients $\langle n | f \rangle$.
This is an extrapolation procedure in the space of wavelet coefficients.

In Section~\ref{sec:power}, I use the wavelet extrapolation to derive a scaling relation for how quickly the wavelet transformation converges towards the true value of $f(x)$ in the large $n$ limit. 
A related result (``wavelet interpolation'') makes it possible to extract derivatives $f'$, $f''$, $f^{(3)}$, etc.~directly from the wavelet coefficients in the $n \gtrsim 1/\Delta x$ limit.
All of these benefits follow from the local nature of large $n$ wavelets: as the wavelet bases of support shrink with increasing $\lambda$, the values of $\langle f | n \rangle$ depend only the local properties of the function $f$.

First, Section~\ref{sec:mathI} demonstrates that the spherical wavelets, or any other piecewise-constant radial function, allow $\mathcal I^{(\ell)}$ to be integrated analytically from \eqref{def:mathcalI}.

\subsection{Analytic Result for Kinematic Scattering Matrix}  \label{sec:mathI}

The scattering rate depends on three objects which might require numeric integration: $\langle g_\chi | \phi \rangle$, $\mathcal I^{(\ell)}$ and $\langle \varphi | f_S^2 \rangle$. 
Of these objects, the kinematic scattering matrix $\mathcal I^{(\ell)}$ appears to be the most expensive: it is a four dimensional array over $\ell, n, j, n'$, with coefficients that depend on $m_\chi$ and $F_\text{DM}$ via the 3d integral \eqref{def:mathcalI}.
Considering the large number of $\mathcal I^{(\ell)}_{n, jn'}$ coefficients that must be evaluated for each DM model, it is imperative that we choose radial basis functions for which \eqref{def:mathcalI} can be integrated analytically. 
In this section I show that wavelet-harmonic basis functions satisfy this requirement, as do any other piecewise-constant radial functions.

Consider any basis where $r_n(v)$ and $R_{jn'}(E, q)$ are piecewise-constant functions of $v$ and $q$. 
In the $(v, q)$ plane at some fixed value of $E$, the product $r_n(v) R_{jn'}(E, q)$ is piecewise-constant on some grid of rectangular regions. 
Let us focus on just one of these rectangular regions, bounded by $v_a \leq v \leq v_b$, $q_a \leq q \leq q_b$, where
\begin{align}
\tilde r_{jn'}(E, q) \equiv \bar r_{j}(E) \cdot A^{(q)}_{n'},
&&
r_n(v) = A^{(v)}_n.
\end{align}
Separating the $E$ integral from $d^3 v\, d^3 q$, define:
\begin{align}
\mathcal I^{(\ell)}_{n, j n'} &\equiv \frac{q_0^3 / v_0^3 }{2 m_\chi \mu_\chi^2 }\int_0^\infty \! \frac{dE}{E_0} \bar r_{j}(E) \cdot A^{(q)}_{n'}  A^{(v)}_n
\times \left( \frac{q_\star^2 v_\star^2}{q_0^2 v_0^2} \, I_\star^{(\ell)}(E) \right) ,
\\
I_\star^{(\ell)}(E) &\equiv \int_{q_a}^{q_b} \!  \frac{qdq}{q_\star^2} \int_{v_a}^{v_b} \! \frac{vdv}{v_\star^2} F_\text{DM}^2(q,v)  \, P_\ell\! \left( \frac{v_\text{min} }{v} \right) \Theta(v -v_\text{min}(q, E) ).
\label{def:Istar}
\end{align}
To simplify the upcoming algebra, I define $q_\star(m_\chi, E)$ and $v_\star(m_\chi, E)$ as follows:
\begin{align}
q_\star \equiv \sqrt{ 2 m_\chi E} ,
&&
v_\star \equiv \frac{q_\star }{m_\chi},
&&
E = \frac{1}{2} q_\star v_\star ,
&&
v_\text{min} = \frac{v_\star}{2} \left( \frac{q}{q_\star} + \frac{q_\star }{q} \right) .
\label{def:stars}
\end{align}
The minimum possible value of $v_\text{min}(q)$ occurs at $q = q_\star$:
\begin{align}
\text{min}\big( v_\text{min}(q) \big) &= v_\text{min}(q=q_\star) = v_\star.
\end{align}

The function $P_\ell$ is a polynomial of degree $\ell$, so the velocity integral can be completed easily, as long as $F_\text{DM}^2(q, v)$ is a sufficiently simple function. 
For example, if 
\begin{align}
F_\text{DM}^2(q, v) &= \sum_{\beta \gamma} c_{\beta \gamma}\left( \frac{q}{\alpha m_e } \right)^\beta v^\gamma, 
\label{eq:Fdm2}
\end{align}
for some dimensionless constants $c_{\beta \gamma}$, 
Appendix~\ref{sec:Istar} demonstrates that $I_\star$ has the solution 
\begin{align}
I_\star^{(\ell)} &= \sum_{\beta \gamma} c_{\beta \gamma} \left( \frac{q_\star}{\alpha m_e} \right)^\beta v_\star^\gamma \, I_{\beta \gamma}^{(\ell)} , 
\\
I_{\beta \gamma}^{(\ell)}( [v_1, v_2], [q_2, q_3] ) &= U^{(\ell)}_{\beta \gamma}( v_2, [q_1, q_2] )  +  T^{(\ell)}_{\beta \gamma}( [v_1, v_2], [q_2, q_3] )  +  U^{(\ell)}_{\beta \gamma}( v_2, [q_3, q_4] ) .
\label{solnIstar}
\end{align}
as summarized in \eqref{Istar:UTU}. 
Here $U$ and $T$ are the analytic functions of $v_\star$ and $q_\star$ provided in Appendix~\ref{sec:Istar}, and $q_{1 \ldots 4}$ are given by $q_a$, $q_b$, or by the solutions to $v_a = v_\text{min}(q_\pm)$ and $v_b = v_\text{min}(\tilde q_\pm)$.

For tree-level DM--SM scattering through a mediator $\phi$ of mass $m_\phi$, the light- and heavy-mediator limits $m_\phi \ll q$ and $q \ll m_\phi$ are well approximated by the $\beta = -4$ and $\beta = 0$ cases of \eqref{eq:Fdm2}, respectively, with $\gamma = 0$ in the leading order terms. Higher order terms in the $v$ and $q/m_e$ expansions for tree level scattering (e.g.~\cite{Catena:2022fnk}) introduce corrections with $\gamma = 2$. 
When the mediator mass is comparable to the momentum transfer, $m_\phi \sim q$, $F_\text{DM}^2$ is not well approximated by any single term of the form \eqref{eq:Fdm2}. 

\medskip 

In the case of discrete final states, $\bar r_j(E) \rightarrow E_0\,\delta(E - \Delta E_j)$, 
\begin{align}
\mathcal I^{(\ell)}_{n, jn'}(E = \Delta E_j) &= \frac{q_0/v_0^5 }{2 m_\chi \mu_\chi^2 }\, A^{(q)}_{n'}  A^{(v)}_n \times 4 (\Delta E_j)^2 I_\star^{(\ell)}(\Delta E_j) .
\label{mathcalI:discrete}
\end{align}
For continuum final states, this $I^{(\ell)}_\star(E)$ is sufficient for finding the differential rate $dR/dE$. In applications where the integral over $E$ must also be completed, following \eqref{rate:continuum}, $\mathcal I^{(\ell)}$ is found by integrating $I_\star^{(\ell)}(E)$ together with the energy basis functions $\bar r_j(E)$. Both $q_\star^2$ and $v_\star^2$ are proportional to ${E}$; and with the exception of a dilogarithm $\propto \text{Li}_2(1/E)$, most of the terms in $T^{(\ell)}$ and $U^{(\ell)}$ are polynomials or logarithms in $E$ or $1/E$, as long as $\beta$ is integer-valued. As long as simple basis functions are used for $\tilde{r}_j(E)$, e.g.~Haar wavelets, this final integral can also be completed analytically.

In the spherical wavelet basis, $\mathcal I_{n, jn'}^{(\ell)}$ generally receives contributions from four regions (if $n\neq 0$ and $n'\neq 0$), 
with $A_{n'}^{(q,v)} \rightarrow [A_{n'}^{(q,v)}, - B_{n'}^{(q,v)}]$, with $A$ and $B$ defined in \eqref{haar:A} and \eqref{haar:AB} in terms of the bases of support for the $(q)$ and $(v)$ wavelet functions.
More generally, $\mathcal I$ can be assembled for any piecewise-constant basis functions from the expression given in \eqref{solnIstar}.

In conclusion, if the radial basis functions $r_n(v)$ and $\tilde r_{n'}(q)$ are piecewise constant, the elements of the kinematic scattering matrix $\mathcal I^{(\ell)}$ can be evaluated analytically. 
This is one of the two criteria listed at the start of Section~\ref{sec:method}.
Appendix~\ref{sec:Istar} provides the full result for $F_\text{DM}^2 \propto q^\beta v^\gamma$ for any $\beta,\gamma$ (including non-integer values). 
Next, in Section~\ref{sec:power}, I show that the spherical Haar wavelet expansion also converges sufficiently quickly as a function of the number of coefficients in the expansion.

\subsection{Wavelet Extrapolation} \label{sec:extrapolation}

Haar wavelets have a special property in the large $n$ limit. As a wavelet becomes arbitrarily narrow, it approaches the first derivative of the Dirac $\delta$ function, $H_{\lambda \mu}(x) \rightarrow 2^{-\lambda/2} \delta'(x - x_2)$, where $x_2 = 2^{-\lambda} (\mu + \frac{1}{2})$ marks the center of the wavelet's base of support. 
In this extreme limit, where the width of the wavelet $\Delta x = 2^{-\lambda}$ vanishes, $\langle H_{n \rightarrow \infty} | f \rangle$ is just proportional to $-(\Delta x) f'(x_2)$.
If $f$ jumps discontinuously by a finite amount $\Delta f$ in the neighborhood of $x_2$, then $\langle H_{n \rightarrow \infty} | f \rangle \propto -\Delta f$ for any wavelets that span the discontinuity.
Functions with infinitely large discontinuities at isolated points require special treatment; of course, this was already true for the original \eqref{rate:continuum}.

\medskip

For wavelets with narrow but finite width, the values of $\langle H_{\lambda \mu} | f \rangle$ can be estimated directly from the derivatives of $f$.
Consider a wavelet in the $\lambda$th generation, with width $\Delta = 2^{-\lambda}$, having a base of support $[x_1, x_3]$ centered at a point $x_2$. 
Let $x$ be normalized so that $f(x)$ lies on the interval $[0,1]$. 
Suppose that within the region $[x_1, x_3]$, a function $f(x)$ is well approximated by its third-order Taylor series centered at $x_2$, 
\begin{align}
f(x) \simeq f(x_2) +  f'_0(x-x_2) + \frac{f''_0}{2}(x-x_2)^2  + \frac{ f^{(3)}_0}{3!} (x-x_2)^3 ,
\label{fu:cubic}
\end{align}
where $f^{(k)}_0 \equiv f^{(k)}(x_2)$. The imprecision of this expression is expected to scale like $(\Delta/2)^4 f^{(4)}$ at the edges of the interval.
 
For (non-spherical) Haar wavelets, $\ket{\lambda \mu}$ is normalized so that it takes the values $\pm  2^{\lambda/2}$ when it is nonzero. In terms of  $\Delta$, the wavelet coefficient $\langle f | \lambda \mu \rangle$ is: 
\begin{align}
\left\langle f \big| H_{\lambda \mu} \right\rangle &\simeq - \frac{ 2^{-\lambda/2} }{4} \left( f'_0 \Delta + \frac{1}{48} f^{(3)}_0 \Delta^3 \right) .
\label{haar:f00}
\end{align}
The subsequent generation of wavelets, defined on the intervals $[x_a, x_2]$ and $[x_2, x_b]$, are sensitive to the second derivative of $f$:
\begin{align}
\left\langle f \big| H_{\lambda+1, 2 \mu}  \right\rangle &\simeq - \frac{(\sqrt{2})^{-\lambda + 1}  }{16} \left[ \Delta f'_0 - \frac{1}{4} \Delta^2 f''_0 + \frac{7 }{192} \Delta^3 f^{(3)}_0 \right],
\label{haar:f10}
\\
\left\langle f \big|  H_{\lambda+1, 2 \mu +1}  \right\rangle &\simeq - \frac{(\sqrt{2})^{-\lambda + 1}   }{16} \left[ \Delta f'_0 + \frac{1}{4} \Delta^2 f''_0 + \frac{7 }{192} \Delta^3 f^{(3)}_0 \right].
\label{haar:f11}
\end{align}
So, given the values of the $\lambda$ and $\lambda+1$ wavelet coefficients, it is possible to extract the values of $f_0^{(p)}$ for $p=1,2,3$ by applying some simple linear algebra to Eqs.~(\ref{haar:f00}--\ref{haar:f11}).
Alternatively, if the derivatives of $f$ are already known (e.g.~if $f(x)$ is given analytically) then the large $\lambda \geq \lambda_\star$ coefficients can be calculated directly from \eqref{haar:f00}, as long as $\lambda_\star$ is large enough that the difference between $\langle f| H_{\lambda \mu} \rangle$ and its cubic-order estimate is acceptably small.
In this way, the values of three coefficients ($\langle f | \lambda \mu \rangle$ and the two $\lambda + 1$ wavelets that overlap with it) can be used to generate a much larger set of $\lambda' \geq \lambda + 2$ coefficients, with a precision controlled by the size of the $\Delta^4 f_0^{(4)}$ term that has been dropped from \eqref{fu:cubic}. 

I refer to this method as ``wavelet extrapolation.'' The extrapolation is in the space of coefficients, $\ket{n}$, not extrapolation in the domain of $x$. It can be used to refine the inverse wavelet transformation $f(x) \simeq \sum_n \langle n | f \rangle \ket{n}$ or to construct interpolating functions for $f(x)$ and its derivatives within the original range $0 \leq x \leq 1$.

To take a specific example, suppose that $f(x)$ in the region of $x_2$ is oscillatory, with some characteristic wavelength $2\pi \, \delta x$. This $\delta x$ satisfies $\delta x\, f' \sim (\delta x)^2 f'' \sim (\delta x)^p f^{(p)}$, indicating that the cubic Taylor series expansion begins to converge once $\Delta < \delta x$. 
Defining $\lambda_\delta \equiv \log_2(1/\delta x)$, the error of the cubic approximation for $\langle f | \lambda \mu \rangle$ scales as
\begin{align}
\frac{\langle f | H_{\lambda \mu} \rangle - \langle f | H_{\lambda \mu}^{\text{cubic} } \rangle }{\langle f | H_{\lambda \mu} \rangle} 
\propto \frac{\Delta^5  f_0^{(5)} }{\Delta  f_0^{(1)} } \sim \left( \frac{\Delta}{\delta x} \right)^4 \sim 16^{\lambda_\delta - \lambda} ,
\label{haar:extrap16}
\end{align} 
while the error of Eqs.~(\ref{haar:f10}--\ref{haar:f11}) scales as $\Delta^3 f^{(4)} / f^{(1)} \sim (\Delta/\delta x)^3 \sim 8^{\lambda_\delta - \lambda}$.
So, if the derivatives $f^{(p)}$ are known exactly, then the approximation \eqref{haar:f00} improves with a factor of 16 in accuracy with each subsequent generation $\lambda \rightarrow \lambda+1$. 
Alternatively, if $f^{(p)}$ are extracted via Eqs.~(\ref{haar:f10}--\ref{haar:f11}) from an initial $\lambda \leq \lambda_\star$ set of integrated wavelet coefficients, then the accuracy of the wavelet extrapolation procedure improves by a factor of 8 when $\lambda_\star$ is incremented by one.

In a hypothetical application where the wavelet transformation and its inverse must be calculated to extremely high precision, this wavelet extrapolation procedure is highly useful. 
For basis functions with no extrapolation method (e.g.~Fourier or Bessel), one would otherwise calculate $\langle f| n \rangle$ for all $n$ such that $ \langle f| n \rangle \geq \epsilon$ , for some precision goal $\epsilon$. 
With the $8^{\lambda_\star}$ scaling of the Eqs.~(\ref{haar:f10}--\ref{haar:f11}) relative error, the cubic wavelet extrapolation method can accurately approximate all of these coefficients from a $\lambda \leq \lambda_\star$ initial set,
where
\begin{align}
\lambda_\star \sim \lambda_\delta + \frac{1}{3} \log_2 \frac{1}{\epsilon} .
\label{haar:extrap8}
\end{align}
Only $n_\text{coeffs} = 2^{\lambda_\star}$ of the coefficients need to be evaluated directly from the inner products $\langle f | n \rangle$, so the precision in $\epsilon$ improves as $\epsilon \propto 1/n_\text{coeffs}^3$.

\paragraph{General Method for Spherical Wavelets:}

An analogous extrapolation method exists for the spherical wavelets, though the adaptation of \eqref{haar:f00} involves the $\lambda\mu$ dependent factors $A_{\lambda\mu}$ and $B_{\lambda \mu}$. 
For this reason alone it is convenient to define some intermediate expressions, so I may as well provide the general $k$th order version of the extrapolation.
For a spherical wavelet $\ket{\lambda \mu}$ centered at $x = x_2$, with base of support $\Delta = x_3 - x_1 = 2^{-\lambda}$, define: 
\begin{align}
F_{p}(\lambda, \mu) &\equiv \frac{\Delta^p f_0^{(p)}}{2^p \, p!} = 2^{-p(\lambda + 1) } \frac{f_0^{(p)} }{p!}  ,
\label{def:Fp}
\\
D_p(\lambda, \mu) &\equiv \frac{\Delta^3  }{8} \left[ \frac{(-1)^p A_{\lambda \mu} - B_{\lambda \mu}}{p+3} - 4 \left(\mu + \tfrac{1}{2}  \right) \frac{ (-1)^p A_{\lambda \mu} + B_{\lambda \mu}  }{p+2 } + 4 \left(\mu + \tfrac{1}{2}  \right)^2 \frac{ (-1)^p A_{\lambda \mu} - B_{\lambda \mu}  }{p+1} \right] ,
\label{def:Hp}
\end{align}
with $A_{\lambda \mu}$ and $B_{\lambda \mu}$ defined in \eqref{haar:AB}.
For a $k$th order Taylor series, the projection of $f$ onto a $\lambda \geq 1$ spherical  wavelet is given by:
\begin{align}
\langle f | h_{\lambda \mu} \rangle & \simeq  \sum_{p=1}^k F_p(\lambda, \mu)\, D_p(\lambda, \mu) .
\label{sphH:cubic}
\end{align}
Note that the $p=0$ term vanishes for all $\lambda \geq 0$, due to the orthogonality of $\langle n=0 | \lambda \mu \rangle$. 
Unlike the Haar wavelets, though, the even derivative terms $f'', f^{(4)}$ etc.~do not vanish.
From \eqref{def:Fp}, the largest subleading term missing from the series is proportional to $2^{-(k+1)(\lambda + 1)}$: so, as $\lambda$ increases, the absolute accuracy of \eqref{sphH:cubic} improves as $2^{(k+1)\lambda}$.

\medskip

For both spherical and regular Haar wavelets, the coefficients $\langle f | n \rangle$ converge predictably once $\Delta = 2^{-\lambda}$ is small enough that the function $f$ is well described by its Taylor series within the base of support of the $n$th wavelet. 
Given the values of some coefficients $\langle f | n \rangle$, the derivatives $f^{(p)}$ can be found algebraically from \eqref{sphH:cubic}. This can be used to define $k$th order interpolating functions for $f(x)$, for example if an analytic version of $f(x)$ is not available. 
Appendix~\ref{sec:interpolation} provides explicit solutions for cubic interpolation.
The relative precision $\epsilon$ scales with the number of integrated coefficients $n_\text{coeffs} = 2^{\lambda_\star}$ as in \eqref{haar:extrap16} and \eqref{haar:extrap8}: in the latter case, for the $k$th order extrapolation on spherical wavelets,
\begin{align}
\epsilon \propto 1/n_\text{coeffs}^k .
\end{align}
If the derivatives $f^{(p)}$ are known exactly, then the precision of \eqref{sphH:cubic} scales as $\epsilon \propto 1/n_\text{coeffs}^{k+1}$.
The generic extrapolation procedure uses the values of the final $j$ generations in the wavelet transformation, $\lambda = \lambda_\star, \lambda_\star - 1, \ldots, \lambda_\star - j + 1$, where $k \leq 2^{j} - 1$ is the polynomial order of the wavelet extrapolation.
For example, a set of three wavelet generations can support a seventh-order extrapolation procedure.

\paragraph{Application to 3d functions:}

The discussion so far involves a 1d function $f(x)$, expanded in Haar or spherical wavelets. 
In the context of $g_\chi(\vec v)$ or $f_S^2(\vec q)$, the relevant $f(x)$ is a radial function $f_{\ell m}$ defined for every angular mode $(\ell, m)$:
\begin{align}
f_{\ell m}(u) =  \langle \ell m | f \rangle  \equiv \int\! d\Omega\, Y_{\ell m}(\Omega) \, f(\vec u) ,
\label{def:f_lm}
\end{align} 
where to match the normalization of this section, $x = u/u_\text{max}$ for some $u_\text{max}$. 
Reverting to the $n = 2^\lambda + \mu$ indexing, the inverse wavelet transformation is given by:
\begin{align}
f_{\ell m}(u) \simeq \sum_{n = 0}^{n_\text{max}} \langle n \ell m | f \rangle \ket{n}.
\end{align}
Given the values of three coefficients, $\langle n \ell m | f \rangle $ for $n = n_\star, 2 n_\star, 2n_\star + 1$, the coefficients for the subsequent generations of wavelets overlapping with $\ket{n}$ can be estimated using the cubic wavelet extrapolation, \eqref{sphH:cubic}.

\paragraph{Concluding Comments:}
To underscore the unique nature of the wavelet extrapolation, consider an analogy to Fourier or Bessel expansions. If something like the cubic extrapolation method were valid for more general basis functions, then under some generic assumption (e.g.~that $f$ varies slowly compared to, say, the $100$th basis function), the first $n < 100$ Fourier coefficients could in this example be used to precisely estimate the coefficients of all of the subsequent $n \lesssim 10^6$ high frequency modes. Unfortunately, no such method exists: the information contained in the first few Fourier coefficients is insufficient for estimating the large-$n$ modes.

On the other hand, similar wavelet extrapolation methods should be available for any other basis functions that approach locality in the large $n$ limit; that is, if the compact base of support $\Delta_n$ for the $n$th function vanishes in the limit
\begin{align}
\lim_{n \rightarrow \infty} \Delta_n = 0.
\end{align}
Other families of wavelets with compact support, e.g.~Daubechies~\cite{doi:10.1137/1.9781611970104.ch6}, have this locality property. 
From the values of the $p \leq k$ central moments $\langle n | (x - x_2)^p \rangle$ in the limit $\Delta_n \rightarrow 0$ (for $| x - x_2| \leq \frac{1}{2} \Delta_n$, the base of support of the $n$th basis function), one could derive an analogous $k$th order extrapolation method by generalizing \eqref{sphH:cubic}. 

So, the fast convergence demonstrated in the higher-order Daubechies wavelets can now be achieved with the easily integrated Haar wavelets.
This simple result has wide-ranging consequences for any numerical methods that involve wavelets, especially in high precision calculations.

\subsection{Global Convergence} \label{sec:power}

There are multiple measures of how quickly a series converges. 
The residual difference between a function $f(\vec u)$ and its basis expansion $\sum_{\phi} \langle \phi | f \rangle \ket{\phi}$ provides a local measurement of the accuracy, for example.
For $L^2$ normalized basis functions, e.g.~wavelet-harmonics, the distributional ``norm-energy'' offers a particularly convenient global measure of accuracy.

A complete, orthogonal, $L^2$ normalized basis preserves the norm-squared of a distribution, which is a functional defined for a real $f( \vec u)$ as
\begin{align}
\mathcal E[f] \equiv \int\! d^3 u \, [f(\vec u)]^2.
\end{align}
Expanding  $f$ in the basis spanned by $\ket{\phi}$, 
\begin{align}
\mathcal E[f] &= \int\! d^3  u \left[ \sum_{\phi} \langle \phi | f \rangle \, \phi(\vec u) \right] f(\vec u) 
= u_0^3 \sum_{\phi} \Big( \langle \phi | f \rangle \Big)^2 ,
\label{def:energy}
\end{align}
it is clear that $\mathcal E$ is preserved in the sum over squared $\langle f | \phi \rangle$ coefficients. 
(That is, $f$ lives in a Hilbert space, which can be spanned by the basis functions $\{\phi\}$.)
The conserved $\mathcal E$ is sometimes called the ``energy'' of a distribution, and partial sums of $\langle \phi | f \rangle^2$ can be referred to as the ``power.'' 
Every term in the $\mathcal E$ sum is nonnegative, so a sum over finitely many basis functions $\ket{\phi}$ approaches $\mathcal E$ from below.
Labeling the $\ket{\phi}$ functions with $i = 0, 1, 2, \ldots$ and truncating the series at $i_\text{max}$, 
\begin{align}
\sum_{i = 0}^{i_\text{max} }  u_0^3  \langle \phi_i | f \rangle^2 \leq  \mathcal E, 
\label{energy:sum}
\end{align}
with the inequality saturated in the $i_\text{max} \rightarrow \infty$ limit.
The difference  $\mathcal E - \sum_{i}^{i_\text{max}} u_0^3 \langle \phi_i | f \rangle^2$ tracks the global accuracy of the basis expansion,
\begin{align}
\Delta \mathcal E(i_\text{max}) &\equiv \mathcal E - \sum_{i=0}^{i_\text{max}} u_0^3 \langle \phi_i | f \rangle^2,
\label{def:MET}
\end{align}
and provides a strict upper bound on the size of any of the unmeasured coefficients:
\begin{align}
\Big| \langle \phi | f \rangle \Big|_{i > i_\text{max}}  \leq \sqrt{ \Delta \mathcal E / u_0^3} .
\end{align}
The bound is saturated only if all unevaluated coefficients except for a single $\langle \phi | f \rangle$ are zero. 

Because the sum \eqref{energy:sum} approaches $\mathcal E$ monotonically, it is a convenient global measurement of the precision of an expansion. 
Other useful quantities include the angular power distribution,
\begin{align}
\mathcal P_{\ell m}(u) &\equiv u^2 \left[ f_{\ell m}(u) \right]^2,
\label{def:power}
\end{align}
where $f_{\ell m}(u) = \langle \ell m | f \rangle$ is the projection of $f(\vec u)$ onto the $\ket{\ell m}$ harmonic, \eqref{def:f_lm};
and the integrated angular power, 
\begin{align}
\mathcal E_{\ell m} &\equiv \int\! du \, \mathcal P_{\ell m}(u),
\end{align}
which is the amount of $\mathcal E$ stored in the ${\ell m}$ harmonic.
Each quantity corresponds to a different partial sum over $\langle f| n\ell m \rangle^2$ coefficients, and can be used to track the convergence of the radial and angular parts of the expansion separately.

\paragraph{Application to Wavelets:}

 In the special case of wavelets, we can estimate the scaling of $\Delta E$ as a function of $n_\text{max}$, in the limit of large $n_\text{max}$. 
First, note from \eqref{haar:f00} that for later generations ($\lambda \gg 1$), the value of $\langle \lambda \mu | f \rangle$ is well approximated by the value of $f'(x)$.
Defining $\mathcal E_\lambda$ as the energy contained in a single wavelet generation $\lambda$, 
\begin{align}
\mathcal E_\lambda \equiv 
\frac{\delta \mathcal E}{\delta \lambda} \bigg|_{\lambda} & = \sum_{\mu = 0}^{2^\lambda - 1} \langle f | \lambda \mu \rangle^2 
\simeq  \frac{4^{-\lambda } }{16} \sum_{\mu} (\Delta x) \, (f_\mu')^2,
\end{align}
where $\Delta x = 2^{-\lambda}$, and where $f_\mu' \equiv f'(x = x_2)$, for $x_2$ defined in \eqref{def:x123} as the center of the $\ket{\lambda \mu}$ wavelet's base of support.
In the $\Delta x \rightarrow 0$ limit, the sum over $(\Delta x) (f_\mu')^2$ approaches a constant, independent of $\Delta x$ and $\lambda$:
\begin{align}
\lim_{\lambda \rightarrow \infty} \left[ \sum_\mu (\Delta x) (f_\mu')^2 \right] = \int_{0}^1 \! dx \, [f'(x)]^2 .
\label{lim:genpower}
\end{align}
So, $\delta \mathcal E/ \delta \lambda$ decreases as $4^{-\lambda}$ for large $\lambda$, indicating that the generational power is concentrated at small $\lambda$, once $\Delta x$ is small enough that the Taylor series of \eqref{haar:f00} converges. 
In this regime, the $\sum_\mu (\Delta x) (f_\mu')^2$ sum is approximately constant for $\lambda \geq \lambda_\star$, 
and $\Delta \mathcal E$ can be found by summing over all $\lambda \geq \lambda_\star$: 
\begin{align}
\Delta \mathcal E &\equiv \sum_{\lambda = \lambda_\star}^{\infty} \frac{\delta \mathcal E}{\delta \lambda} 
\simeq \left(  \frac{ 4^{-\lambda_\star } }{12} \right)  \left[  \sum_{\mu = 0}^{2^\lambda_\star - 1} 2^{-\lambda_\star}  \, (f_\mu')^2  \right] 
\simeq \left(  \frac{ 4^{-\lambda_\star } }{12} \right)  \int_{0}^1 \! dx \, [f'(x)]^2 .
\label{eq:METfmu}
\end{align}
The term in brackets is the same approximately-constant term whose large $\lambda$ limit is given in \eqref{lim:genpower}.

Noting that the $\lambda < \lambda_\star$ wavelet expansion includes a total of $n_\text{coeffs} = 2^\lambda_\star$  coefficients, the missing energy scales as
\begin{align}
\Delta \mathcal E \propto \left(\frac{1}{n_\text{coeffs}} \right)^2 .
\label{eq:METhaar}
\end{align}
In conclusion, the Haar wavelet expansion converges very quickly as $\lambda$ is increased.
The spherical wavelets converge similarly fast with $n_\text{coeffs}$, though with a few additional factors in \eqref{eq:METfmu}. 

In applications where very high precision is required, the ${n_\text{coeffs}}$ of \eqref{eq:METhaar} can still be rather large. 
However, thanks to the wavelet extrapolation procedure of Section~\ref{sec:extrapolation}, only a small fraction of these coefficients need to be calculated from the inner products $\langle f | n \rangle$. 
A $k$th order method can predict $(n_\text{coeffs}^\text{int.})^k$ coefficients from an initial set of $n_\text{coeffs}^\text{int.}$ integrated coefficients, at the same level of precision in $\Delta \mathcal E$:
\begin{align}
\Delta \mathcal E \propto \left( \frac{1}{n_\text{coeffs}^\text{int.}} \right)^{2k} .
\end{align}

For comparison, in the analogous limit where the function $f(x)$ varies slowly compared to $\sin(n \pi x)$, the individual coefficients in a Fourier-type series scale as $1/n$. So, from $\sum_{n > n_\text{coeffs}} n^{-2} \approx 1/n_\text{coeffs}$ in the limit of large $n$, the harmonic-type series converge as $\Delta \mathcal E \propto 1/n_\text{coeffs}$.

\subsection{Summary}

There are now four good reasons to use the spherical wavelet-harmonic basis:
\begin{enumerate}
\item The kinematic scattering matrix $\mathcal I^{(\ell)}$ can be integrated analytically for piecewise-constant basis functions.
\item As measured by the ``missing norm-energy'' in \eqref{def:MET}, the error in the wavelet expansion scales as $1/n_\text{coeffs}^2$, where $n_\text{coeffs}$ is the number of coefficients included in the expansion. (Other examples, e.g.~the Fourier series, converge more slowly as $1/n_\text{coeffs}$.)
\item The wavelet extrapolation method allows the large $n$ coefficients to be evaluated from the derivatives of the function, via \eqref{sphH:cubic}, rather than from integrating the inner products $\langle \phi_{n \ell m} | f \rangle$. 
\item From \eqref{eq:logNeval}, the computational difficulty in evaluating $\langle n \ell m | f \rangle$ actually drops for large $n$ wavelets, so that the integration time grows logarithmically with the number of coefficients to be calculated. For other basis choices, the integration time scales as some power of $n_\text{coeffs}$ (e.g.~linearly or quadratically).
\end{enumerate}
Combining the $1/n_\text{coeffs}^2$ convergence of $\mathcal E$ with the $k$th order extrapolation method, 
the integration time not only scales logarithmically with the precision goal $\Delta  \mathcal E$, it does so with an additional numeric factor:
\begin{align}
N_\text{evals}^\text{(wavelet)} \propto \frac{1}{4^{k} } \log_2 (1/\Delta \mathcal E), 
\end{align}
with e.g.~$k=3$ for the cubic extrapolation method. 
For comparison, the integration time for an oscillatory radial function (Fourier/Bessel/etc.) scales as
\begin{align}
N_\text{evals}^{(\text{harmonic})} \propto n_\text{coeffs}^p \propto  (\Delta \mathcal E)^{-p} ,
\end{align}
with $p=2$ unless special efforts are made to handle the oscillatory part of the Fourier integrals.

\section{Demonstrations for Direct Detection} \label{sec:demo}

In this section I test the convergence of the spherical wavelet expansion and perform a mock direct detection analysis, for a particle-in-a-box $f_S^2$ example, and a $g_\chi$ that is a sum of four gaussians, 
\begin{align}
g_\chi(\vec v) &= 0.4\, g_{(0)}(\vec v) + 0.3 \,g_{(1)}(\vec v) + 0.2\, g_{(2)}(\vec v) + 0.1\, g_{(3)}(\vec v) ,
\label{eq:model4}
\\
g_{(i)}(\vec v, \vec v_i, \bar v_i) &= \frac{1}{\pi^{3/2} \bar v_i^3} \exp\left( - \frac{|\vec v - \vec v_i |^2}{\bar v_i^2 } \right),
\label{gaussianform}
\end{align}
with parameters given in the following table:
\begin{align}
\begin{tabular}{| c | l | c |  } \hline
index& $\vec v_i$			~[km/s]			& $\bar v_i$ \\ \hline
(0)	&  $(-230 \hat z)$						& 220\,km/s	\\
(1)	&  $(+80 \hat x - 80 \hat z)$				&  70\,km/s 	\\
(2)	&  $(-120 \hat x - 250 \hat y - 150 \hat z)$		& 50\,km/s	\\
(3)	&  $(+50 \hat x + 30 \hat y - 400 \hat z)$		& 25\,km/s	\\ \hline
\end{tabular}
\label{eg:gaussian}
\end{align}
This roughly approximates a galaxy with a DM halo, $g_{(0)}$, and three large and increasingly narrow streams, $g_{(1,2,3)}$.
The narrowest Stream~(3) confronts the spherical harmonic expansion with an especially sharp feature, just a few degrees wide.
Incidentally, this is the type of model that might call for repeating the analysis for many similar $g_\chi$ distributions: 
each gaussian is described by 5 parameters, $(c_i, \vec v_i, \bar v_i)$, any of which might be varied. 

To demonstrate the scattering rate part of the calculation, a $g_\chi$ model must be paired with a detector form factor, $f_S^2(\vec q, E)$. 
For the detector form factor, $f_S^2(\vec q, E)$, I use a model where a single particle of mass $m$ is confined to a rectangular box with sides of length $(L_x, L_y, L_z)$, with position space wavefunction
\begin{align}
\Psi_{\vec n} = \frac{2^{3/2} }{\sqrt{V_\text{cell}}} \sin\frac{\pi n_x x}{L_x}  \sin\frac{\pi n_y y}{L_y}  \sin\frac{\pi n_z z}{L_z} ,
\label{eq:wavebox}
\end{align}
where $V_\text{cell} = L_x L_y L_z$ is the volume of the microscopic unit cell.
The macroscopic detector target consists of $N_T$ of these unit cells, with a total volume of $V_T = N_T V_\text{cell}$.
The final states are excitations above the $n_{x,y,z}=1$ ground state, with discrete energies
\begin{align}
\Delta E_{\vec n} &= \frac{\pi^2 }{2m} \left( \frac{n_x^2 - 1}{L_x^2} + \frac{n_y^2 - 1}{L_y^2 } + \frac{n_z^2 -1}{L_z^2} \right) .
\end{align}
In the notation of \eqref{compare:fSfs}, $f_S^2(\vec q, E) = \delta(E - \Delta E_{\vec n}) f_s^2(\vec q)$ can be written in terms of a 3d momentum form factor,
\begin{align}
f_s(\vec q) &= \left\langle \Psi_{\vec n}(\vec r) \Big| e^{i \vec q \cdot \vec r } \Big| \Psi_0(\vec r) \right\rangle
= \left\langle \tilde{\Psi}_{\vec n}(\vec k + \vec q) \Big| \tilde{\Psi}_0(\vec k) \right\rangle,
\end{align}
which in this example can be evaluated analytically.  
The result:
\begin{align}
f_s(\vec q) &=  e^{i \vec q \cdot \vec L /2 } \prod_{j = x,y,z} \frac{e^{-i q_j L_j /2} + (-1)^{n_j} e^{i q_j L_j /2} }{-2 i} \left[ \frac{2 q_j L_j}{(q_j L_j)^2 - \pi^2 (n_j-1)^2 } - \frac{2 q_j L_j }{(q_j L_j)^2 - \pi^2 (n_j+1)^2 } \right] ,
\label{fS:box}
\end{align}
where $\vec L \equiv( L_x , L_y , L_z)$. 

As a sanity check, consider the $\vec q \rightarrow \vec 0$ limits of this form factor. When the final state matches the ground state, $n_j = 1$, then $f_s(\vec 0) = \left\langle \Psi_0 | \Psi_0 \right\rangle = 1$. 
This is true for \eqref{fS:box}: each term in the product  reduces to $\sinc(q_j L_j/2)$ in the $q_j \ll L_j^{-1}$ limit, so the product approaches $1$ as $q\rightarrow 0$. 
In the $\vec n \neq (1, 1, 1)$ case, where the final state is not the ground state, the orthogonality of the energy eigenstates implies $f_s(\vec 0) = 0$.
From the product in \eqref{fS:box}, we see that taking $q_j = 0$ with $n_j \neq 1$ sets this term to zero, so that the product vanishes as expected. 

What matters for the rate calculation is $|f_s|^2$.
Noting that $\cos^2 (\phi ) = \sin^2(\phi \pm \frac{\pi}{2} (n \pm 1) )$ for even $n$, and $\sin^2(\phi) = \sin^2(\phi \pm \frac{\pi}{2}(n \pm 1) )$ for odd $n$,
each term  can be rewritten to be explicitly finite for all $\vec q$:
\begin{align}
f_s^2(\vec q) &= \!\!\prod_{j = x, y, z} \left[ \frac{\sinc\left( \frac{|q_j L_j| - \pi (n_j - 1) }{2} \right) }{1 + \pi(n_j - 1)/|q_j L_j|  }  +  \frac{\sinc\left( \frac{|q_j L_j| - \pi (n_j + 1) }{2} \right) }{1 + \pi(n_j + 1)/|q_j L_j|  }  \right]^2. 
\label{fS2:box}
\end{align} 
In the large $q_j \gg L_j^{-1}$ limit, $f_j^2(q_j) \propto 1/q_j^2$. So, along an arbitrary direction $\hat q$ not aligned with any of the axes, $f_s^2(\vec q) \propto 1/q^6$ once  $q_j \gg L_j^{-1}$ for all three $j = x,y,z$. However, on-axis, i.e.~$\hat q \approx \pm \hat x, \pm \hat y, \pm \hat z$, the form factor falls off more slowly, as $f_s^2(\vec q) \propto 1/q^2$.

\begin{figure}
\centering
\includegraphics[width=0.98\textwidth]{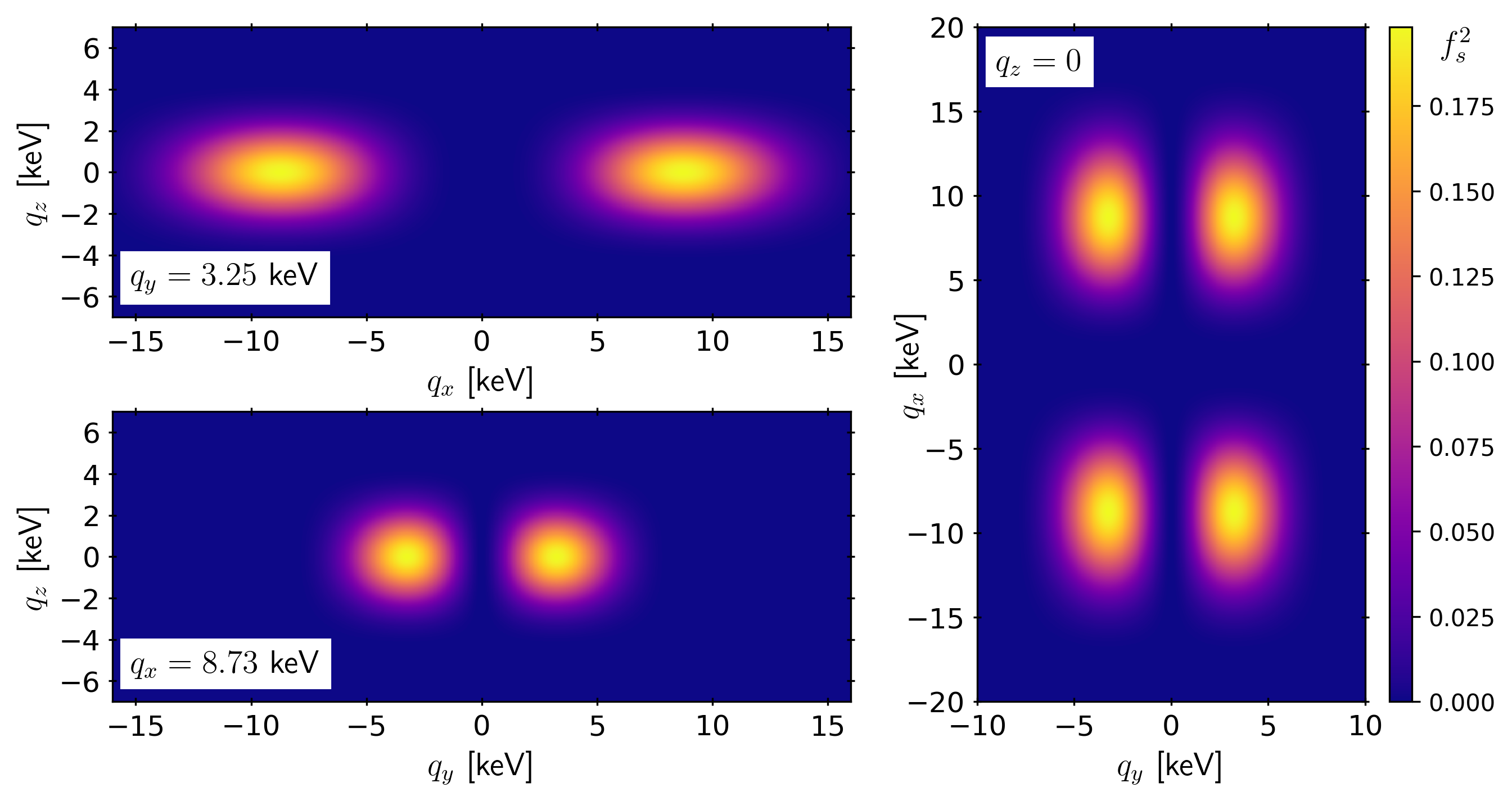}
\caption{ Values of $f_s^2(\vec q)$ for the $\vec L = (4 a_0, 7 a_0, 10 a_0)$ particle-in-a-box example in the $\vec n = (3, 2, 1)$ excited state, evaluated on the planes $q_x=8.73\,\text{keV}$, $q_y =3.25\,\text{keV}$, and $q_z =0$ (clockwise from lower left corner) for scattering to the $n_z = 2$ excited state. 
Each plane passes through the maxima of $f_s^2$, where $f_s^2(\vec q) = 0.198$.
}
\label{fig:fs2}
\end{figure}

In the general case with distinct $L_x$, $L_y$, and $L_z$, the box has three $\mathbbm{Z}_2$ symmetries, i.e.~$q_j \rightarrow - q_j$ for any of $j=x,y,z$. 
As a result, many of the $\langle \ell m | f_s^2 \rangle$ coefficients vanish: those with odd $m$, odd $\ell$, or $m <0$.
For the special case of a square prism, $L_x = L_y$, the symmetry group expands to $\mathbbm{Z}_2 \times \mathbbm{Z}_4 \times \mathbbm{Z}_2$, implying that $\langle \ell m | f_s^2 \rangle = 0$ unless $m$ is a multiple of 4.
These $m$-related simplifications are orientation-specific, occurring only when the $\hat z$ direction is aligned with one of the symmetry axes. 
On the other hand, the restriction of $\ell$ to even values is generic: it is a consequence of the central inversion symmetry $\vec q \rightarrow - \vec q$ of the crystal, which remains a symmetry of the coordinate system even when the detector is rotated.

\medskip

To demonstrate the vector space version of the rate calculation, I take  $m = m_e$ for the particle mass, in an asymmetric box with $(L_x, L_y, L_z) = (4 a_0, 7 a_0, 10 a_0)$, where $a_0 = 1/(\alpha m_e)$ is the Bohr radius. 
The first excited state is $\vec n = (1, 1, 2)$, with $\Delta E \simeq 4.03\, \text{eV}$: its $f_s^2(\vec q)$ is maximized at  $q_z = \pm 4.15\, \text{keV}$ with $q_x = q_y =0$. 
This $f_s^2(\vec q)$ has no small angular features, and so it converges quickly with $\ell$. For a slightly more challenging target, I also model the $\vec n = (3, 2, 1)$ excited state.
Figure~\ref{fig:fs2} shows three cross sections of $f_s^2(\vec q)$ for the $(3, 2, 1)$ final state, evaluated at fixed $q_x = 8.73\, \text{keV}$, $q_y = 3.25 \,\text{keV}$, and $q_z = 0$, respectively. 
These $q_{x, y}$ intersect with one of the four degenerate global maxima of $f_s^2$, where $f_s^2 = 0.198$.

\paragraph{Projections of $\ket{g_\chi}$ for gaussian functions:}

To accelerate the evaluation of the inner products $\langle g_\chi | n \ell m \rangle$ for the velocity model \eqref{eg:gaussian}, I used properties of spherical harmonics to perform the angular integrals analytically. This method is described briefly in Section~\ref{sec:basisGaussian}, and in greater detail in Appendix~\ref{sec:gaussian}. 
The result: for a function $g_\chi$ that is the sum of $k$ gaussians,
\begin{align}
g_\chi(\vec v) &= \sum_{i = 1}^k  \frac{ c_i}{\bar v_i^3 \pi^{3/2} } \exp\left( - \frac{|\vec v - \vec v_i |^2 }{\bar v_i^2} \right), 
\end{align}
the inner product $\langle g_\chi | n \ell m \rangle$ can be simplified to 
\begin{align}
\langle g_\chi | n \ell m \rangle &= \sum_{i = 1}^k c_i \frac{Y_{\ell m }(\hat v_i ) }{v_0^3} \mathcal G_{n \ell}(v_i, \bar v_i) , 
\\
\mathcal G_{n \ell} (v_i, \bar v_i) &\equiv \frac{4}{\sqrt{\pi} } \int_0^\infty\! \frac{v^2 dv }{\bar v_i^3} r_n^{(\ell)}(v) \, e^{ - (v^2 + v_i^2) /\bar v_i^2 } \, i_\ell^{(1)} \! \left(\frac{2 v_i v }{ \bar v_i^2 } \right)  .
\label{gaussian:main}
\end{align}
In this example, $c_i = \{ 0.4, 0.3, 0.2, 0.1\}$; $\vec v_i$ and $\bar v_i$ from \eqref{eg:gaussian} describe the location and width of the gaussians; and, in the equations for $\mathcal G_{n \ell}$, $ v_i \equiv |\vec v_i |$ and $\hat v_i = \vec v_i / v_i$ are the magnitude and unit vector for each $\vec v_i$. 
Here $i_\ell^{(1)}$ is the  $\ell$th modified spherical Bessel function of the first kind, which appears in the spherical harmonic expansion of $\exp(-2 \vec v \cdot \vec v_i/ \bar v_i^2)$; 
and $r_n^{(\ell)}$ and $v_0$ describe the velocity basis functions.
Instead of evaluating 3d integrals for each $(n \ell m)$, we need only to integrate a version of the 1d integral $\mathcal G_{n\ell}$ for each pair of $(n, \ell)$, for each gaussian component of $g_\chi$.
As a result, the evaluation of $\ket{ g_\chi }$ is extremely fast compared to generic $\ket{f_s^2}$.

\subsection{Angular Convergence} \label{sec:convergenceAng}

Narrow features in $g$ map onto wide features in $ | \ell m \rangle$ frequency space. 
In this example,  Streams $(2)$ and $(3)$ are localized on the sky: using $\vartheta_i = \bar v_i /|\vec v_i|$ as an estimate of the angular scale, 
\begin{align}
\vartheta_2 \sim 12^\circ,
&&
\vartheta_3 \sim 3.6^\circ.
\end{align}
From $\vartheta_i$, one expects $\langle g | \ell m \rangle $ to  peak around $\ell_\text{peak} \sim 180^\circ / \vartheta$, with substantial support at neighboring $\ell \sim \mathcal O( \ell_\text{peak})$.
For stream $(2)$, $\ell_\text{peak} \sim 15$; for $(3)$ the expectation is $\ell_\text{peak} \sim 50$. 
To accurately reconstruct either feature, we should continue the harmonic expansion out to some $\ell_\text{max} \gg \ell_\text{peak}$.
Components $(0)$ and $(1)$, on the other hand, do not have narrow angular features, so we can expect them to be accurately reconstructed with relatively small $\ell_\text{max}$. 

\begin{figure}
\centering
\includegraphics[width=0.98\textwidth]{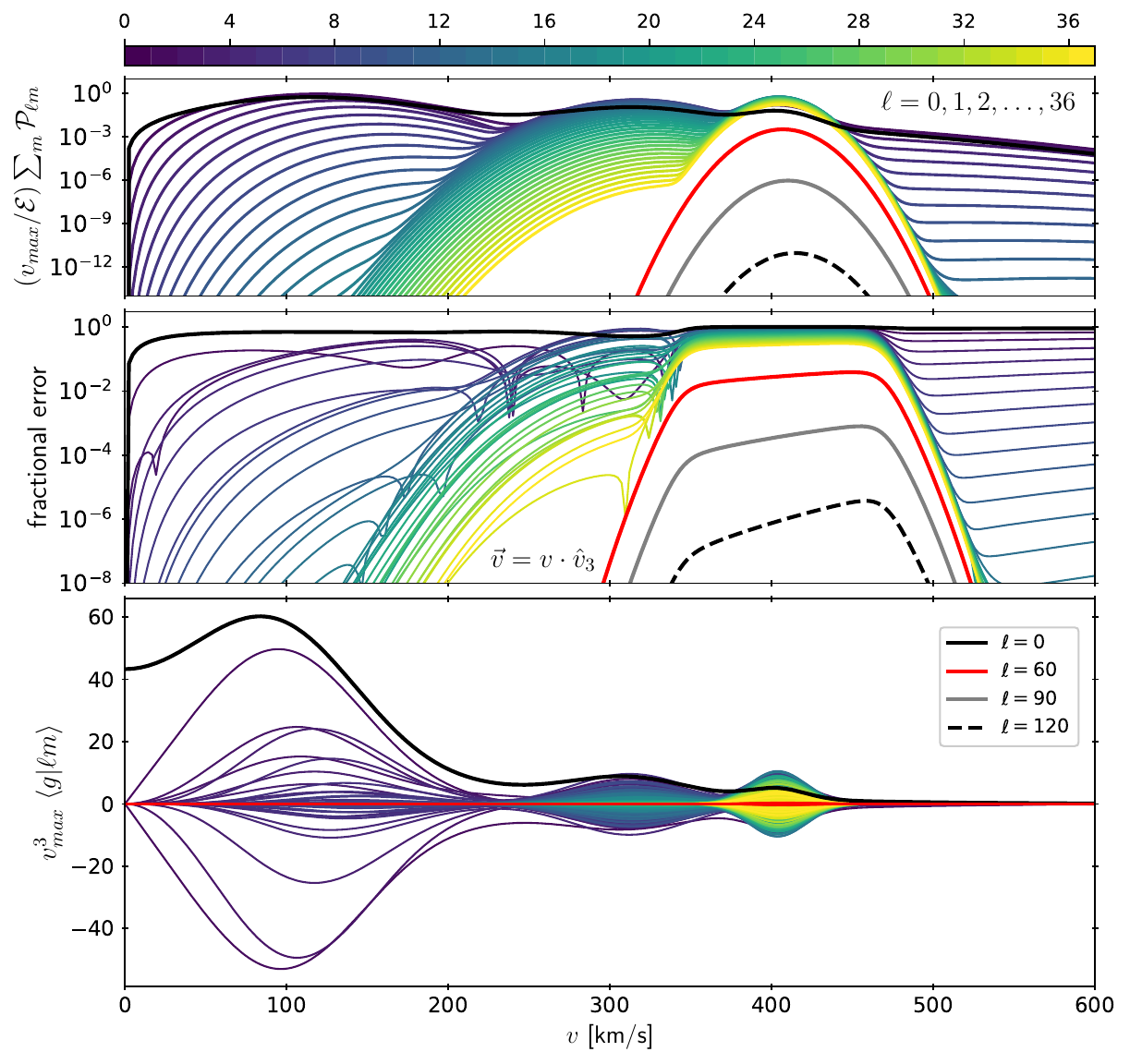}
\caption{A sequence of $g_{\ell m}(v) = \langle g | \ell m \rangle$ radial functions demonstrates the convergence of the spherical harmonic expansion, for $0 \leq \ell \leq 36$ and $- \ell \leq m \leq \ell$ in the \eqref{eg:gaussian} model. 
The lower panel shows $g_{\ell m}(v)$ in units of $v_\text{max}^{-3}$ for all $\ell \geq 36$, with the color bar indicating the value of $\ell$ in each line, except for the special values $\ell = 0$ (in black) and $\ell = 60$ (in red).
The small $v$ behavior, $\langle g | \ell m \rangle \propto v^\ell$, is visible by eye in the $\ell = 0,1,2$ cases.
In the upper panel, the angular power $\sum_m \mathcal P_{\ell m}(v)$ is shown for each $\ell$ in units of $\mathcal E/ v_\text{max}$, with $\ell = 90$ and $\ell = 120$ added to show the convergence in the large~$\ell$ limit. In each case the scale $v_\text{max}$ is arbitrary, set here to $v_\text{max} = 960\, \text{km/s}$. 
In the middle panel, showing the absolute fractional error, $| g(\vec v) - \sum_{\ell m}^{\ell_\text{max}} g_{\ell m}(v) \ket{\ell m} |/g(\vec v)$, the velocity $\vec v$ follows a ray $\vec v = v \hat v_3$ that passes through the center of the narrowest gaussian.
}
\label{fig:angularconvergence}
\end{figure}

Figure~\ref{fig:angularconvergence} shows $\langle g | \ell m \rangle$ for $0 \leq \ell \leq 36$, for the spherical harmonic projection
\begin{align}
g_{\ell m}(v) \equiv
\langle g | \ell m \rangle &= \int \! d\Omega\, Y_{\ell m }(\Omega) \, g(\vec v) 
\end{align}
as a function of $v$. 
The upper panel shows the distributional power from \eqref{def:power},
\begin{align}
\mathcal P_{\ell m}(u) &\equiv  u^2 \, \langle g | \ell m \rangle^2,
&
\sum_{\ell m} \int\! du\, \mathcal P_{\ell m} (u) = \mathcal E[g],
\end{align}
summed over $m = - \ell , -\ell + 1 , \ldots, \ell$: i.e.
\begin{align}
\mathcal P_\ell(u) &\equiv \sum_{m = -\ell}^\ell \mathcal P_{\ell m }(u). 
\end{align}

From Figure~\ref{fig:angularconvergence}, it is clear that the $(0)$ and $(1)$ components of $g(\vec v)$ are well described by the first $\ell \lesssim 10$ angular modes. In the inner $v \lesssim 200\, \text{km/s}$ region of the main panel, every $g_{\ell m}(v)$ with $\ell \gtrsim 10$ is indistinguishable from zero, and $\mathcal P_\ell(v)$ drops quickly for increasing $\ell$. 
At larger $v > 200\, \text{km/s}$, on the other hand, 
Gaussians~(2) and~(3) generate substantial contributions to the $\ell > 10$ harmonics. 
For $(2)$, most of the support is on $10 \lesssim \ell \lesssim 20$, exactly as expected from the characteristic $\bar v_i/v_i \sim 12^\circ$ angular scale. 
For the narrowest gaussian, $(3)$, the power $\mathcal P_{\ell m}$ peaks around $\ell \sim 25$, and is still substantial at $\ell = 36$. 

By $\ell = 60$ the contributions from $g_{\ell m}$ are relatively small, leading to a barely perceptible thickness in the red line on the lower panel. The upper panel of Figure~\ref{fig:angularconvergence} shows that the power has decreased to $\mathcal P_{\ell = 60} \sim 10^{-3} \mathcal E/v_\text{max}$.
At $\ell = 90$, the power near $v\approx 400\, \text{km/s}$ has decreased to one part in $10^6$, suggesting that the local values of $g \simeq \sum_{\ell m} g_{\ell m} \ket{\ell m}$ would be accurate to about one part in $10^3$, i.e.~$\propto \sqrt{\Delta \mathcal E}$. 
Likewise,  $\mathcal P_{\ell = 120} \sim 10^{-12} \mathcal E/v_\text{max}$ suggests that Gaussian~(3) is resolved to a local precision of about $10^{-6}$ at this level of the expansion. 

The middle panel of Figure~\ref{fig:angularconvergence} shows the local relative accuracy, 
\begin{align}
\big|\text{fractional error}\big| &\equiv \frac{1}{g(\vec v) } \left| g(\vec v) - \sum_{\ell m}^{\ell_\text{max}} g_{\ell m}(v) \ket{ \ell m } \right| ,
\end{align}
as a function of $\vec v = v \hat v_3$. This line in $\vec v$ is chosen to pass directly through the center of the narrow Gaussian~(3). 
Figure~\ref{fig:angularconvergence} demonstrates that $\ell = 120$ captures the difficult region of $g(\vec v)$ to a precision of one part in $10^6$, while $\ell = 90$ has a fractional error no larger than $10^{-3}$.
Elsewhere, away from the peak at $400\, \text{km/s}$, the harmonic expansion converges more quickly.
Comparing the large $\ell$ limits of the middle and upper panels confirms that the power $\mathcal P_{\ell m}$ is a useful proxy for the local error: in each case, the fractional error scales like $\sqrt{\mathcal P_\ell }$.

For $\ell \gg 50$, the spherical harmonics oscillate more quickly than any of the features in $g(\vec v)$. This is the regime where the accuracy of the 1d Fourier series scales like $1/N_{\ell m}$, with $N_{\ell m}$ the number of spherical harmonic modes. 
In the middle panel, the $\ell > 60$ fractional errors scale like $1/\ell^2$: and, noting that $N_{\ell m} = (\ell_\text{max} +1 )^2$ when including all  $\ell \leq \ell_\text{max}$, $|m| \leq \ell$, 
we find a scaling 
\begin{align}
\Delta \mathcal E \propto \frac{1}{N_{\ell m}^2},
&&
\text{max}\left( |\text{fractional error}|\right) \propto \sqrt{ \Delta \mathcal E} \sim \frac{1}{N_{\ell m}} 
\end{align}
for resolving narrow isolated sources, e.g.~Gaussian~(3).

\begin{figure}
\centering
\includegraphics[width=0.98\textwidth]{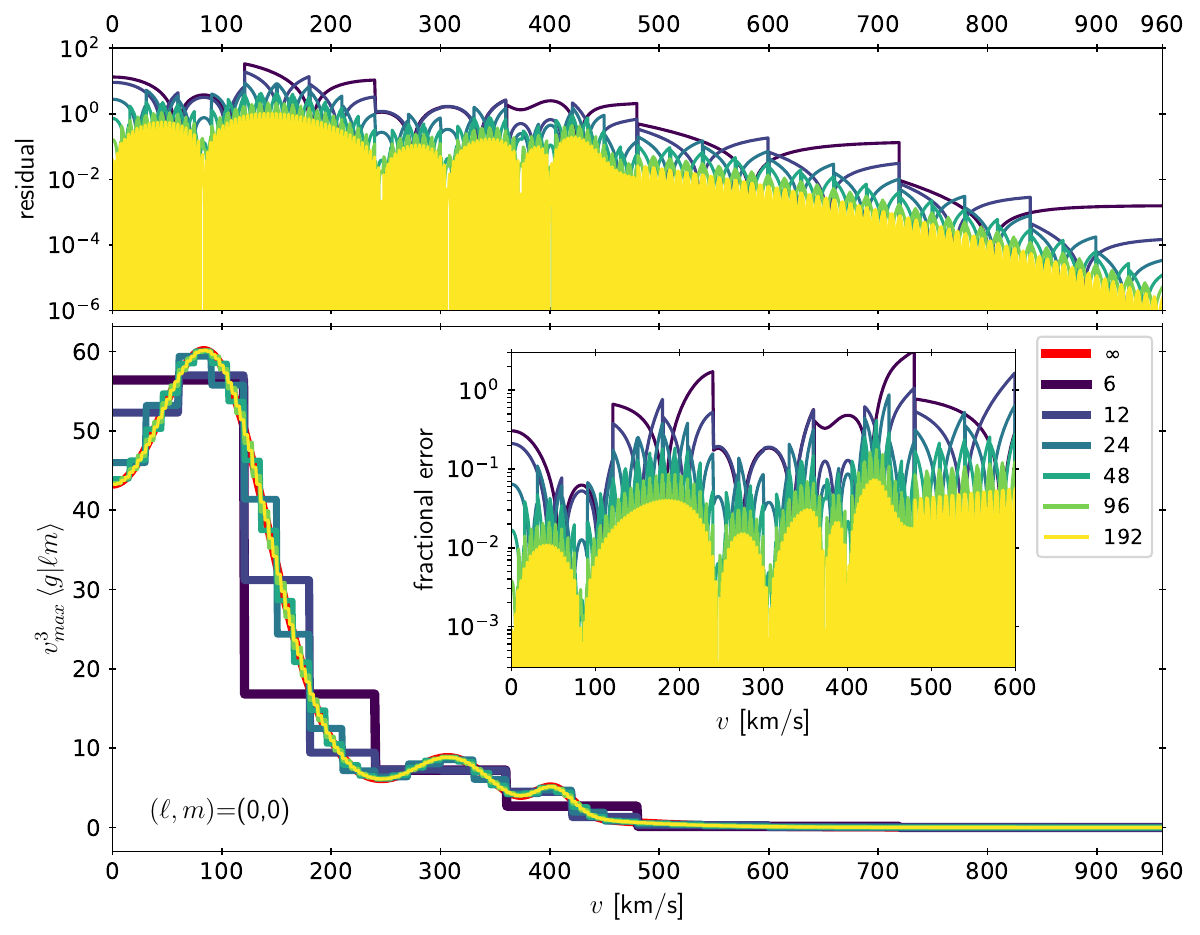}
\caption{Comparing $\sum_n \langle g | n \ell m \rangle \ket{n} $ against the continuous function $g_{\ell m}(v) = \langle g | \ell m \rangle$ to test the convergence of the radial basis expansion, for the four-gaussian function $g(\vec v)$ from \eqref{eg:gaussian}. This example uses $\ell = m = 0$, with $v_\text{max} = 960\, \text{km/s}$ for the basis functions. Because the interesting features of $g(\vec v)$ are concentrated at $v \lesssim 500 \, \text{km/s}$, we take $n_\text{coeffs} = 6, 12, 24, \ldots$ so as to provide finer resolution in the $v \leq \frac{1}{2} v_\text{max}$ region. 
In the main panel, the functions $\sum_n \langle g | n \ell m \rangle \ket{n}$ converge towards the exact result, $g_{\ell m}(v)$ (shown underneath in red). 
The upper panel shows the absolute value of the residual function, $| g_{\ell m}(v) - \sum_n \langle g | n \ell m \rangle \ket{n} |$, while the inset shows this quantity divided by $g_{\ell m}(v)$. 
Starting around $n_\text{coeffs} \gtrsim 24$, each successive generation of wavelets (doubling $n_\text{coeffs}$) reduces the error by a factor of 2, as predicted in Section~\ref{sec:extrapolation}. 
For $n=96$ and $n=196$, the fractional error plot is essentially showing $g_{\ell m}'(v)$, with the smallest error where $g_{\ell m}'(v) \approx 0$. 
}
\label{fig:convergence}
\end{figure}

\subsection{Radial Convergence and Wavelet Extrapolation} \label{sec:radconvergence}

Figure~\ref{fig:convergence} demonstrates the convergence of the radial function expansion, comparing the exact value of $\langle g_\chi | \ell m \rangle$ to the inverse wavelet transformation $\sum_{n=0}^{n_\text{max}} \langle g_\chi | n \ell m \rangle \ket{n}$. The absolute value of the residual, $g_{\ell m}(v)  - \sum_{n=0}^{n_\text{max}} \langle g_\chi | n \ell m \rangle \ket{n} $, is shown in the upper panel, while the fractional error is shown in the inset. 
Each generation doubles the number of coefficients included in the sum, with $n_\text{coeffs} = n_\text{max} + 1$ given values of $6, 12, 24, \ldots, 192$. 
This half-integer number of generations, $n_\text{coeffs} = \frac{3}{2} 2^\lambda$, was chosen so as to provide better precision at $v \leq 480\, \text{km/s}$, where $g_\chi(\vec v)$ is largest.

With $n_\text{coeffs} \leq  24$ the wavelet reconstruction is not particularly accurate, though it does reflect the coarsest features of the function.
To consistently resolve $g_{\ell m}(v)$ with better than $10\%$ relative precision, one must use $n_\text{coeffs} = 96$, or $n_\text{coeffs} = 192$ in the vicinity of the narrow gaussian at $v \sim 400 \, \text{km/s}$. 
Regions with $g_{\ell m}'(v) \approx 0$ are easier to model accurately: see the dips in fractional error where $g_{\ell m}(v)$ reaches a local maximum or minimum. 

Starting around $n_\text{coeffs} = 48$, the fractional error of each successive wavelet generation is reduced by factors of $2$, exactly as Section~\ref{sec:extrapolation} leads us to  expect. Following the notation of \eqref{haar:extrap16}, and noting the ``oscillations'' in this $g_{\ell m}(v)$ on scales of about 100\,km/s,  
one might select $\delta x \sim (100/960) / 2\pi \approx 1/60$ as the scale at which the Taylor series expansion becomes precise, i.e.~$\lambda \approx 6$. 
For $\lambda \geq 7$ (i.e.~$n_\text{coeffs} = 96, 192, \ldots$) the value of $\langle g_{\ell m} | n \rangle$ should be fairly well approximated by the linear term in \eqref{sphH:cubic}.

\begin{figure}
\centering
\includegraphics[width=0.88\textwidth]{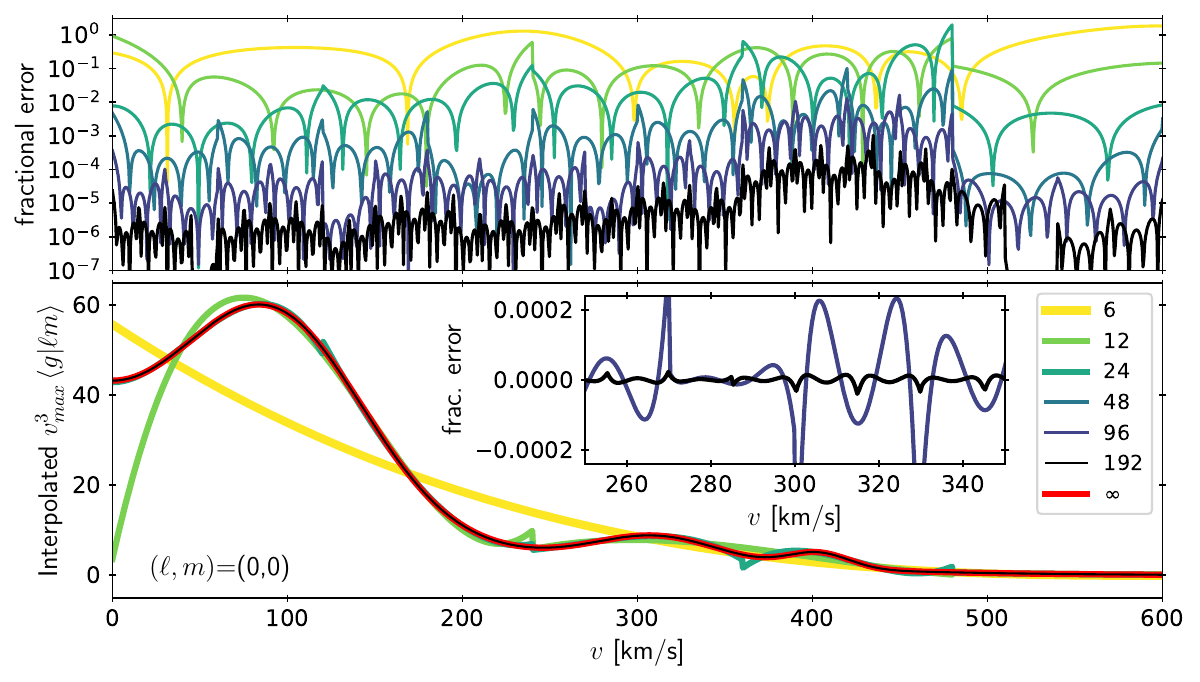}
\caption{The derivatives $g_{\ell m}^{(p)}$ are extracted from the $n_\text{coeffs}$ using the cubic method, and used to define interpolating functions $\bar g_{\ell m}(v)$. At larger $n_\text{coeffs}$, the local accuracy improves $\propto n_\text{coeffs}^3$, as shown in the upper panel. 
The inset reveals the residual functions $g_{\ell m}(v) - \bar g_{\ell m}(v)$ to have cubic or quartic profiles within each bin. 
For $n_\text{coeffs} \geq 24$, $\bar g_{\ell m}$ matches $g_{\ell m}$ up to a few percent (or better).
}
\label{fig:extrapolation}
\end{figure}

This is the regime in which the cubic extrapolation method outlined in Section~\ref{sec:extrapolation} becomes precise.
Figure~\ref{fig:convergence}, for example, suggests that two additional wavelet generations ($n_\text{coeffs} \rightarrow 4 \times 192$) would reduce the fractional error near $v\sim 200\, \text{km/s}$ down to $1\%$, 
while to reach a precision of $10^{-3}$ would require about five new generations, i.e.~$n_\text{coeffs} \rightarrow 32 \times 192 = 6144$. 
Rather than integrating $\langle g_{\ell m} | n \rangle$ for each coefficient, 
wavelet extrapolation uses \eqref{sphH:cubic} to estimate all of these coefficients from the latter generations of the wavelet expansion, e.g.~$48 \leq n < 192$ for the cubic method. 
Relying as it does on simple linear algebra, the wavelet extrapolation is essentially instantaneous, when compared to numeric integration.

Figure~\ref{fig:extrapolation} demonstrates the great improvement in accuracy made possible by the cubic wavelet extrapolation method. In these plots the local values of the $g_{\ell m}^{(1,2,3)}$ derivatives are extracted from the $ n \leq n_\text{coeffs}$ wavelet coefficients, following the method of Section~\ref{sec:extrapolation}, and used to define interpolating functions $\bar g_{\ell m}(v)$ within each wavelet's base of support. 
The $n_\text{coeffs} = 6, 12$ interpolations are (unsurprisingly) rather bad, while the $n_\text{coeffs} > 24$ versions are increasingly precise.
The exact value from \eqref{gauss:lm} is shown in red; it is barely distinguishable even from the $n_\text{coeffs} = 24$ version, except at $v=0$ and a couple isolated points. For $n \geq 48$ we must rely on the upper panel, which shows the fractional error on a logarithmic scale, to identify any discrepancy between the function and its extrapolation-assisted inverse wavelet transformation. The inset shows the linear-scale fractional error for just $n=96$ and $n=192$. By $n=192$, the relative error is on the order of $10^{-4}$, and decreasing by an order of magnitude with each new generation of coefficients.
This is a substantial improvement from Figure~\ref{fig:convergence}: in that example, which did not use any extrapolation methods, the relative local precision at $n=192$ was only $10^{-2}$--$10^{-1}$.

Even better precision can be achieved through higher-order methods, e.g.~the seventh-order \eqref{fdelta:seventh}, or by using marginally more sophisticated cubic methods (e.g.~spline functions with continuous $\bar g_{\ell m}$ interpolations).
Section~\ref{sec:future} discusses an alternative: for particularly fine precision, the wavelet expansion can be smoothly capped off by expanding the basis $\{\phi \}$ to include a few orthogonal polynomials defined within each piecewise-constant interval of the inverse wavelet transformation.
The order (linear, cubic, etc.) is controlled by the number of polynomials included in the series.
This concentrates the information from the $k$th order extrapolation into $k$ additional sets of basis functions, rather than some exponentially larger number of Haar wavelets.

\subsection{Local Accuracy} \label{sec:accuracy}

As a final measure of convergence, we can check how well the wavelet-harmonic expansion matches the original function $f(\vec u)$, for some collections of points $\vec u$. Local accuracy of the inverse wavelet transformation is a sufficient, though not necessary, condition for the accuracy of the scattering rate calculation. For example, accurate representations of $g_\chi$ and $f_S^2$ may require different values of $\ell_\text{max}$, and the angular diagonalization $\mathcal M \rightarrow \mathcal I^{(\ell)}$ allows us to truncate the expansions at the smaller of the two $\ell_\text{max}$.
In any case, it is prudent to ensure local accuracy of $\ket{ f_S^2}$ and $\ket{g_\chi}$, so that the tabulated values can later be used in other settings (e.g.~paired with functions with greater support at large $\ell$).

\begin{figure}
\centering
\includegraphics[width=0.98\textwidth]{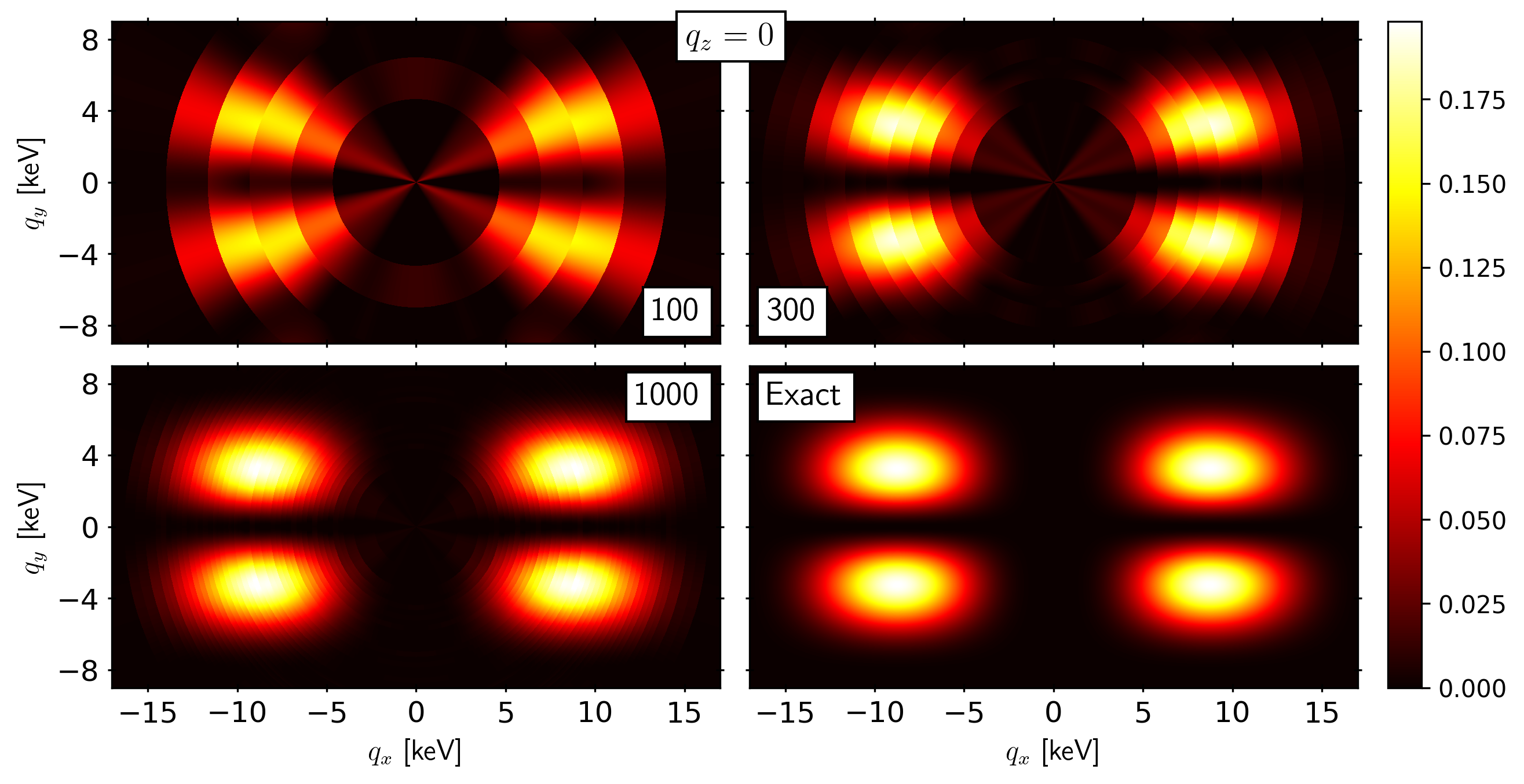}
\caption{
Comparing the $\vec n = (3, 2, 1)$ momentum form factor $f_s^2(\vec q)$ (``Exact'') to its inverse wavelet-harmonic transformation $\sum_{\varphi} \langle \varphi | f_s^2 \rangle \ket{\varphi}$ using the $N$ most important coefficients, for $N = 100, 300, 1000$, evaluated on the $q_y = 0$ and $q_x = 0$ planes (left and right columns, respectively). 
Although $N=100$ correctly identifies the locations of the important features in $f_s^2$, resolving the shapes of each feature requires $N=300$ or $N=1000$. 
}
\label{fig:accuracy}
\end{figure}

A one-dimensional version of this local accuracy test was performed in the middle panel of Figure~\ref{fig:angularconvergence} for $g_\chi(\vec v)$, for a ray $\vec v = v \hat{v}_3$ passing through the center of the narrowest gaussian stream.
For this demonstration, I switch to the particle-in-a-box form factor $f_s^2(\vec q)$ from \eqref{fS2:box}, expanded in a wavelet basis cut off at $q_\text{max} = 10 (\alpha m_e) \simeq 37.3\, \text{keV}$. As in Figure~\ref{fig:fs2}, I focus on fitting the $\vec n = (3,2,1)$ excited state, rather than the lowest-lying $(1,1,2)$ excited state, because its structure is less trivial: there are more nodes, and the larger $n_x$ pushes the maxima towards larger $q$.
Due to the rectangular symmetries of the box, it is still the case that $\langle f | n \ell m \rangle = 0$ for odd $\ell$, odd $m$, or $m < 0$.

After finding the coefficients for all $n < 2^{10}$ and all even $\ell \leq 36$, Figure~\ref{fig:accuracy} shows the inverse wavelet-harmonic transform calculated from only the $N = 100, 300, 1000$ largest $n\ell m$  coefficients, plotted here on the $q_z = 0$ plane.
With $N=100$, the maxima at $q_x = \pm 8.73\, \text{keV}$, $q_y = \pm 3.25\, \text{keV}$ are reproduced with the correct magnitude, but the $q \approx 0$ region and the peaks themselves are poorly resolved.
At $N=300$, all of the secondary features away from the origin are resolved distinctly, but the small-$q$ region retains a spurious conical shape. 
By $N=1000$ the accuracy is much improved. Each peak has the correct shape, even on the margins. 

\begin{figure}
\centering
\includegraphics[width=0.98\textwidth]{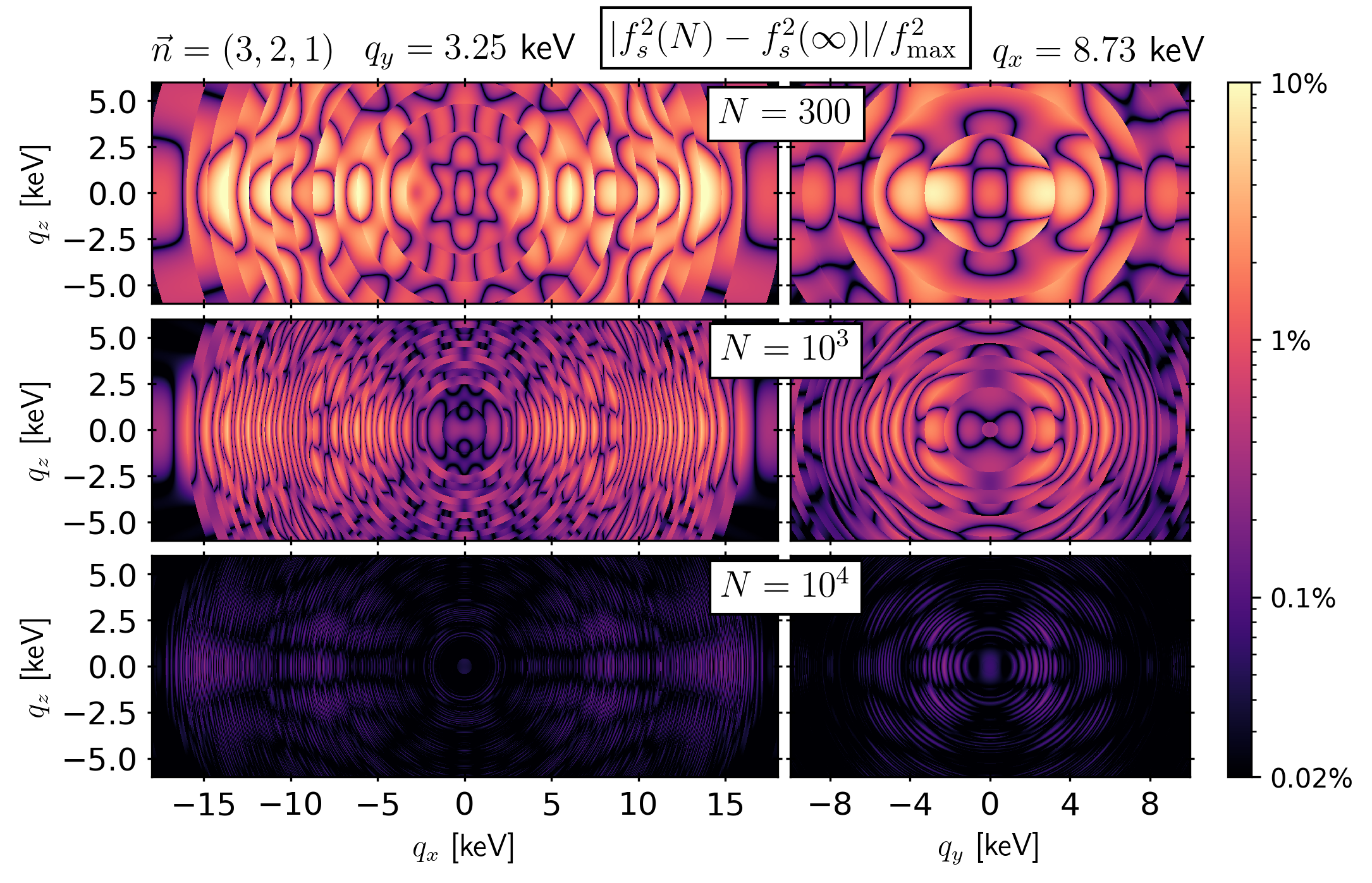}
\caption{
Absolute difference between $\sum_{\phi} \langle \phi | f_s^2 \rangle \ket{\phi}$ and the exact $f_s^2(\vec q)$, normalized by $f_\text{max}^2 = 0.1984$, using only the largest $N = 300, 10^3, 10^4$ coefficients $\langle \phi | f_s^2 \rangle$, for the transition to the $\vec n = (3, 2, 1)$ excited state. 
Each panel shows a 2d slice in momentum space that passes through the point $\vec q = (3.25\,\text{keV}, 8.73\,\text{keV}, 0)$, with the left and right columns (respectively) keeping $q_y$ and $q_x$ constant.
}
\label{fig:relaccuracy}
\end{figure}

Figure~\ref{fig:relaccuracy} shows the difference between $f_s^2(\vec q)$ and its wavelet-harmonic representation, 
$|f_s^2(\vec q) - \sum_{n\ell m} \langle n \ell m | f_s^2 \rangle \ket{n\ell m}|/ f_\text{max}^2$,
normalized by $f_\text{max}^2 \equiv \text{max}(f_s^2(\vec q)) = 0.1984$.
With $N=300$ the largest absolute difference is $15.6\% f_\text{max}^2$ (at points in the $q_y = 3.25\, \text{keV}$ plane), 
while at $N=10^3$ the maximum difference decreases to $4.9\%$. Not pictured here, the $N=100$ example from Figure~\ref{fig:accuracy} has absolute differences as large as $34.8\% f_\text{max}^2$.
At $N=10^4$, the difference between the original function and its wavelet-harmonic expansion is everywhere smaller than $0.31\% f_\text{max}^2$.

\paragraph{Angular Convergence:}

\begin{figure}
\centering
\includegraphics[width=0.98\textwidth]{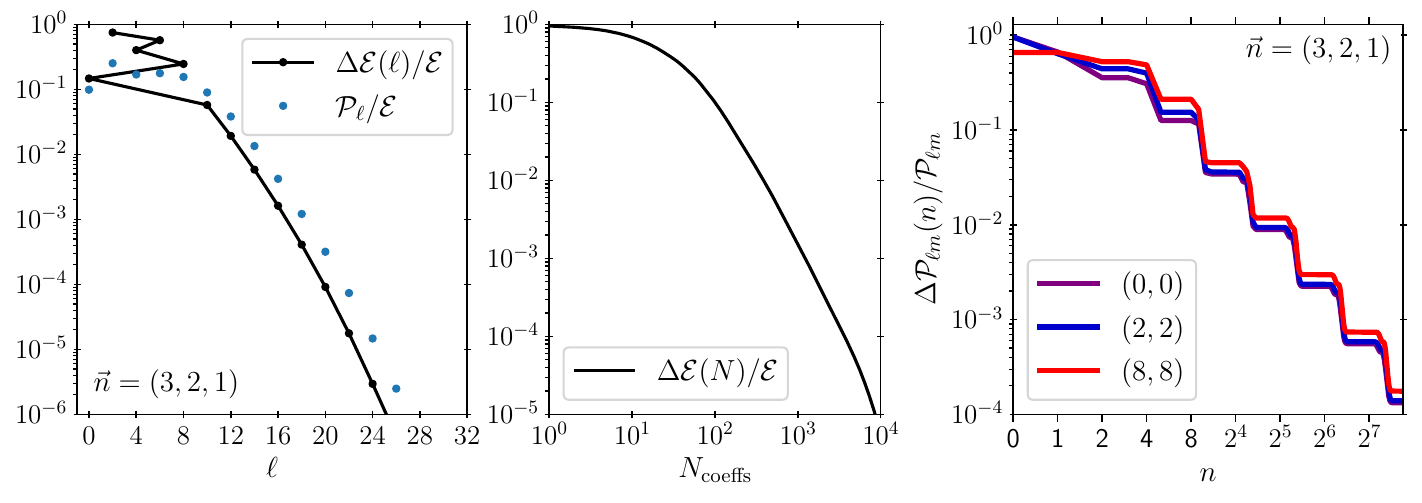}
\caption{Convergence of the spherical harmonic and wavelet expansions, for the $\vec n = (3, 2, 1)$ transition form factor $f_s^2(\vec q)$. \textbf{Left:}  the angular power $\mathcal P_\ell$ is shown (in blue) for each $\ell$, in units of $\mathcal E$.
In black, $\Delta \mathcal E(\ell)$ is defined as the difference between $\mathcal E$ and the sum $\sum_{\ell} \mathcal P_{\ell}$. Here the $\ell$ sum is ordered based on the size of $\mathcal P_\ell$, beginning with $\ell = 2, 6, 4, 8, 0, 10, \ldots$.
\textbf{Center:} An alternative version of $\Delta \mathcal E(N)$, summing over the $N_\text{coeffs}$ largest values of $\langle f_s^2 | n \ell m \rangle^2$. Here the coefficients in the sum are ordered by decreasing magnitude. 
\textbf{Right:} The rate of convergence in $n$ is shown for three of the $(\ell, m)$ angular modes: $(2, 2)$, which has the greatest total $\mathcal P_{\ell m}$; the isotropic mode, $\ell =0$; and $(8, 8)$, chosen because its convergence is slightly delayed compared to the others.
Each line shows $\Delta \mathcal P_{\ell m}(n)$  defined in \eqref{deltaPlmn}, normalized by the total angular power $\mathcal P_{\ell m} \simeq \sum_{n=0}^{1023} \mathcal P_{n \ell m}$. 
}
\label{fig:f2convergence}
\end{figure}

Rather than repeating every aspect of the Section~\ref{sec:convergenceAng} analysis for the particle-in-a-box $f_s^2(\vec q)$,  Figure~\ref{fig:f2convergence} summarizes the primary results. 
Thanks to the analytic form for $f(\vec q)$, the distributional norm can be calculated directly from the integral $\mathcal E = \int\! d^3q \, (f_s^2)^2$. The result, $\mathcal E \simeq 0.10123 \, (\alpha m_e)^3$, can be used to quantify the global convergence  of the wavelet-harmonic expansion.

From the original set of $N_\text{tot} = 2^{10} \cdot 19 \cdot 10 = 194560$ coefficients ($0 \leq n \leq 1023$; $0 \leq \ell \leq 36$ for even $\ell$; $0 \leq m \leq \ell$ for even $m$),
Figure~\ref{fig:f2convergence} shows two versions of the ``missing distributional energy,'' 
\begin{align}
\Delta \mathcal E(N) &= \mathcal E - \sum_{i}^N q_0^3 \langle f_s^2 | \phi_i \rangle^2.
\end{align}
The left panel of Figure~\ref{fig:f2convergence} organizes the $\Delta \mathcal E$ based on the polar $\mathcal P_\ell$ angular power, which sums over both $m$ and $n$:
\begin{align}
\mathcal P_\ell &\simeq \sum_{m = -\ell}^\ell \sum_{n= 0}^{1023} q_0^3 \langle f_s^2 | n \ell m \rangle^2, 
\label{eq:Pell}
\end{align}
i.e.\ $\mathcal E - \Delta \mathcal E(\ell) = \sum_{\ell' = 0}^{\ell}  \mathcal P_{\ell'}$. 
Some small imprecision in \eqref{eq:Pell} comes from terminating the radial mode expansion at $1023 \neq \infty$.
Examining the 2d angular power, $\mathcal P_{\ell m}$, one finds it to be concentrated at small $\ell$, with the largest four coefficients given by:
\begin{align}
\mathcal P_{2,2} &\simeq 0.1462\, \mathcal E, 
&
\mathcal P_{2,0} &\simeq 0.1059\, \mathcal E, 
&
\mathcal P_{0,0} &\simeq 0.0988\, \mathcal E, 
&
\mathcal P_{8,8} &\simeq 0.0836\, \mathcal E.
\end{align}
The plot of $\Delta \mathcal E(\ell)$ reaches a plateau at $\ell > 24$, indicating that all of the higher-$\ell$ modes are irrelevant: to improve the accuracy, one must include some of the $n \geq 2^{10}$ radial modes, e.g.~starting with $\ell \leq 8$.

In the center panel, Figure~\ref{fig:f2convergence} shows the version of $\Delta \mathcal E(N)$ that corresponds most closely with the plots of Figure~\ref{fig:accuracy}. The list of $194560$ coefficients is sorted by decreasing size, and $\Delta \mathcal E(N)$ is found by summing over $\langle f_s^2 | \phi_i \rangle^2$ in that order.
The qualitatively accurate $N=10^3$ expansion from Figure~\ref{fig:accuracy} is shown here to account for 99.8\% of the distributional energy, while by $N= 10^4$ the error is reduced to $\Delta \mathcal E \sim 10^{-5} \mathcal E$.

\paragraph{Radial Convergence:}

The right panel of Figure~\ref{fig:f2convergence} shows the rate of convergence in $n$, for three examples of fixed $(\ell, m)$. This plot shows $\Delta\mathcal P_{\ell m}(n)$, defined as
\begin{align}
\Delta \mathcal P_{\ell m}(n) &\equiv  \mathcal P_{\ell m} - \sum_{n'=0}^n q_0^3 \langle f_s^2 | n'\ell m \rangle^2.
\label{deltaPlmn}
\end{align}
In the plot, $\Delta \mathcal P_{\ell m}(n)$ is normalized by the total $\mathcal P_{\ell m} \approx \sum_{n=0}^{1023} q_0^3 \langle f_s^2 | n\ell m \rangle^2$ independently for each $(\ell, m)$.
Three of the four largest modes are included: $(2, 0)$, the largest; $(0,0)$, the isotropic mode; and $(8,8)$, because it converges somewhat more slowly than the others at small $n \leq 4$. From $n \geq 8$, each $\Delta \mathcal P_{\ell m}$ decreases by a factor of roughly 4 with each new wavelet generation, as anticipated by \eqref{eq:METhaar}.

\subsection{Direct Detection Scattering Rate}

As a final demonstration, the velocity distribution $g_\chi$ from \eqref{eg:gaussian} and the momentum form factor $f_s^2$ from \eqref{fS2:box} are combined in a mock direct detection analysis, for a list of DM models $(m_\chi , F_\text{DM})$ including a light scalar mediator $F_\text{DM} \propto 1/q^2$ and a spin-independent contact interaction, $F_\text{DM}= 1$. 
Two practical questions are answered in this section: How fast does the scattering rate converge with $\ell_\text{max}$?
And how fast, in units of evaluations per second, is the wavelet-harmonic integration method?
A third question (what is the maximal daily modulation amplitude?) is also answered, though as it applies only to the toy models of $g_\chi$ and $f_S^2$ the answer itself is not broadly relevant.

This analysis uses the partial rate matrix $K^{(\ell)}$, which is optimized for handling detector rotations, so the calculation proceeds in two parts. First, the matrices $K^{(\ell)}( g_\chi, f_s^2, m_\chi, F_\text{DM})$ are assembled for each combination of models.
Then, for each combination of models, the total scattering rate is calculated from \eqref{def:partialRate} by summing over $\text{Tr}( G^{(\ell)}(\mathcal R) [ K^{(\ell)}]^T )$ for $\ell \leq \ell_\text{max}$,
for some list of rotations $\mathcal R \in SO(3)$. 

For spin-independent DM scattering with $g_\chi(\vec v)$ given in the lab frame, the scattering matrix is diagonal in $\ell$: so, even though $g_\chi(\vec v)$ is maximally asymmetric, its pairing with a center-symmetric detector material (invariant under $\vec q \rightarrow - \vec q$) permits us to drop all terms with odd $\ell$. 
The rotation matrices $G^{(\ell)}_{m m'}$, on the other hand, generally induce mixing between even and odd values of $m,m'$, so it would be incorrect to drop the negative and/or odd values of  $m$ from $g_\chi$ when assembling $K^{(\ell)}_{m m'}$.

\paragraph{Partial Scattering Rates:}
Once $G^{(\ell)}$ and $K^{(\ell)}$ are known, it is very easy to evaluate the scattering rate $R$.
A precursor result is the partial rate $\tilde{R}^{\ell}$ from \eqref{def:partialRate}, i.e.~the contribution to $R$ coming specifically from spherical harmonics with polar index $\ell$.
This is a helpful indicator of how quickly the spherical harmonic expansion is converging.

\begin{figure}
\centering
\includegraphics[width=0.9\textwidth]{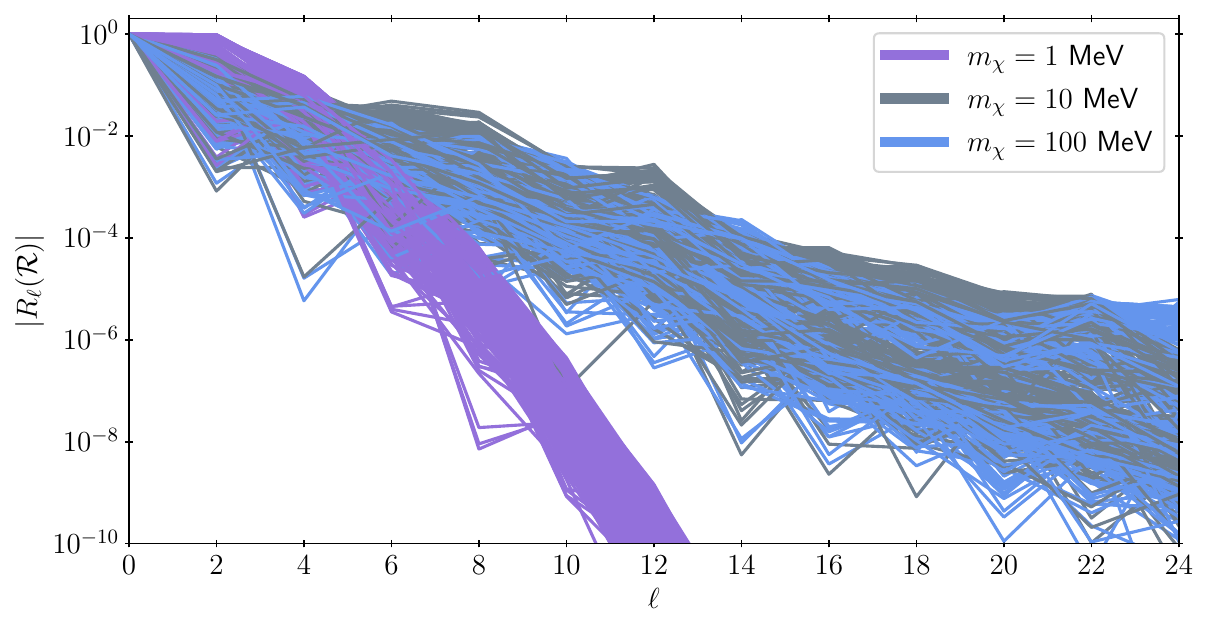}
\caption{
The partial rates $R_\ell$ are shown for $\ell = 0, 2, 4, \ldots, 24$, for an ensemble of $N_{\mathcal R} = 10^2$ detector orientations. The three bands of color correspond to the $m_\chi = 1, 10, 100\, \text{MeV}$ models, with both $F_\text{DM} = 1$ and $F_\text{DM} \propto 1/q^2$ form factors shown together.
Each line shows $|R_\ell| /  \langle R \rangle$, normalized by the angular average rate $\langle R \rangle$, for the four-gaussian $g_\chi(\vec v)$ and the $\vec n = (3, 2, 1)$ form factor, with $\Delta E = 4\,\text{eV}$. 
For each $\ell \geq 2$ there are some rotations $\mathcal R$ that can set $R_\ell(\mathcal R) \rightarrow 0$, so the lower border of each band is a function of $N_{\mathcal R}$. In the $N_{\mathcal R} \rightarrow \infty$ limit, each band would extend to the bottom of the plot.
}
\label{fig:partialrate}
\end{figure}

Figure~\ref{fig:partialrate} shows $|R_\ell| /R$ for a list of 100 rotations.
In this plot each line is colored based on the value of $m_\chi$ for the 1, 10, and $100\,\text{MeV}$ models. No visual distinction is made between the $F_\text{DM} = 1$ and $F_\text{DM} \propto 1/q^2$ form factors, simply because the bands of differing $F_\text{DM}$ largely overlap.
Although the total rate $R = \sum_\ell R_\ell$ and its angular average $R_{\ell = 0}$ must both be positive, the higher $\ell$ modes can be negative.
For all six DM models in the figure, $|R_{\ell}| < 10^{-4} R$ for all of the $\ell \geq 16$ modes.
This is encouraging: even though $\ell_\text{max} \rightarrow 16$ might not be large enough to describe $g_\chi$ and $f_s^2$ individually at the desired precision, it turns out that $\ell_\text{max} = 16$ is perfectly acceptable for the direct detection analysis.

\paragraph{Daily Modulation:}
The wavelet-harmonic method makes it very easy to scan over large numbers of detector orientations, e.g.~to maximize or minimize the expected rate.
For the $g_\chi$ and $f_s^2$ in this demonstration, a scan over $N_{\mathcal R} = 10^3$ orientations finds the following maximum and minimum rates:
\begin{align}
m_\chi = 1 \, \text{MeV},~~F_\text{DM} = 1:& \hspace{1em} 0.41 < R/\langle R \rangle < 1.73 \label{eq:rm1n0}
\\
m_\chi = 10 \, \text{MeV},~~F_\text{DM} = 1:& \hspace{1em} 0.67 < R/\langle R \rangle < 1.36
\\
m_\chi = 100 \, \text{MeV},~~F_\text{DM} = 1:& \hspace{1em} 0.85 < R/\langle R \rangle < 1.23
\\
m_\chi = 1 \, \text{MeV},~~F_\text{DM} \propto 1/q^2:& \hspace{1em} 0.40 < R/\langle R \rangle < 1.85  \label{eq:rm1n2}
\\
m_\chi = 10 \, \text{MeV},~~F_\text{DM} \propto 1/q^2:& \hspace{1em} 0.67 < R/\langle R \rangle < 1.41
\\
m_\chi = 100 \, \text{MeV},~~F_\text{DM} \propto 1/q^2:& \hspace{1em} 0.77 < R/\langle R \rangle < 1.37
\end{align}
Here $\langle R \rangle$ is the exact angular average, 
\begin{align}
\langle R \rangle &= R_{\ell = 0}.
\end{align}
For a detector that is fixed in place in a terrestrial lab, i.e.~co-rotating with the Earth, a more detailed scan over $N_{\mathcal R}$ could uncover which initial orientations of the detector crystal would produce the most statistically significant daily modulation signal. 
A scan over multiple $g_\chi$ models would reveal how sensitive this modulation signal is to each of the four constituent gaussian functions.

Figure~\ref{fig:partialrate} shows that the $1\,\text{MeV}$ models have the largest support at $\ell = 2$ and $\ell = 4$, so one expects these models to have the largest daily modulation amplitudes. This expectation is confirmed by the results of the $N_{\mathcal R} = 10^3$ scan.  The $1\, \text{MeV}$ model is also barely above threshold, requiring at least $v > 850 \, \text{km/s}$ to excite the $4\, \text{eV}$ transition, so its scattering rate is driven by the exponentially small tails of the four-gaussian velocity distribution.
The 10~and $100\, \text{MeV}$ models, on the other hand, are sensitive to a much larger part of $g_\chi(\vec v)$, so the total rate is greater in these cases even though their modulation amplitudes are not as large.

\subsection{Computation Time} \label{sec:comptime}

As described in \eqref{counting:old}, a generic analysis may include $N_{g_\chi}$ and $N_{f_S}$ models of the DM velocity distribution and detector form factor; $N_\text{DM}$ points in the $(m_\chi, F_\text{DM})$ parameter space; and $N_{\mathcal R}$ detector orientations, $\mathcal R \in SO(3)$. 
The matrices $\mathcal I^{(\ell)}$ need to be evaluated for every DM particle model, while $K^{(\ell)}$ is evaluated for every combination of $g_\chi$, $f_s^2$, $m_\chi$ and $F_\text{DM}$. 
In total, the evaluation time for the complete analysis is
\begin{align}
T_\text{total} &= N_{g_\chi} N_{f_S} N_\text{DM} \left( N_{\mathcal R} \cdot  T_\text{eval.R} 
+  T_\text{eval.$K$} \right) 
+ N_\text{DM} \cdot  T_\text{eval.$\mathcal I$} 
+ N_{\mathcal R} \cdot T_\text{eval.$G$} 
+ T_\text{total}^\text{proj} ,
\label{timing:total}
\end{align}
where each $T_\text{eval}$ is the average time needed to evaluate $\text{Tr}( G\cdot K^T)$, $K$, $\mathcal I$, or $G(\mathcal R)$, respectively, 
and where $T_\text{total}^\text{proj}$ is the total time needed to evaluate all relevant $\ket{g_\chi}$ and $\ket{f_S^2}$ vectors that have not been provided.
If none of the $\ket{g_\chi}$ and $\ket{f_S^2}$ coefficients are known, then
\begin{align}
T_\text{total}^\text{proj} &= N_{g_\chi} T_\text{proj.$V$} + N_{f_s} T_\text{proj.$Q$} ,
\end{align}
where $T_\text{proj.$V,Q$}$ are the average computation times for evaluating all relevant $\langle g_\chi | n \ell m \rangle$ or $\langle f_s^2 | n \ell m \rangle$ coefficients in a single example.
The expression $T_\text{eval.$\mathcal I$}$ is for the case of continuum final states.
In the case of $N_j$ discrete final states, or the differential rate $dR/dE$ evaluated at $N_j$ points $E_j$, this can be expressed in terms of $T_\text{eval.$I$}$, the average time needed to evaluate the simpler quantity $I_\star(E_j)$:
\begin{align}
T_\text{eval.$\mathcal I$}  &= N_j   T_\text{eval.$I$} 
\end{align}

Appendix~\ref{sec:timing} provides the evaluation times $T_i$ for various precision goals and values of $\ell_\text{max}$.
Following Figure~\ref{fig:partialrate}, an analysis aiming for $10^{-4}$ precision in the rate can safely drop all $\ell > 16$.
From the Appendix, typical upper values for each $T_i$ for $\ell_\text{max} = 16$ and  $\Delta \mathcal E / \mathcal E \sim 10^{-4}$ are:
\begin{align}
T_\text{eval.R} &\lesssim 5\, \mu\text{s}, 
&
T_\text{eval.$K$} & \lesssim 1\, \text{s} ,
&
T_\text{eval.$G$} &\lesssim 5 \, \text{ms} ,
&
T_\text{eval.$I$} &\lesssim 10^3\, \text{s}.
\label{timing:values}
\end{align}
Only a small subset of the analysis needs to be performed with $\ell_\text{max} = 36$ or $\ell_\text{max} = 24$: once the equivalent of Figure~\ref{fig:partialrate} has been generated for a coarse list of $m_\chi$ values, $\ell_\text{max}$ can be revised downwards to fit the precision goal.
As the number of coefficients in $K^{(\ell)}_{mm'}$ and $G^{(\ell)}_{mm'}$ scale as $\ell^3$, the $N_{\mathcal R}$ evaluation times can be substantially reduced by keeping $\ell_\text{max}$ reasonably small.

This is a relatively conservative estimate for the evaluation time of a realistic analysis, thanks in part to the complexity of $g_\chi(\vec v)$. A more typical analysis using an azimuthally symmetric Standard Halo Model will have a much smaller $T_{eval.K}$, because $K^{(\ell)}_{m m'} \propto \delta_{m, 0}$ vanishes for all nonzero $m$, and because the value of $\ell_\text{max}$ for each precision goal is generally smaller. 

\medskip 

Consider a hypothetical hybrid daily/annual modulation analysis with $20$ different non-SHM $g_\chi$ profiles, $10$ form factors $f_s^2$, and $10^5$ detector orientations, tested at 50 points in $(m_\chi, F_\text{DM})$ parameter space.
Given pre-tabulated $\ket{g_\chi}$ and $\ket{f_s^2}$ vectors,
the analysis can be completed in about
\begin{align}
T_\text{total}(\ell_\text{max} = 16)  \sim 16\, \text{hours},
\label{timing:good}
\end{align}
based on the \eqref{timing:values} values for $T_i$, aiming for $10^{-4}$ precision in the rate. 
Most of the evaluation time is taken up by $\mathcal I^{(\ell)}$. 
For a precision goal of  $\Delta \mathcal E / \mathcal E = 0.3\%$ instead, the evaluation time shrinks to 
\begin{align}
T_\text{total}(\ell_\text{max} = 16)  \longrightarrow  3.0\, \text{hours},
\end{align}
with slightly more than half of the total now coming from the $10^9 \cdot T_\text{eval.R} + 10^5 T_\text{eval.G}$ terms. 
For other precision goals and values of $\ell_\text{max}$, see the tabulated values in Appendix~\ref{sec:timing}. 

Compare this analysis to the alternative: evaluating the five-dimensional integral of \eqref{rate:discrete} for each of the $10^9$ configurations (after using the $\delta$ function to reduce the dimensionality of the original 6d integrand by one).
In this example, the direct integration approach would take
\begin{align}
T_\text{total}^\text{\,slow} &= N_{g_\chi} N_{f_S} N_\text{DM} N_{\mathcal R} \cdot  T_\text{eval.int},
\end{align}
where $T_\text{eval.int}$ is the time needed to integrate \eqref{rate:discrete} once.
Even for this relatively simple example with analytic functions for $g_\chi$ and $f_s^2$, 
Appendix~\ref{sec:timing} finds $T_\text{eval.int} \sim 600\, \text{s}$ when evaluating the rate integral at a similar $\mathcal O(10^{-3})$ precision. 
For the $10^9$ point analysis,
\begin{align}
T_\text{total}^\text{\,slow} &\sim 19\,000\, \text{cpu-years},
\label{timing:bad}
\end{align}
larger by a factor of about
\begin{align}
T_\text{total}^\text{\,slow} /
T_\text{total}^\text{\,wavelet}(\ell_\text{max} = 16) 
\sim 
\text{10--56} \ \text{million},
\end{align}
compared to the $\ell_\text{max} = 16$ wavelet-harmonic analysis.
In the limit of very many detector orientations, $N_{\mathcal R} \gg  10^5$, this ratio asymptotes to $T_\text{eval.int} / T_\text{eval.R}$, which is as large as
\begin{align}
T_\text{eval.int} / T_\text{eval.R}(\ell_\text{max} = 16) \sim 10^8 .
\end{align}
In cases where $\ell_\text{max} < 16$ still provides a good approximation to the total rate, or where the rate integral itself is more complicated, this ratio becomes even larger.

\section{Discussion} \label{sec:future}

Sections~\ref{sec:method}--\ref{sec:demo} have established that the wavelet-harmonic integration method provides an accurate and fast alternative to the old approach for calculating scattering rates (i.e.~integrating \eqref{rate:continuum}).
Comparing \eqref{timing:good} with \eqref{timing:bad}, the wavelet-harmonic version is the obvious choice for any analysis that involves more than a few detector orientations.

Looking to the future, there are several aspects of the wavelet-harmonic method that call for further study.
In particular, the ability to express the orientation-dependent scattering rate in terms of the partial rate matrix has a number of physics implications, several of which are highlighted in Section~\ref{sec:implications}. 
The wavelet-harmonic method can also be generalized to any problem involving integrands which depend linearly on multiple input functions.
Section~\ref{sec:generalization} provides this general version of the vector space integration method,
followed in Section~\ref{sec:gen:formfactor} by a specific application to the detector form factor calculation. 

Section~\ref{sec:basisGaussian} describes another topic for future work: transformations between different bases, $\langle \phi' | \phi \rangle$.
A transformation of this type was used for $g_\chi$ in Section~\ref{sec:demo}, and it reduced the integration time for evaluating $\ket{g_\chi}$ by some orders of magnitude. Annual modulation analyses are another obvious target for basis-related simplification.
Section~\ref{sec:squarepositive} comments briefly on how to ensure that $\ket{g_\chi}$ and $\ket{ f_s^2}$ remain explicitly positive,
and Section~\ref{sec:polycap}  describes how sets of continuous basis functions can be added to the wavelet expansion to achieve an exponential improvement in the precision.

\subsection{Generalization} \label{sec:generalization}

As pointed out in ref.~\cite{Lillard:2025ixi}, the vector space integration method works because the rate integrand depends linearly on the inputs $g_\chi$ and $f_S^2$.
Any integral of this type can be completed this way.
For a generic functional $S$ depending on $k$ input functions $f_i$ and some set of parameters $\vartheta = \{\vartheta_j \}$,
\begin{align}
S(\{f_i\}, \vartheta) &\equiv  \int \! d\Pi \, f_1(\vec x_1) f_2(\vec x_2)  \ldots f_k(\vec x_k) \cdot \hat{\mathcal O}_S(\vec x_1, \vec x_2, \ldots, \vec x_k ; \vartheta), 
\label{eq:genOk}
\end{align}
the integral is factorized by projecting each $f_i$ onto basis functions $\phi^{(i)}$:
\begin{align}
S(\{f_i\}, \vartheta) &= \sum_{\phi^{(1)}} \sum_{\phi^{(2)}} \ldots \sum_{\phi^{(k)}} \langle \phi^{(1)} | f_1 \rangle \langle \phi^{(2)} | f_2 \rangle \ldots \langle \phi^{(k)} | f_k \rangle 
\cdot
\langle \hat{\mathcal O}_S( \vartheta) |  \phi^{(1)}   \ldots \phi^{(k)} \rangle ,
\label{eq:Sfactorizes}
\\
\langle \hat{\mathcal O}_S(\vartheta) |  \phi^{(1)}    \ldots \phi^{(k)} \rangle
&\equiv \int \! d\Pi \, \phi^{(1)} \phi^{(2)}  \ldots \phi^{(k)} \, \hat{\mathcal O}_S(\vartheta) ,
\label{eq:Ointegral}
\end{align}
Here $d\Pi$ represents integration over the phase space for $\vec x_{1 \ldots k}$, and the vectors $\vec x_i$ can be of any dimension.
If the linear operator $\hat{\mathcal O}_S$ has any symmetries, then a good basis choice can block-diagonalize the rank-$k$ tensor $\hat{\mathcal O}_S(\vartheta)$. 

If $S(\{ f_i\}, \vartheta)$ must be evaluated for many different versions of the functions $f_i$, or for a large number of parameter choices $\vartheta$, 
then the factorization inherent to \eqref{eq:Sfactorizes} can greatly simplify the evaluation of $S$. 
This is especially true if the basis functions $\phi^{(i)}$ permit \eqref{eq:Ointegral} to be integrated analytically.

Some classes of problems require not just $S$, but also its derivatives: either with respect to the parameters, i.e.\ $\partial S/\partial \vartheta_j$, or functional derivatives, $\delta S/ \delta f_i$.
In multivariate statistical inference, for example, a measurement of $S$ provides best-fit points in the $\{f_i\}, \vartheta$ spaces, with confidence intervals given by derivatives of the likelihood function with respect to the parameters. 
Mixed partial derivatives, e.g.~$\partial^2 S/ \partial \vartheta_i \partial \vartheta_j$ or $\partial^2 S / \partial f_i \partial \vartheta_j$, indicate the correlations between the inputs to $S$.

Derivatives with respect to $\delta f_i$ can be extracted from the $\hat{O}_S$ tensor directly,
simply by dropping the $f_i$ related sum in \eqref{eq:Sfactorizes}.
That is, 
\begin{align}
\frac{\delta S}{\delta \langle \phi^{(i)}_n | f_i \rangle } &= \langle f_1 f_2 \ldots f_{i - 1} | \phi^{(i)}_n \hat{\mathcal O}_S(\vartheta) | f_{i+1} \ldots f_k \rangle,
\end{align}
where the index $n$ indicates a derivative with respect to the $n$th basis function of type $\phi^{(i)}$. 
Derivatives of the form $\partial \hat{\mathcal O}_S / \partial \vartheta_j$ are given by
\begin{align}
\frac{\partial}{\partial \vartheta_j} \langle \hat{\mathcal O}_S(\vartheta) |  \phi^{(1)}    \ldots \phi^{(k)} \rangle
&\equiv \int \! d\Pi \, \phi^{(1)} \phi^{(2)}  \ldots \phi^{(k)}  \frac{\partial \hat{\mathcal O}_S(\vartheta) }{\partial \vartheta_j} .
\label{eq:partialthetaO}
\end{align}
In some cases this may even be analytically tractable.  Otherwise, it can be evaluated numerically, either from \eqref{eq:partialthetaO} or by using finite-difference methods on $ \mathcal O(\vartheta + d\vartheta)$. 
Drawing the connection to the DM direct detection example, it is straightforward to calculate $\partial \mathcal I^{(\ell)} / \partial m_\chi$, and therefore $\partial R/\partial m_\chi$,  given the analytic form for $I_\star(E)$ derived in Appendix~\ref{sec:Istar}.
The mild tedium involved is still much preferable to the high-dimensional finite-difference methods that would otherwise be required, e.g.~in a global analysis of $(m_\chi, F_\text{DM}, g_\chi)$ models in the case of DM direct detection.

\paragraph{Hyperspherical Wavelet-Harmonic Basis:} 

If the operator $\mathcal O$ transforms in some simple way under rotational symmetries of the arbitrary-dimensional coordinates $\vec{x}_i$, then a generalized hyperspherical wavelet-harmonic basis may provide an optimal basis.
For some orthogonal hyperspherical harmonics $\mathcal Y$ defined on the $S^{p-1}$ sphere, a $p$ dimensional spherical wavelet-harmonic basis function defined on $|\vec u | \leq u_\text{max}$ is given by
\begin{align}
\phi(\vec u) &\equiv h_{\lambda \mu}^{(p)}(u/u_\text{max} ) \, \mathcal Y_{\{\ell\}}(\hat u),
\end{align}
where $\{\ell\}$ represents the $p-1$ angular indices, and where $h_{\lambda \mu}^{(p)}$ is normalized such that
\begin{align}
\int_0^1\! x^{p-1} dx\, h_{n}^{(p)}(x) \, h_{n'}^{(p)}(x) &= \delta_{n n'}.
\end{align}
Here I use the same $(\lambda\mu) \rightarrow n$ integer mapping from \eqref{nLambdaMu}. 
For an explicit form of $\mathcal Y_{j \ell m}(\hat x)$ in four dimensions, see the $Z_{j \ell}^{m}$ of ref.~\cite{Domokos:1967fgx}, or related results in nuclear physics (e.g.~\cite{Viviani:2004vf}).
Just as in \eqref{def:sphericalHaar}, the generalized radial functions are given by:
\begin{align}
{h}_{\lambda \mu}^{(p)} (x) = 
\left\{
\begin{array}{r c l}
+A_{\lambda \mu}^{(p)} && 2^{-\lambda} \mu \leq x <  2^{-\lambda} (\mu + \frac{1}{2} ) ,
\\[4pt]
-B_{\lambda \mu}^{(p)} &&  2^{-\lambda} (\mu + \frac{1}{2} )  < x \leq 2^{-\lambda} (\mu+ 1) ,
\\[4pt]
0 && \text{otherwise} ,
\end{array}
\right.
\end{align}
where
\begin{align}
x_1 = 2^{-\lambda} \mu,
&&
x_2 = 2^{-\lambda} (\mu + \tfrac{1}{2} ), 
&&
x_3 = 2^{-\lambda} (\mu + 1) ,
\end{align}
and 
\begin{align}
A_{\lambda \mu}^{(p)} &= \sqrt{ \frac{p}{x_3^p - x_1^p} \frac{x_3^p - x_2^p }{x_2^p - x_1^p} } \, ,
&
B_{\lambda \mu}^{(p)} &= \sqrt{ \frac{p}{x_3^p - x_1^p} \frac{x_2^p - x_1^p}{x_3^p - x_2^p } } \, .
\end{align}
The constant $n=0$ wavelet takes the value
\begin{align}
h_{n=0}^{(p)}(x) &= \sqrt{ p}.
\end{align}

\subsection{Application to the Momentum Form Factor}  \label{sec:gen:formfactor}

The momentum form factor $f_S^2$ is another difficult problem that can be simplified with wavelet-harmonic integration.  
It depends linearly on initial and final state wavefunctions for the SM particle; and its $\hat{\mathcal O}(\vec q)$ is a simple momentum transfer operator.
Taking the \eqref{def:fgs} expression, with position space wavefunctions $\Psi_i$,
\begin{align}
f_s(\vec q) &= \langle \Psi_s(\vec r) | e^{i \vec q \cdot \vec r} | \Psi_0(\vec r) \rangle ,
\end{align}
and expanding the wavefunctions $\Psi$ in a basis with (complex) spherical harmonics, 
\begin{align}
f_s(\vec q) &= \sum_{n \ell m } \sum_{n' \ell' m'} \langle \Psi_s | n' \ell' m'  \rangle  \langle n \ell m | \Psi_0 \rangle
\sum_{\lambda \mu} 4\pi  i^\lambda \, Y_{\lambda}^{\mu}(\hat q)
\cdot \langle n' \ell' m' | Y_{\lambda}^{\mu}(\hat r)  \, j_\lambda(q r) | n \ell m \rangle .
\label{eq:fslm}
\end{align}
Notice that  $Y_\lambda^\mu(\hat q)$ simply factors out of the $d^3 \vec r$ integral. This is extremely helpful, if the eventual goal is to project $f_s^2(\vec q)$ onto a similar basis of spherical harmonic $\varphi$ functions.

The angular integrals $\langle \ell' m' | Y_\lambda^\mu(\hat r)  | \ell m \rangle$ are given by the Wigner $3j$~symbols,
\begin{align}
\int\! d\Omega \,  Y_{\ell'}^{m' \star} Y_\lambda^\mu Y_{\ell}^m 
&= (-1)^{m'} \sqrt{ \frac{(2 \ell' + 1)(2 \lambda + 1)(2 \ell + 1) }{4\pi} } 
\left( \begin{array}{c c c} \ell' & \lambda & \ell \\ 0 & 0 &0 \end{array} \right)\left( \begin{array}{c c c} \ell' & \lambda & \ell \\ -m' &\mu &m \end{array} \right) .
\label{eg:3j}
\end{align} 
The $3j$ symbols have certain selection rules, which enforce $-m' + \mu + m = 0$ and set $(\ell' + \lambda + \ell)$ to be even.
Calling the right side of \eqref{eg:3j} $c_{\ell' \lambda \ell}^{m' \mu m}$ for conciseness, the form factor is:
\begin{align}
f_s(\vec q) &= \sum_{\lambda \mu} 4\pi i^\lambda 
\sum_{n \ell m } \sum_{n' \ell' m'} c_{\ell' \lambda \ell}^{m' \mu m} \langle \Psi_s | n' \ell' m'  \rangle  \langle n \ell m | \Psi_0 \rangle
\,Y_{\lambda}^{\mu}(\hat q) \cdot \langle n' | j_\lambda(qr) | n \rangle.
\end{align}
The last term with $j_\lambda$ as an operator is a function of $q$, analogous to the $\mathcal I^{(\ell)}$ integrand. 
For radial basis functions $\ket{ n \ell m } = Y_\ell^m(\hat r) \, w_n(r)$, 
this function of $q$ is given by
\begin{align}
\langle n' | j_\lambda(qr) | n \rangle &= \int r^2 dr\, w_{n'}(r) \, j_\lambda(q r) \, w_n(r) .
\label{eq:besseljOp}
\end{align}
If the basis functions $w_n$ are simple enough (e.g.~piecewise-constant), this integral can be completed analytically.
So, the problem of calculating $f_s(\vec q)$ has been replaced with the simpler problem of projecting the wavefunctions $\Psi_i$ onto the $\ket{n \ell m } $ basis functions. 

In many contexts an analytic form for $\Psi_i$ may be known, at least approximately. 
This is the case for the organic chemistry examples of Refs.~\cite{Blanco:2019lrf,Blanco:2021hlm}, where the molecular wavefunctions are approximated as linear combinations of $2p_z$ atomic orbitals, 
and in more precise modern physical chemistry methods where $\Psi_{g, s}(\vec r)$ are approximated by sums of Gaussian functions~\cite{Taketa_1966,Roberts_1967,Dunlap_1990,VAHTRAS1993514}.

Projecting $f_s$ onto a $\ket{\varphi} = \ket{n \ell m}$ basis is now fairly easy, especially because $f_s(\vec q)$ in \eqref{eq:fslm} is already expanded as a sum in real spherical harmonics. 
For the remaining $\ket{n}$ part, one simply adds some $f_s$ radial momentum basis functions  $\tilde{r}_\nu(q)$ to the integral of \eqref{eq:besseljOp}:
\begin{align}
\langle \nu \lambda \mu | f_s \rangle &= 4\pi i^\lambda 
\sum_{n \ell m } \sum_{n' \ell' m'} c_{\ell' \lambda \ell}^{m' \mu m} \langle \Psi_s | n' \ell' m'  \rangle  \langle n \ell m | \Psi_0 \rangle
\,  \mathcal J^{(\lambda)}_{\nu n' n},
\\
\mathcal J^{(\lambda)}_{\nu n' n} &\equiv   \langle \nu n' | j_\lambda(qr) | n \rangle =  \int \! q^2 dq\, r^2 dr\, \tilde{r}_\nu(q) \,w_{n'}(r) \, j_\lambda(q r) \, w_n(r) .
\end{align}
Just like the kinematic scattering matrix $\mathcal I^{(\ell)}$,  $\mathcal J^{(\lambda)}$ can be evaluated analytically for simple position-space $w_n$ functions.

Finally, to get $\langle \varphi | f_S^2 \rangle$ from $\langle \varphi | f_s \rangle$, one needs to supply the algebraic properties of the basis functions, i.e.~the coefficients $c_{ij}^k$ in
\begin{align}
\ket{\phi_i} \ket{\phi_j} &= \sum_{k} c_{ij}^k \ket{\phi_k} .
\label{eq:phialgebra}
\end{align}
This is straightforward for spherical harmonics, and for wavelets it is even easier. If the wavelet basis functions $r_{i}$ and $r_{j}$ do not overlap, then $c_{ij}^k = 0$. If $i \neq j$ overlap, then $c_{ij}^{k} \propto \{ \delta_{ik} \text{ or } \delta_{jk}\}$, for whichever of $i,j$ has the smaller base of support. 
If $i = j$, then $\ket{\phi_i}^2$ includes a sum over all of the $k \leq i$ wavelets with bases of support that overlap with $r_i$.

\paragraph{Connection to the Resolution-of-Identity Approximation:}
Triple integrals of this type arise frequently in physical chemistry, e.g.~when applying the resolution of the identity approximation to simplify the four-center gaussian integrals in Hartree-Fock and density functional theory calculations~\cite{Taketa_1966,Roberts_1967,Dunlap_1990,VAHTRAS1993514,Strout_1995,White_1996,Weigend_1997,WEIGEND1998143,SHAO2000425,Kussmann_2021}. 
For a generic choice of basis functions $\{ \phi_i \}$, the number of nonzero $c_{ij}^k = \langle \phi_k | \phi_i \phi_j \rangle$ coefficients is expected to scale as $N_\phi^3$, where $N_\phi$ is the number of basis functions in the set. 
Especially in precision studies of large molecules, this $N_\phi^3$ factor can become inconveniently large, leading to ongoing efforts in the field of physical chemistry to streamline this type of calculation~\cite{Strout_1995,White_1996,Weigend_1997,WEIGEND1998143,SHAO2000425,Kussmann_2021}. 
For the wavelet-harmonic basis this is less of a problem, because the number of nonzero triple products is much smaller than $N_\phi^3$. 
With $N_\ell$ the number of $\ell$ modes (e.g.~$N_\ell = \ell_\text{max} + 1$)  and $N_n$ the number of radial basis functions $r_n$ included in the expansion, the number of 3d basis functions scales as
\begin{align}
N_\phi \sim N_\ell^2 N_n.
\end{align}
When calculating \eqref{eq:phialgebra} in the wavelet-harmonic basis, the radial and angular parts of the triple product factorize. 
Thanks to the locality of the large $n$ wavelet basis functions, the number of radial triple products scales as $N_n \log_2 N_n$. From the spherical harmonic selection rules, on the other hand, $\langle \lambda \mu | \ell m \rangle \ket{\ell' m'} = 0$ unless $\mu = m + m'$ and $|\ell - \ell'| \leq \lambda \leq \ell + \ell'$.  
Consequently the number of $\langle \lambda \mu | \ell m \, \ell' m' \rangle$ inner products scales as $N_\ell^5$ (divided by a further $\mathcal O(1)$ factor) rather than $(N_\ell^2)^3$.
Together, taking $N_n \sim N_\ell \sim N_\phi^{1/3}$, the number of nonzero $\langle \nu \lambda \mu | n \ell m \, n' \ell' m' \rangle$ triple products scales as 
\begin{align}
N_n \log_2 N_n \cdot N_\ell^5 / \mathcal O(1) \sim N_\phi^{2} \log N_n.
\end{align}
Compared to the $N_\phi^3$ scaling of generic sets of basis functions, this is smaller by a factor of approximately $N_\ell N_n^2$, an improvement that is typically several orders of magnitude. 

For a detailed discussion of the fundamental differences between the wavelet-harmonic integration method and the standard resolution of the identity method in computational chemistry, see the Supplemental Material in ref.~\cite{Lillard:2025ixi}.

\subsection{Basis Transformations} \label{sec:basisGaussian}

Most of the results of this paper follow from representing $\ket{g_\chi}$ and $\ket{f_S^2}$ as vectors in the Hilbert spaces spanned by one set of $\{ \phi \}$ and one set of $\{ \varphi \}$ basis functions. 
In the language of \eqref{eq:genOk}, the Choices~\ref{choice:spherical}--\ref{choice:FDM} made in Section~\ref{sec:genSummary} were designed to optimize the evaluation of the tensor $\phi^{(1)} \ldots \phi^{(k)} \hat{\mathcal O}_k$ --- possibly at the expense of evaluating $\langle \phi^{(i)} | f_i \rangle$ in a less-than-optimal basis.

A more complete description of transformations from one basis to another would be highly useful.
Suppose a function $f$ is well described by basis functions $\psi$. The transformation from the $\{\psi \}$ basis to a $\{\phi\}$ basis is given by:
\begin{align}
| f \rangle &= \sum_\psi \langle \phi | \psi \rangle \langle \psi | f \rangle \ket{\phi} 
\label{eq:basisphi}
\end{align}
where $\langle \phi | \psi \rangle$ are the coefficients of a basis transformation matrix, $\mathcal B^\phi_\psi$, acting on the $\ket{\psi}$ functions. 
This way, $\ket{f}$ and $\hat{\mathcal O}$ can be represented using different, independently optimal basis functions. 

For example, the rectangular box of the $f_S^2(\vec q)$ model in Section~\ref{sec:demo} might lend itself more naturally to a description in Cartesian coordinates (e.g.~\eqref{fS2:box}). 
That is, the $\ket{f} = \langle \psi | f \rangle \ket{\psi}$ series could in principle converge more quickly with oscillatory Cartesian basis functions, rather than wavelet-harmonics.
If the coefficients  $\mathcal B^\phi_{\psi_i} = \langle n \ell m | \psi_i \rangle$ can be calculated analytically, then it may be faster to find $\langle \phi | f \rangle$ from \eqref{eq:basisphi} rather than calculating the inner product with $\phi$ directly. 

There are two especially promising applications: families of functions $f_i$ related to each other by simple operations; and cases where $f$ is already provided as a sum of other functions, e.g.~as an analytic model, or as a result of the numeric method used to calculate $f$.
Both scenarios arise for $g_\chi$ and $f_S^2$ in direct detection.

\paragraph{Annual Modulation:}

Choice~\ref{choice:labframe}, to use lab-frame velocity distributions $g_\chi$, simplifies the scattering operator in \eqref{rate:vectors} by making it spherically symmetric for spin-independent scattering, permitting the $\mathcal M_{n \ell m}^{j n' \ell' m'} \propto \delta_{l}^{l'} \delta_{m}^{m'}$ block diagonalization in \eqref{def:mathcalI}. 
However, the lab-frame velocity distribution is time-dependent, thanks to the Earth's revolution around the Sun: 
so, any analysis sensitive to annual modulation must evaluate $\ket{g_\chi(t) }$ repeatedly over the course of the year. 

It would be more appealing to define $\ket{g_\chi}$ once, in the rest frame of the galaxy or of the Sun, and to use a time-dependent basis transformation to extract the lab-frame coefficients.
The partial rate matrix from \eqref{def:mathcalK} would given by:
\begin{align}
 K^{(\ell)}_{m m'}(g_\chi(t), f_S^2 )  &\equiv \sum_{n=0}^\infty \sum_{ j, n' = 0}^\infty \sum_\psi   \langle v_0^3 g_\chi | \psi \rangle \cdot \langle \psi (t)  |  n \ell m \rangle \cdot  \mathcal I^{(\ell)}_{n , jn'} \cdot   \langle j n' \ell m' | E_0 f_S^2 \rangle,
\end{align}
where $\mathcal B^\phi_\psi(t) \equiv \langle \psi (t)  |  n \ell m \rangle$ is a time-dependent boost matrix that corrects for the changing Earth velocity.
This is completely analogous to the utility of defining $G^{(\ell)}(\mathcal R)$ for another continuously-variable parameter $\mathcal R \in SO(3)$, in the context of detector rotations. 
If $\mathcal B^\phi_\psi(t)$ can be expressed as an analytic function of $\vec v_E(t)$, then the vector space integration method allows the rate $R(t)$ to be evaluated on a continuum of times $t$. 
Alternatively, if the lists of coefficients $N_\psi$ and $N_\phi$ are not too large, $\mathcal B^\phi_\psi(t)$ could be evaluated (numerically, if necessary) for some grid of values $t_i$, and tabulated for future reference.

\paragraph{Gaussian Basis Functions:}

As an example of the latter type of application, there are several contexts where it makes sense to use a basis of gaussian functions for $\{\psi\}$. 
It is a common way to model DM streams, as in the Section~\ref{sec:demo} demonstration.  
In physical chemistry systems too complicated for the LCAO model~\cite{Blanco:2022pkt}, the initial and final state wavefunctions for $f_s(\vec q)$ may be calculated numerically using a basis of gaussian functions~\cite{Taketa_1966,Roberts_1967,Dunlap_1990,VAHTRAS1993514}. 
There are also simple systems (e.g.~harmonic oscillators) which have gaussian wavefunction profiles (e.g.~$\Phi \sim H_n(x) e^{-x^2}$ for some Hermite polynomials $H_n$). 

Each of these cases can be simplified using the $\langle \phi | \psi \rangle$ transformation matrices for gaussian functions $\psi$.
Defining a unit-normalized 3d gaussian centered at $\vec u_i$ with width $\sigma_i$, 
\begin{align}
{g}_i(\vec u, \vec u_i, \sigma_i ) &=  \frac{e^{-|\vec u - \vec u_i |^2/2\sigma_i^2 } }{\sigma_i^3 \, (2\pi)^{3/2} },
&
\int\! d^3 u \, {g}_i(\vec u, \vec u_i, \sigma_i )  \equiv  1,
\label{gauss:norm}
\end{align}
the exponential of the dot product $\vec u \cdot \vec u_i$ can be expanded in spherical harmonics, making the angular integrals trivially easy. The remaining radial integral takes the form:
\begin{align}
 g_i  &= \frac{4\pi  }{(2\pi)^{3/2} \sigma_i^3}  \exp\left( - \frac{u^2 + u_i^2 }{2\sigma_i^2} \right) \sum_{\ell = 0}^\infty \sum_{m = - \ell}^{\ell}  i_\ell^{(1)}\!\left( \frac{ u_i u }{ \sigma_i^2} \right) Y_{\ell m}(\hat u) Y_{\ell m } (\hat u_i) ,
\label{gi:nlm}
\\
\langle  g_i(\vec u_i, \sigma_i) | \nlm \rangle &= \sqrt{\frac{2}{\pi}} \frac{Y_{\ell m } (\hat u_i)  }{ u_0^3}   \int_0^\infty \frac{u^2 du}{  \sigma_i^3} r_n^{(\ell)}(u) \,e^{-(u^2+u_i^2)/2\sigma_i^2}  \,i_\ell^{(1)}\!\left( \frac{ u_i u }{ \sigma_i^2} \right) ,
\label{gi:integral}
\end{align}
where $i_\ell^{(1)}$ is the spherical modified Bessel function of the first kind, and where $r_n^{(\ell)}$ is the radial basis function for $\ket{\phi}= \ket{n \ell m} $. 
The derivation and further details are in Appendix~\ref{sec:gaussian}. 

For a function $f = \sum_i c_i g_i$ that is a sum of gaussians, the coefficients $\langle n \ell m | f \rangle$ are given by a sum of 1d radial integrals of the form \eqref{gi:integral}. As with $\mathcal I^{(\ell)}$, a suitably simple choice of $r_n(u)$ permits \eqref{gi:integral} to be integrated analytically, bypassing numeric integration entirely.
This makes it much easier to perform scans over the parameters of the as-yet-undetected streams in the DM velocity distribution, for example.

\paragraph{Other Basis Transformations:}

Finally, the $\langle \psi | \phi \rangle$ matrices can assist with more mundane tasks, e.g.~rescaling the basis parameters.
Section~\ref{sec:demo} uses a basis with $q_\text{max} = 10 \alpha m_e \simeq 37.3\, \text{keV}$, and $v_\text{max} = 960\, \text{km/s}$. These values are arbitrary, as long as they are large enough to encompass all or nearly all of $g_\chi$ and $f_S^2$. 
A different detector material might require $q_\text{max} \rightarrow 50 \, \text{keV}$, however, in which case a different set of $\ket{\varphi}$ must be used. 
Rather than repeating all of the $\langle \varphi' | f_S^2 \rangle$ calculations in a new basis, it may be easier to evaluate $\langle \varphi' | \varphi \rangle$ once and to apply it, as needed, to the previously-calculated $f_S^2$ models,
especially if the rescaling must be applied to large sets of tabulated $\ket{f_S^2}$.

One of the great promises of the vector space integration method is the ability to build up catalogs of $\ket{g_\chi}$ and $\ket{f_S^2}$ functions. Consistency is a key requirement for this project. Fast basis transformations make it easy to compare results generated from different basis choices.

\subsection{Squares, Roots, and Positivity} \label{sec:squarepositive}

The functions $g_\chi$ and $f_S^2$ are explicitly positive. The basis functions $\phi$ and $\varphi$ are not: so, it is possible that an inverse wavelet-harmonic transformation (with finitely many terms) can yield negative values, by underestimating the original function in regions where it is close to zero. 
When it is important to forbid negative values in $\ket{g_\chi}$ and $\ket{f_S^2}$, it may be better to expand the functions $\sqrt{g_\chi}$ and $f_S$ instead, and to obtain $g_\chi$ and $f_S^2$ from  $\ket{\sqrt{g_\chi}}^2$ and $\ket{f_s}^\star \ket{f_s}$. 
This way, every finite inverse wavelet transformation will be manifestly nonnegative. 
Evaluating the squares of $\ket{\sqrt{g_\chi}}$ and $\ket{f_S}$ is simple, once the coefficients $c_{ij}^k$ from \eqref{eq:phialgebra} are known. 

The usefulness of $\mathcal E \sim \langle f | \phi \rangle^2$ for tracking the convergence of the basis function expansion provides another motivation for evaluating $\ket{\sqrt{g_\chi}}$ instead.
With this approach, $\mathcal E$ corresponds to a physical quantity, 
\begin{align}
\mathcal E_v &= \int\! d^3 v\, \left( \sqrt{ g_\chi} \right)^2 \equiv 1.
\end{align}
Similarly, even though $f_s^2(\vec q)$ is not specifically $L^1$ normalized, the value of 
\begin{align}
\mathcal E_q &= \int\! d^3 q\, \left| f_s(\vec q) \right|^2
\end{align}
is still more closely related to the physical quantity that we are trying to fit.
These functions, not $g_\chi^2$ or $f_s^4$, are what appear in the rate integrand.

\medskip

Rotations can also be handled in this explicitly positive way by applying the Wigner $G^{(\ell)}$ matrix (or the original $D^{(\ell)}$, for complex-valued $f_s$) before squaring, noting that $\mathcal R \cdot \ket{f_S^2} = \ket{\mathcal R \cdot f_s}^2$.
The cost is that the vectorized scattering rate no longer factorizes into $G^{(\ell)}$ and the partial rate matrix $ K^{(\ell)}$ of \eqref{def:mathcalK}, when the rotations are applied this way.
So, unless avoiding negative values is absolutely mandatory, it is simpler to pursue positivity through accuracy, keeping enough terms in the expansion so that the inverse wavelet-harmonic transformation is positive even when the function is small.
Figure~\ref{fig:convergence} with $n_\text{coeffs} > 24$ provides such an example.

\subsection{Polynomial Cap for the Wavelet Expansion} \label{sec:polycap}

Section~\ref{sec:extrapolation} describes in detail how the information from a small number of somewhat-narrow wavelets ($\lambda \gg 1$) can be used to fill in the values of a much larger number of wavelet coefficients $(n_\text{coeffs}^k)$. 
This is quite useful: but, if a large number of coefficients can be evaluated with a small amount of information, it implies that there should be some other basis that organizes that information more compactly. 
As the form of \eqref{fu:cubic} makes clear, there is such a basis: a set of $k$ orthogonal polynomials would contain all of the information of the $k$th order wavelet extrapolation within each interval.

For the $n$th piecewise-constant bin from the wavelet transformation with $n_\text{coeffs}$ coefficients, one can define new orthogonal basis functions $\ket{n, p}$ of polynomial order $p =1, 2, \ldots, k$. This basis is mutually orthogonal, as long as no higher-generation $p=0$ wavelets (e.g.~$2 n$, $2n+1$, $4n$, etc.) are included. 
This hybrid form is much better suited to generating interpolating functions: instead of adding together all $\mathcal O(n_\text{coeffs}^k)$ coefficients from the wavelet extrapolation, only $k \cdot n_\text{coeffs}$ terms are required in the hybrid wavelet--polynomial method.
Whether or not it is practical for the rate calculation depends on how long it takes to integrate \eqref{eq:Istar} with the new basis functions.

On the subject of hybrid wavelet basis functions, there may be situations in which cutting off $g_\chi$ and $f_S^2$ at $v_\text{max}$ and $q_\text{max}$ is undesirable. While models of $g_\chi$ often fall off exponentially as $\exp(-v^2/v_0^2)$, the momentum form factor sometimes falls off as a power law instead: this is the case for atomic or molecular form factors, for example~\cite{Blanco:2019lrf,Blanco:2021hlm}, or the particle-in-a-box from Section~\ref{sec:demo}.
For these cases the orthogonal basis can be extended to include sets of radial functions defined on $[u_\text{max}, \infty)$. 
In other contexts it may be helpful to have a continuous description of the $\vec u \approx \vec 0$ origin. Figure~\ref{fig:accuracy}, for example, shows blips at the origin of the inverse wavelet transformation even with $N=10^3$ coefficients. 
Some versions of $F_\text{DM}^2$ enhance the contributions from the small $q$ region,  so it can be disproportionately important to model the origin accurately, e.g.~by defining orthogonal functions on $[0, u_\text{min}]$ that scale as $u^\ell Y_{\ell m}$ in the limit of small $u$.

A generic hybrid wavelet-harmonic basis might then include small-$u$ functions in $[0, u_\text{min}]$; wavelets, for the main $[u_\text{min}, u_\text{max}]$ region; narrow-bin polynomials, to replace the $k$th order extrapolation within $[u_\text{min}, u_\text{max}]$; and a class of large-$u$ basis functions, defined on $[u_\text{max}, \infty)$. 
An investigation of which analytic results can be extended to these more complicated basis functions is left to future work.
For direct detection this is primarily useful in the limit of large DM masses: otherwise, the divergence in $v_\text{min}(q, m_\chi)$ cuts off the $q \rightarrow 0$ portion of the rate integral, making the near-origin behavior of $f_S^2(\vec q)$ irrelevant. 

\subsection{Library of Functions and Numeric Implementation}

For generic functions $g_\chi(\vec v)$ or $f_S^2(\vec q, E)$, the projections onto the vector spaces $g_\chi \rightarrow \ket{g_\chi}$, $f_S^2 \rightarrow \ket{f_S^2}$ usually take longer than the newly-easy rate calculation, unless $N_\text{DM} N_{\mathcal R}$ is exceptionally large.  The results are also easily saved, especially considering that the wavelet extrapolation methods of Section~\ref{sec:extrapolation} can generate a large number of coefficients from a relatively small initial list.
The complete list of $\ell \leq 36$, $n < 2^8$ coefficients of $\langle g_\chi | n \ell m \rangle$, for example, which when including odd $\ell$ consists of about $3.5\cdot 10^5$ coefficients, is still small enough to fit in an email attachment. A selected list of the $10^3$ or $10^4$ most important coefficients could be printed out and sent through the post.

In this spirit, the numeric implementation of the wavelet-harmonic integration method, Vector Spaces for Dark Matter (\texttt{vsdm}), is available at
\begin{align*}
\texttt{https://github.com/blillard/vsdm.git} \, .
\end{align*}
The repository includes the Python code needed to generate the $\langle g_\chi | n \ell m \rangle$ and $\langle n \ell m | f_s^2 \rangle$ coefficients; the analytic expressions for $I_\star(E)$; a routine for assembling $K^{(\ell)}_{m m'}$; and the rate calculation $R(\mathcal R)$, following \eqref{def:partialRate}, for discrete final states or for spectra $dR/dE$. 
The repository also includes CSV files of the $\ket{g_\chi}$ and $\ket{f_S^2}$ values used in Section~\ref{sec:demo}.
In the future, the hope is to expand this list to include many physically motivated models for $g_\chi$ and for $f_S^2$.

The Python implementation of \texttt{vsdm} is available on the Python Package Index (PyPI), and can be installed via
\begin{align*}
\texttt{pip install vsdm} \ .
\end{align*}

\subsection{Physics Implications} \label{sec:implications}

Formulating the scattering rate  in terms of the partial rate matrix 
has a number of physical implications for the search for dark matter, beyond its utility in speeding up the rate calculation.

\paragraph{Model-Independent Constraints:}
The partial rate matrix $K^{(\ell)}_{m m'}$ captures all of the physically observable properties of the detector--DM system, for spin-averaged systems and/or spin-independent scattering.
By measuring the scattering rate in various orientations $\mathcal R \in SO(3)$, 
\begin{align}
R(\mathcal R) \propto \sum_{\ell mm'} K^{(\ell)}_{m m'} \cdot G^{(\ell)}_{m m'}(\mathcal R), 
\end{align}
the $R(\mathcal R)$ data can be translated into an experimental best fit and correlation matrix for the coefficients of the partial rate matrix. This is a convenient and entirely model-independent way to present constraints from anisotropic detectors.
These constraints can then be propagated into the fundamental physics inputs, e.g.~to uncover the DM particle model $F_\text{DM}^2$ and $m_\chi$ given some assumptions about $g_\chi$. 

Re-calculating the constraints on $F_\text{DM}^2$ and $m_\chi$ given new assumptions about $g_\chi$ then becomes significantly easier: rather than repeating the whole statistical analysis using the raw data, one need only calculate the new $K^{(\ell)}_{m m'}$ coefficients and compare them to the best fit values, using the correlation matrix to quantify the deviation from the best fit. 
Among other benefits, this makes it much more practical to propagate the systematic uncertainties from the astrophysical DM distribution into the predictions for the scattering rate.
Likewise, the constraints on $K^{(\ell)}$ can be used directly to probe alternative $F_\text{DM}^2$ models (e.g.~beyond the $F_\text{DM} = 1$ and $F_\text{DM} \propto 1/q^2$ models most often discussed in the literature).

\paragraph{Daily Modulation:}

The partial rate matrix is also highly useful for combining daily modulation measurements with the same type of crystal in different initial orientations. Considering that $\mathcal R(t)$ traces a one-dimensional curve through the three-dimensional space of $SO(3) \simeq S^3/Z_2$, it is otherwise quite challenging to construct a ``full sky'' picture of $R(\mathcal R)$ for all possible orientations. Even a fairly large set of generic initial orientations (e.g.~10--100) might have only a few points of approximate intersection, where $\mathcal R_1(t_1) \approx \mathcal R_2(t_2)$ for two distinct daily modulation cycles $\mathcal R_1(t) \neq \mathcal R_2(t)$. 
On the other hand, the data from multiple $\mathcal R(t)$ cycles can be combined to determine a global best fit for $K^{(\ell)}_{m m'}$, whether or not the different $\mathcal R(t) \in SO(3)$ curves have any points of intersection.

\paragraph{Detector Symmetries and the Velocity Distribution:}

From \eqref{eq:angdiag}, the scattering operator $\mathcal M  = \langle n \ell m | F_\text{DM}^2 \, \delta( E \ldots) | n' \ell' m'\rangle$ is diagonal in the angular indices, $\mathcal M \propto \delta_{\ell \ell'} \delta_{m m'}$.
So, a given $\ell m$ mode only  contributes to the scattering rate if both $g_\chi(\vec v)$ and $f_S^2(\vec q, E)$ have support on that harmonic ($\langle \ell m | g_\chi \rangle \neq 0$ and $\langle \ell m | f_S^2 \rangle \neq 0$). 

Many crystalline target materials have discrete symmetries that set $\langle f_S^2 | \ell m \rangle = 0$ for certain $\ell$ and $m$. For real spherical harmonics, these symmetries include: 
\begin{itemize}
\item central inversion: if $f(\vec u) = f( - \vec u)$, then $\langle f | \ell m \rangle = 0$ for all odd $\ell$
\item discrete $Z_k$ rotation about $\hat z$: if a function $f(\vec u)$ is invariant when rotated by an angle $2\pi / k$ about the $\hat z$ axis, then $\langle f | \ell m \rangle = 0$ unless $m$ is divisible by $k$
\item reflection in $\hat y$: if $f(u_x, u_y, u_z) = f(u_x, -u_y, u_z)$, then $\langle f | \ell m \rangle = 0$ for all negative $m$.
\end{itemize}
Central inversion symmetry is a rotationally-invariant property of $f(\vec u)$: i.e.~if $\langle f | \ell m  \rangle = 0$ for all odd $\ell$, then the same is true for $\langle \mathcal R \cdot f | \ell m \rangle$. The rotation operator can mix $\ket{\ell m} \rightarrow \ket{\ell m'}$, but it does not mix spherical harmonics of different $\ell' \neq \ell$. 
Neither does the spin-independent kinematic operator $\mathcal M \propto \delta_{\ell \ell'} \delta_{m m'} \mathcal I^{(\ell)}_{n n'}$ of \eqref{eq:angdiag}.
Consequently, a centrosymmetric anisotropic detector material will be insensitive to the odd $\ell = 1, 3, \ldots$ components of $g_\chi(\vec v)$.

For example, trans-stilbene has been proposed as one candidate material for an anisotropic detector~\cite{Blanco:2021hlm}, 
due to its relatively large daily modulation, its scintillation efficiency, and because it is easy to grow into crystals. 
It is also centrosymmetric, meaning that the modulation predicted in ref.~\cite{Blanco:2021hlm} is generated entirely by the $\ell = 2, 4, 6, \ldots$ modes. 
A non-centrosymmetric target material with substantial support at $\ell = 1$ could have an even larger modulation amplitude, by coupling to the large $\langle g_\chi | 1, 0 \rangle$ component of the SHM velocity distribution.

Detector materials with different symmetry properties will have fundamentally distinct (though not uncorrelated) systematic uncertainties from the unknown DM velocity distribution. 
By itself, this is one reason for the experimental direct detection program to include multiple types of target materials.
Non-centrosymmetric materials in particular are worth further investigation: for spin-independent DM--SM scattering rates, these are the only materials that can generate a nonzero $K^{(\ell = 1)}$ partial rate matrix. 
The importance of non-centrosymmetric target materials for direct detection has not been previously recognized, but the structure of the partial rate matrix makes it obvious.

\paragraph{Detecting Spin-Dependent Interactions:}

The partial rate matrix formalism introduced in ref.~\cite{Lillard:2025ixi} assumes either that the DM--SM interaction is spin-independent, or that the DM and SM systems have unpolarized spins, so that DM--SM scattering does not transfer angular momentum into the SM sector. 
Under these assumptions, \eqref{eq:angdiag} shows that the scattering operator is diagonal in the spherical harmonic basis.
With different assumptions, e.g.~polarized SM spins, some of the spin-dependent operators of ref.~\cite{Fitzpatrick:2012ix} would add off-diagonal terms, e.g.~$\langle \ell \pm 1, m \pm1 | \hat{\mathcal O} | \ell m \rangle$.

By searching for these interactions with a polarizable detector target, an experiment could directly constrain specific spin-dependent operators, possibly revealing the spin of the particle that mediates the DM--SM interaction.
Centrosymmetric target materials may be particularly useful: the odd $\ell$ moments of $g_\chi$ would only contribute to the scattering rate if there is some nontrivial spin--momentum or spin--velocity coupling.

\paragraph{Direct Detection Astronomy:}

Measurements of $K^{(\ell)}_{ m m'}$ elements can also be used to confirm or reject the Standard Halo Model, even without knowing the DM mass and form factor precisely. 
Thanks to the azimuthal symmetry of the SHM, $K^{(\ell)}_{0m'} = 0$ for all $m'$, assuming $\vec v$ is defined with its $\hat z$ axis parallel to the instantaneous lab velocity $\vec v_E$. 
Evidence for any nonzero $K^{(\ell)}_{m m'}$ coefficients at $m \neq 0$ would indicate that the lab-frame $g_\chi$ is not azimuthally symmetric, and therefore that $g_\chi$ includes some beyond-the-SHM components.

If dark matter is discovered in a detector system with substantial support at multiple values of $\ell$ and $m$, then a subsequent experimental run with larger exposure $\text{mass} \times \text{time}$ could measure a large set of $K^{(\ell)}_{m m'}$ coefficients with some precision.
In this happy future, this direct detection data can be used to infer 
the values of $\langle g_\chi | n \ell m \rangle$ directly from the data. This potential ability to measure $g_\chi(\vec v)$ from direct detection data has been referred to as DM astronomy~\cite{Peter:2011eu,Peter:2013aha,Lee:2014cpa}.

\medskip 

A single experiment operating with fixed excitation energy $\Delta E$ is largely insensitive to the radial part of $g_\chi(\vec v)$,
because, when $K^{(\ell)}$ is found by summing over the radial modes $\ket{n}, \ket{n'}$ of $g_\chi$ and $f_S^2$ in \eqref{def:mathcalK}, the information about the radial (speed) profile of $g_\chi$ is lost. Even if $m_\chi$ and $F_\text{DM}^2$ were known exactly, there is not enough information in $K^{(\ell)}_{m m'}$ to reconstruct the speed functions $\langle \ell m | g_\chi\rangle = g_{\ell m}(v)$. 

To break this degeneracy in the unknown $\langle g_\chi | n \ell m \rangle$ coefficients, the measurement of $K^{(\ell)}_{m m'}$ must be repeated with different materials and/or different excitation energies.\footnote{Annual variation, by boosting $g_\chi(v)$ a modest amount, would also provide some insight into the radial component of $g_\chi(\vec v)$, lifting some of the degeneracy in the statistical inference of $\langle g_\chi | n \ell m \rangle$. 
}
Each new $f_S^2$ generates a different set of predictions for $K^{(\ell)}_{m m'}$ according to the values of $\langle g_\chi | n \ell m \rangle$, so a coarse speed profile $g_{\ell m}(v)$ could be reconstructed from the ensembles of $K^{(\ell)}_{m m'}$ measurements for each of the largest $\ell m$ modes. 
A large ensemble of different detector materials, with distinct $\langle \ell m | f_S^2 \rangle = f_{\ell m}^2(q, E)$ radial profiles in their response functions, would be able to constrain the 3d velocity distribution $g_\chi(\vec v)$ in this way.

\paragraph{Ideal Detector Properties:}
In conclusion, an ideal experimental program would utilize multiple types of target materials, with different symmetry properties  and excitation energies, so that a discovery of DM daily modulation could be quickly followed up with measurements of $m_\chi$ and $F_\text{DM}^2$. 
Some of these detector targets should be center-asymmetric, to take advantage of the large expected $\ell = 1$ anisotropy in $g_\chi(\vec v)$. 
The ``inverse problems'' (i.e.~identifying $m_\chi$ and $F_\text{DM}^2$ from the data, or extracting the values of $\langle g_\chi | n \ell m \rangle$) 
may be easier for non-centrosymmetric materials, simply because there are roughly twice as many nonzero values of $K^{(\ell)}_{ m m'}$ that could potentially be measured (compared to a centrosymmetric alternative with a similar limit on the largest measurable $\ell_\text{max}$).

A parallel detection of DM scattering in multiple detector materials with well-understood momentum response functions $f_S^2(\vec q, E)$ would make it possible to separate the uncertainties in the DM particle model (described by $\mathcal I^{(\ell)}$) from the uncertainties in $g_\chi(\vec v)$. Extracting the DM mass from the scattering rate data will be easiest if the different $f_S^2(\vec q, E)$ functions have their greatest support at distinct values of $E$ or $|\vec q|$.

\section{Conclusion}

The vector space integration method is extremely good at its job, speeding up the evaluation time for a multivariate orientation-dependent analysis by factors of $10^7$--$10^8$ when using the partial rate matrix.  
This drastic improvement makes some formerly impossible analyses possible, while analyses that previously were merely difficult are now almost trivially easy.
The only part of future calculations requiring substantial numerical work is the projection of new astrophysical or detector models onto their respective vector spaces of velocity or momentum basis functions.
This method is highly adaptable, and can be generalized to any functional that depends linearly on some set of input functions.

A rotationally-dependent scattering rate has a convenient, compact representation in terms of the partial rate matrix $K^{(\ell)}_{m m'}(g_\chi, f_S^2, m_\chi, F_\text{DM})$ defined in \eqref{alt:mathcalK}. 
Not only does $K^{(\ell)}_{m m'}$ drastically speed up the rate calculation, the partial rate matrix  is a physically significant object worth calculating for its own sake. 
The zeros of $K^{(\ell)}_{m m'}$, which encode the symmetries of the detector and the velocity distribution, determine which components of $g_\chi$ do and do not contribute to the scattering rate. In particular, the partial rate matrix reveals that centrosymmetric and non-centrosymmetric target materials would provide complementary information about the DM particle physics model and the details of the local velocity distribution, suggesting that the experimental program of anisotropic DM direct detection would benefit by utilizing multiple types of detector targets.

\section*{Acknowledgements}
I thank
Pouya Asadi,
Leonardo Badurina,
Carlos Blanco,
Matthew Buckley,
David Curtan, 
Valerie Domcke, 
Peizhi Du,
Paddy Fox,
Yonatan Kahn,
David~E. Kaplan,
Graham Kribs,
Mariangela Lisante,
Samuel McDermott,
Robert McGehee,
Surjeet Rajendran,
David Shih,
Dave Soper,
Scott Thomas,
Ken Van Tilberg,
Tien-Tien Yu, 
and Zhengkang Kevin Zhang
for helpful and thought-provoking conversations, 
and Xu-Xiang Li and Pankaj Munbodh for noticing typos in the Appendix.  
I especially thank Patrick Draper for collaboration in an early phase of this project, and for related discussions of orthogonal polynomial functions; and Aria Radick, for insights on the numeric implementation. 
I thank Ken Van Tilberg for hospitality at the Center for Computational Astrophysics, where some of this work was performed.
This work was supported in part by the U.S. Department of Energy under Grant Number DE-SC0011640.
This work was performed in part at the Aspen Center for Physics, which is supported by National Science Foundation grant PHY-2210452.
This work did not benefit from any access to high-performance computing facilities.

\appendix

\section{Spherical Harmonics and Rotations}

\subsection*{Real Spherical Harmonics} \label{appx:realspherical}

In term of the complex spherical harmonics, 
\begin{align}
Y_\ell^m(\theta, \phi) &=  \sqrt{ \frac{2\ell + 1}{4\pi} \frac{(\ell - m)!}{(\ell+m)!} } P_\ell^m (\cos\theta) e^{i m \phi} ,
\end{align}
the real spherical harmonics are defined so that  $m \geq  0$ corresponds to $\cos(m\phi)$, and $m < 0$ to $\sin(m\phi)$:
\begin{align}  
Y_{\ell m}(\theta, \phi) &\equiv
\left\{ \begin{array}{l c c}
\sqrt{2} \, (-1)^m \text{Im}\, Y_{\ell}^{|m|}(\theta, \phi)  
&& \text{for } m < 0,
\\[\medskipamount]
Y_\ell^0(\theta, \phi)  && \text{for } m = 0,
\\[\medskipamount]
\sqrt{2} \,  (-1)^m \text{Re}\,Y_\ell^m (\theta, \phi) 
&& \text{for } m > 0,
\end{array}  
\right.
\end{align}
with the explicit expression:
\begin{align}
Y_{\ell m}(\theta, \phi) &=
\left\{ \begin{array}{l c c}
\sqrt{2}  (-1)^m    \sqrt{ \dfrac{2\ell + 1}{4\pi} \dfrac{(\ell - |m|)!}{(\ell+|m|)!} } P_\ell^{|m|} (\cos\theta) \sin(|m|\varphi)
&& \text{for } m < 0,
\\[\bigskipamount]
\sqrt{ \dfrac{2\ell + 1}{4\pi} } P_\ell(\cos\theta)  && \text{for } m = 0,
\\[\bigskipamount]
\sqrt{2}   (-1)^m   \sqrt{ \dfrac{2\ell + 1}{4\pi} \dfrac{(\ell - m)!}{(\ell+m)!} } P_\ell^m (\cos\theta) \cos(m\varphi)
&& \text{for } m > 0.
\end{array}  
\right.
\end{align}
Here $P_\ell^m$ are the associated Legendre polynomials, defined as
\begin{align}
P_\ell^m(x) &\equiv \frac{(-1)^m}{2^\ell \ell!}  (1 - x^2)^{m/2} \frac{d^{\ell+m}}{dx^{\ell+m}}  (x^2 - 1)^\ell, \\
P_\ell^{-m}(x) &= (-1)^m \frac{(\ell - m)!}{(\ell + m)!} P_\ell^m(x) ,
\end{align}
with $P_\ell(x) = P_\ell^{m=0}(x)$ given by the $m=0$ case. The polynomial in $x$ can be written explicitly as:
\begin{align}
P_\ell^m(x) &=  (-1)^m 2^\ell (1 - x^2)^{m/2} \sum_{k= m}^\ell \frac{k! }{(k- m)!} x^{k - m} \left( \begin{array}{c} \ell \\ k \end{array} \right) \left( \begin{array}{c} \frac{1}{2}(\ell + k - 1) \\ \ell \end{array} \right) ,
\label{appx:plm} 
\end{align}
for $m \geq 0$.
As part of the usual orthogonality relation, the $P_\ell^m$ with fixed $m$ obey
\begin{align}
\int_{-1}^1 \! d\cos\theta \, P_\ell^m(\cos\theta) P_{\ell'}^m(\cos\theta) &= \frac{  (\ell + m)!  }{(\ell - m)! } \frac{2\, \delta_{\ell'\ell} }{2\ell + 1} .
\end{align}
The inverse mapping from real to complex harmonics is given by
\begin{align}
Y_\ell^m(\Omega) &= 
\left\{ \begin{array}{l c c}
\dfrac{1}{\sqrt{2} } (Y_{\ell |m| } - i Y_{\ell, -|m|} )
&& \text{for } m < 0,
\\[\bigskipamount]
\dfrac{(-1)^m}{\sqrt{2} } (Y_{\ell |m| } + i Y_{\ell, -|m|} )
&& \text{for } m > 0.
\end{array}  
\right.
\label{Ylm:inverse}
\end{align}
The complex and real spherical harmonics have the same completeness relation:
\begin{align}
\delta(\cos \theta_1 - \cos\theta_2) \delta(\phi_1 - \phi_2) &
= \sum_{\ell m} Y_{\ell m } (\theta_1, \phi_1) Y_{\ell m } (\theta_2, \phi_2) ,
\end{align}
and the Legendre polynomial of the dot product between unit vectors $\hat x$ and $\hat y$ can be expanded as
\begin{align}
P_\ell(\hat x \cdot \hat y) &= \frac{4\pi}{2\ell + 1} \sum_{m = - \ell}^\ell Y_{\ell m}(\hat x) Y_{\ell m}(\hat y) .
\end{align}

\subsection*{Rotations and Spherical Harmonics} \label{sec:wignerG}

The $Y_\ell^m$ of fixed $\ell$ transform under $\mathcal R \in SO(3)$ as a $2\ell+1$ dimensional irreducible representation of the rotation group:
\begin{align}
Y_\ell^m(\vec{v'} ) = \sum_{m' = - \ell}^\ell D_{m' m}^{(\ell)} (\mathcal R) \, Y_\ell^{m'}(\vec{v}) ,
\end{align}
where $D^{(\ell)}$ is the Wigner D-matrix,  
\begin{align}
D^{(\ell)}_{m' m} &= \bra{ \ell m'} \mathcal R \ket{\ell m} ,
\label{WignerD:def}
&
D_{m' m}^{(\ell)} = (-1)^{m' - m} D_{-m', -m}^{(\ell) \star}.
\end{align}
The entries of each $D^{(\ell)}(\mathcal R)$ can be calculated without integrating \eqref{WignerD:def}; for example, $D^{(\ell)}(\mathcal R)$ has explicit solutions in terms of the Euler angles $(\alpha, \beta, \gamma)$ parameterizing $\mathcal R$,
given in terms of the Jacobi polynomials $P^{(a, b)}_j(\cos \beta)$. 
In the special case $D_{m0}^{(\ell)}$, the $P^{(a, b)}$ simplify to the associated Legendre polynomials, with 
\begin{align}
D_{m0}^{(\ell)\star } ( \alpha, \beta, \gamma) & = \sqrt{ \frac{(\ell - m)! }{(\ell + m)!} } e^{i m \alpha } P_\ell^{m}(\cos\beta) = \sqrt{ \frac{4\pi}{2\ell + 1} } Y_\ell^{m}(\beta, \alpha).
\end{align}
This is particularly relevant for systems with an azimuthal symmetry. In such an example, $\langle j n \ell m | f_{S}^2 \rangle \propto \delta_{m 0}$, and generic rotations of the detector require only the $D_{m' 0}$ coefficients of the $D$ matrix.\footnote{This expression uses the z-y-z type of Euler angle, where $\beta$ corresponds to an intermediate rotation about the $\hat y$ axis. 
}

For the real spherical harmonics, I define a related $G_{m'm}^{(\ell)}$ to encode the action of $\mathcal R \in SO(3)$:
\begin{align}
\mathcal R \cdot Y_{\ell m} \equiv \sum_{m' = - \ell}^\ell G^{(\ell)}_{m' m} Y_{\ell m'} ,
\label{Gmatrix:def}
\end{align}
where $G$ is given explicitly as a real function of the Wigner $D$ matrix:
\begin{align}
G_{m'm}^{(\ell)} &= 
\left\{ \begin{array}{l c c}
\text{Re}\left[ D^{(\ell)}_{-m', -m} - (-1)^{m'} D^{(\ell)}_{m', -m}   \right]
 &&  m' < 0  , m < 0
\\[\bigskipamount]
-\sqrt{2} \,\text{Im}\left[ D^{(\ell)}_{-m', 0}   \right]
 &&  m' < 0  , m = 0
\\[\bigskipamount]
- \, \text{Im} \left[ D^{(\ell)}_{-m', m} - (-1)^{m'} D^{(\ell)}_{m', m}   \right]
 &&  m' < 0  , m > 0
\\[\bigskipamount]
\sqrt{2} \,\text{Im}\left[ D^{(\ell)}_{0,-m}   \right]
 &&  m' = 0  , m < 0
\\[\bigskipamount]
\text{Re} \left[D_{0,0}^{(\ell)} \right]
  &&  m' = 0  , m = 0
\\[\bigskipamount]
\sqrt{2} \, \text{Re} \left[ D^{(\ell)}_{0, m} \right]
 &&  m' = 0  , m >0
\\[\bigskipamount]
\text{Im} \left[ D^{(\ell)}_{m', -m} + (-1)^{m'} D^{(\ell)}_{-m', -m}  \right]
 &&  m' > 0  , m < 0
\\[\bigskipamount]
\sqrt{2} \, \text{Re}  \left[ D^{(\ell)}_{m', 0} \right]
 &&  m' > 0  , m = 0
\\[\bigskipamount]
\text{Re} \left[ D^{(\ell)}_{m', m} + (-1)^{m'} D^{(\ell)}_{-m', m}   \right]
 &&  m' > 0  , m > 0
\end{array}  
\right.
\label{Gmatrix}
\end{align}
With this organization, note that the right hand side of \eqref{Gmatrix} uses coefficients $D_{a, b}$ with $a = \pm m'$ but $b = + m$.

\section{Analytic Expression for Scattering Matrix} \label{sec:Istar}

In many simplified models of DM--SM scattering, the particle interaction form factor $F_\text{DM}^2(q, v)$ can be written as a series expansion in velocity and momentum,
\begin{align}
F_\text{DM}^2(q, v) &= \sum_{n,m} c_{n,m} \left( \frac{q}{q_0}\right)^n \left( \frac{v}{c} \right)^m ,
\end{align}
particularly when the interaction is mediated by a particle whose mass $m_\text{med}$ is irrelevantly heavy or irrelevantly light compared to the momentum transfer $q$, $m_\text{med} \gg q$ or $m_\text{med} \ll q$, respectively. 
For form factors of this type, \eqref{def:Istar} can be integrated analytically:
\begin{align}
I_\star^{(\ell)} &= \sum_{n,m} c_{nm} \left( \frac{q_\star}{q_0} \right)^n \left( \frac{v_\star}{c}\right)^m \, I_{nm}^{(\ell)} , 
\end{align}
for an analytic function $I_{nm}^{(\ell)}(q_\star, v_\star)$ to be derived in this section.

Begin by defining momentum-dependent bounds on the velocity integral:
\begin{align}
v_1(q) &\equiv \text{max}\left(v_a, v_\text{min}(q) \right),
&
v_2(q) &\equiv \text{max}\left(v_b, v_1(q) \right),
\end{align}
and define
\begin{align}
I_{nm}^{(\ell)} &\equiv \int_{q_a}^{q_b}\! \frac{qdq}{q_\star^2} \left( \frac{q}{q_\star} \right)^n \int_{v_1}^{v_2} \! \frac{ v dv}{v_\star^2} \left( \frac{v}{v_\star} \right)^{m} P_\ell\! \left( \frac{v_\text{min} }{v} \right) . 
\end{align}
This reshapes the rectangular $v_a \leq v \leq v_b$, $q_a \leq q \leq q_b$ region of integration to cut out any part where $v < v_\text{min}(q)$. 
Next, define a coordinate $w \equiv v_\text{min}(q)/v$, so
\begin{align}
I_{nm}^{(\ell)} &=  \int_{q_a}^{q_b} \!  \frac{qdq}{q_\star^2}  \left( \frac{q}{q_\star} \right)^n \left( \frac{v_\text{min}}{v_\star} \right)^{2+m}  \int_{w_2(q)}^{w_1(q)} \! \frac{dw}{w^{3+m}} P_\ell(w) ,
\\
&= \int_{q_a}^{q_b} \!  \frac{qdq}{q_\star^2}   \left( \frac{q}{q_\star} \right)^n \left( \frac{q}{q_\star}+ \frac{q_\star}{q} \right)^{m+2}  \sum_{k=0}^\ell   \!   \frac{ 2^{\ell-m-2} }{k! (\ell - k)!} \frac{ \Gamma\left( \frac{ k + \ell + 1 }{2} \right) }{ \Gamma\left( \frac{ k - \ell + 1 }{2} \right) } 
\int_{w_2(q)}^{w_1(q)} \! dw\, w^{k-m-3} ,
\label{eq:Istar}
\end{align}
where $w_i(q) = v_\text{min} / v_i(q)$ for $i=1,2$, 
applying the monomial expansion of $P_\ell(w)$ in the  second line.
Note that when $(k + \ell)$ is odd, the summand vanishes. 
Although some radial basis choices include the $v \rightarrow \infty$ limit, the expression $I_\star^{(\ell)}$ is valid only for piecewise constant basis functions, which by definition do not include $v\rightarrow \infty$ (assuming that the basis functions are normalizable).
So $w$ does not approach zero. For finite $E$ and $m_\chi$, $v_\star \neq 0$, and so $w$ does not approach $\infty$ either.
(This analysis is not designed for the $E \rightarrow 0$ limit: here I assume that $E$ is bounded from below at some finite positive $\Delta E_\text{min}$.)

So, 
\begin{align}
I_{nm}^{(\ell)} &= \int_{q_a}^{q_b} \!  \frac{qdq}{q_\star^2}  \left( \frac{q}{q_\star} \right)^n
\Bigg\{
 \sum_{k \neq m+2}^\ell   \!   \frac{ 2^{\ell-k} \left( \frac{q}{q_\star} + \frac{q_\star}{q} \right)^{k} }{k! (\ell - k)!  } \frac{ \Gamma\left( \frac{ k + \ell + 1 }{2} \right) }{ \Gamma\left( \frac{ k - \ell + 1 }{2} \right) } \left[ \frac{v_2^{m+2-k} - v_1^{m+2-k} }{v_\star^{m+2-k} (m+2-k)}  \right]  
\nonumber\\&\hspace{2em} +
\left[ \frac{2^{\ell -m-2} }{(m+2)! (\ell -m - 2)!} \left( \frac{q}{q_\star} + \frac{q_\star}{q} \right)^{m+2} \frac{ \Gamma\left( \frac{3 +m + \ell}{2} \right) }{\Gamma\left( \frac{3 +m - \ell }{2} \right) }\, \log\left( \frac{v_2}{v_1 } \right) \right]_{\text{if } \ell \geq m+2}
\Bigg\} .
\label{Istarv}
\end{align}
The $k=m+2$ term is evaluated separately, if $\ell \geq m+2$, as it produces a logarithm rather than powers $v_{1,2}$.

For the $q$ integral, we must consider the $q$ dependence of the $v_1(q)$ and $v_2(q)$ boundaries of integration in the $(q,v)$ plane. 
Some cases are simple: for example, if $v_a > v_\star$ for all $q \in [q_a, q_b]$, then the integrand is just a sum of terms $q^n (q/q_\star + q_\star/q)^k$.
(Even simpler, if $v_b \leq v_\star$, then the integrand vanishes, and $I_{mn}^{(\ell)} = 0$.)
On the other hand, if the line $v = v_\text{min}(q)$ passes through the rectangular region $[q_a, q_b] \times [v_a, v_b]$, then at least part of the integration boundary is not flat. 
Analytic results can be derived for $I_{mn}^{(\ell)}$ in every case.

\begin{figure}
\centering
\includegraphics[height=0.45\textwidth]{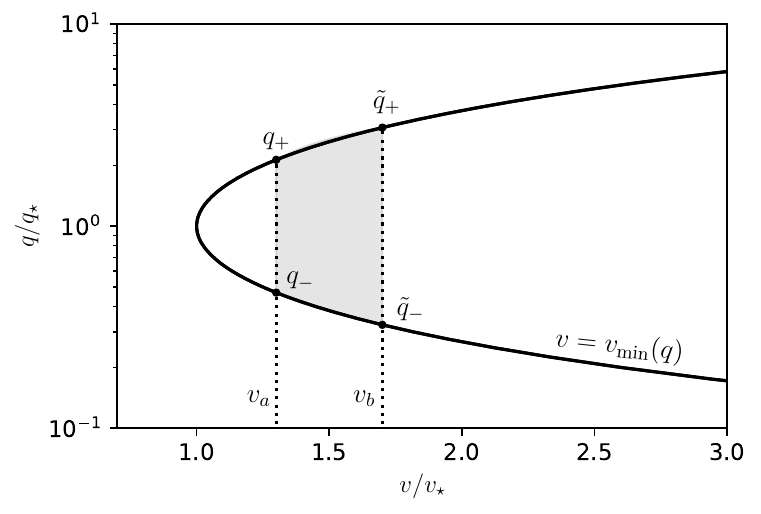}
\caption{The solutions for $q_\pm$ and $\tilde{q}_\pm$ are shown together with $v_\text{min}(q)$, with vertical dashed lines marking $v=v_a$ and $v = v_b$ as the boundaries of an integration region (shaded). For $[q_a, q_b]$ limits of integration lying entirely within this shaded region, only the $T_N^{(\ell)}$ form of the integral is needed; but if $q_a$ or $q_b$ intersects the $v = v_\text{min}(q)$ line, then one uses the $U_N^{(\ell)}$ form instead.}
\label{fig:vqstar}
\end{figure}

Given the constants $v_a$ and $v_b$, let us define a $q_\pm$ and a $\tilde q_\pm$ as the real solutions to 
\begin{align}
v_\text{min}(q_\pm) &\equiv  v_a, 
&
\frac{q_\pm}{q_\star} &= \frac{v_a}{v_\star} \pm \sqrt{ \frac{v_a^2}{v_\star^2} - 1 } 
\\
v_\text{min}(\tilde q_\pm) &\equiv  v_b, 
&
\frac{\tilde q_\pm}{q_\star} &= \frac{v_b}{v_\star} \pm \sqrt{ \frac{v_b^2}{v_\star^2} - 1 } ,
\end{align} 
with $q_\star$ and $v_\star$ defined in \eqref{def:stars}.
These points mark the transitions between rectangular $v$ integrals, with the lower bound set by $v = v_a$, and curved regions bounded instead by $v \geq v_\text{min}(q)$. 
If $v_a < v_\star$ but $v_b > v_\star$, then $v \geq v_\text{min}$ sets the lower bound throughout the full range of $q$. 
Figure~\ref{fig:vqstar} depicts $q_\pm$ and $\tilde{q}_\pm$ for an example with $v_\star < v_a < v_b$. 

A generic rectangular region can be split up as follows:
\begin{align}
q \leq \tilde q_- &\longrightarrow \text{integrand vanishes}
\\
\tilde q_- \leq q \leq q_- &\longrightarrow \text{$v_\text{min}(q)$ sets lower bound}
\label{qrange:left}
\\
q_- \leq q \leq q_+ &\longrightarrow \text{$v_a$ sets lower bound, if $v_a > v_\star$}
\\
q_+ \leq q \leq \tilde q_+ &\longrightarrow \text{$v_\text{min}(q)$ sets lower bound}
\label{qrange:right}
\\
\tilde q_+ \leq q \phantom{_+} &\longrightarrow \text{integrand vanishes}.
\end{align}
So, there are at most three nonzero regions, which depend on two types of integral.

For the rectangular regions, define a
\begin{align}
T^{(\ell)}_{nm}([v_1, v_2], [q_1, q_2]) &\equiv
\int_{q_1}^{q_2} \!  \frac{qdq}{q_\star^2}\left( \frac{q}{q_\star} \right)^n \int_{v_1}^{v_2} \! \frac{vdv}{v_\star^2} \left( \frac{v}{v_\star} \right)^m P_\ell\! \left( \frac{v_\text{min} }{v} \right),
\end{align}
and related integral
\begin{align}
B_{n, k}(x) &\equiv  \frac{1}{2}  \int^{x}_{x_0} \! dy\, y^{\frac{n-k}{2}} (1 +y)^k   ,
\end{align}
defined in terms of an arbitrary but positive reference point $x_0 > 0$. 
$B_{n, k}$ can be written in terms of a hypergeometric ${_2 F_1}$ function, or explicitly as
\begin{align}
B_{n, k}(x) &=  \frac{1}{2} 
\sum_{j = 0}^{k} \left( \! \begin{array}{c} k \\ j \end{array}\! \right)
 \left[  \left\{ \frac{x^{ \frac{n-k}{2} + 1 + j} }{ \frac{n-k}{2} + 1 + j} \right\}_{\text{if } j \neq  \frac{k-n}{2} - 1} + \delta_{  \frac{k-n}{2} - 1}^{j} \, \log(x) \right] .
\end{align}
An explicit expression for $T^{(\ell)}$ can be read from \eqref{Istarv}:
\begin{align}
T^{(\ell)}_{nm}([v_1, v_2], [q_1, q_2]) &= 
\Bigg\{
 \sum_{k \neq m+2}^\ell   \!   \frac{ 2^{\ell-k} \, B_{n,k}(x) }{k! (\ell - k)!  } \frac{ \Gamma\left( \frac{ k + \ell + 1 }{2} \right) }{ \Gamma\left( \frac{ k - \ell + 1 }{2} \right) } \left( \frac{v_2^{m+2-k} - v_1^{m+2-k} }{v_\star^{m+2-k} (m+2-k)}  \right)  
\nonumber\\&\hspace{2em} +
\left[ \frac{2^{\ell - m- 2}\, B_{n,m+2}(x)  }{(m+2)! (\ell -m - 2)!}\frac{ \Gamma\left( \frac{3 + m + \ell}{2} \right) }{\Gamma\left( \frac{3 +m - \ell }{2} \right) }\, \log\left( \frac{v_2}{v_1 } \right) \right]_{\text{if } \ell \geq m+2}
\Bigg\} \Bigg|_{x = q_1^2/q_\star^2}^{q_2^2/q_\star^2} 
\end{align}

For the non-rectangular regions, bounded from below by $v \geq v_\text{min}$, define an analogous
\begin{align}
U^{(\ell)}_{nm} (v_2, [q_1, q_2]) &\equiv
\int_{q_1}^{q_2} \!  \frac{qdq}{q_\star^2} \left( \frac{q}{q_\star} \right)^n  \int_{v_\text{min}}^{v_2} \! \frac{vdv}{v_\star^2} \left( \frac{v}{v_\star} \right)^m P_\ell\! \left( \frac{v_\text{min} }{v} \right).
\end{align}
Integrating the $k\neq m+2$ terms in \eqref{Istarv} remains simple: but, the replacement $v_1 \rightarrow v_\text{min}$ introduces a logarithmic dependence on $v_\text{min}(q)$, requiring new integrals.
Define:
\begin{align}
S_{nm}(x)\Big|_{q_1^2/q_\star^2}^{q_2^2/q_\star^2}  &\equiv \int_{q_1}^{q_2}\! \frac{dq}{q_\star} \left( \frac{q}{q_\star} \right)^{n -m -1} \left( 1 + \frac{q^2}{q_\star^2} \right)^{m+2} \, \log\left( \frac{v_\star/2}{ v_\text{min}}  \right) .
\end{align}
Integrating by parts, and defining $t \equiv q/q_\star$,
\begin{align}
dV_{nm} &= dt\, t^{n-1}\left( 1 + t^2 \right)^{2+m} ,
&
u &= - \log\left( t + \frac{1}{t} \right),
&
du &= \frac{dt}{t} \frac{1 -t^2}{1+t^2} ,
\end{align}
where $V_{nm}$ is given in terms of $x = t^2$ by
\begin{align}
V_{nm}(x) &=  \sum_{j = 0}^{m+2} \frac{1}{2} \left( \begin{array}{c} m+2 \\ j \end{array} \right) \left[\left( \frac{x^{j + \frac{n-m}{2} } }{j + \frac{n-m}{2} } \right)_\text{if $j \neq \frac{n-m}{2}$} 
+ \delta_{ \frac{n-m}{2} }^j \log x
\right]
\end{align}
for generic $m,n$. If $n-m$ is even and $0 \leq n-m \leq 2m+4$, then $V_{nm}(x)$ includes a logarithm in $x$. Some common examples include:
\begin{align}
V_{0,0}(x) &= \frac{1}{2} \log x + x + \frac{1}{4} x^2 ,
\\
V_{-2,0}(x) &= -\frac{1}{2x} + \log x + \frac{x}{2}  ,
\\
V_{-4,0}(x) &= - \frac{1}{4x^2} - \frac{1}{x} + \frac{1}{2} \log x.
\end{align}
In terms of $V_{nm}(x)$,
\begin{align}
S_{nm}(x) &= \left( V_{nm}(x) \, \log\left(\frac{\sqrt{x}}{1+x}  \right) \right) - \int_{}^{x} \! \frac{dx'}{2x'} \left( \frac{ 1 -x'}{1+x'} \right) V_{nm}(x') .
\end{align}
To complete the remaining integral,
define
\begin{align}
C_\alpha(x) &\equiv  -\int^{x} \! \frac{dx'}{2x'} \left( \frac{1-x'}{1+x'} \right) \frac{x'^{\alpha}}{\alpha} 
\end{align}
for $\alpha \neq 0$, and
\begin{align}
C_0(x) &\equiv  - \int^{x} \! \frac{dx'}{2x'} \left( \frac{1-x'}{1+x'} \right) \log(x)
= - \frac{(\log x)^2 }{4} + (\log x) \log(1+x) + \text{Li}_2(-x) ,
\end{align}
for any logarithmic ``$\alpha = 0$'' terms in $V_N$. 
Here $\text{Li}_2(-x)$ is the dilogarithm: it is real and negative for $x > 0$. It is related to the Spence function via $\text{Spence}(z) = \text{Li}_2(1-z) $. 
For generic values of $\alpha \neq 0$, one can use
\begin{align}
C_\alpha(x) &=  \frac{x^\alpha}{\alpha^2} \left[ \frac{1}{2} - {_2 F_1}\!\left.\left( \begin{array}{c} 1, ~~\alpha \\ 1+\alpha \end{array} \right| -x \right) \right] 
\end{align} 
directly. 

For negative and positive integers $\alpha$, it is faster to express $C_\alpha$ as a power series plus a logarithm.
For example, when $m=0$, the only values of $\alpha$ appearing in $V_{mn}$ are $\alpha = n/2, n/2+1, n/2+2$, so only $\alpha = -2, -1, 0, 1, 2$ are needed for the typical heavy/light mediator models ($n=0, -4$, respectively):
\begin{align}
C_{-2}(x) &=-  \frac{1}{8 x^2} + \frac{1}{2x}  - \frac{1}{2} \log \frac{1+x}{x} ,
\label{eq:Cm2}
\\
C_{-1}(x) &= - \frac{1}{2x} + \log\frac{1+x}{x} ,
\\
C_1(x) &=  \frac{x}{2} - \log(1 + x) ,
\\
C_2(x) &= \frac{1}{8} \left( x^2 - 4x + 4 \log(1+x) \right) .
\label{eq:Cp2}
\end{align}
More generally, for any positive integer $\alpha \geq 1$, 
\begin{align}
C_{\alpha \geq 1} (x) &= \frac{(-1)^\alpha }{\alpha} \log(1 + x) + \frac{(1 + x)^\alpha }{2 \alpha^2} + \sum_{j=1}^{\alpha - 1} \frac{(-1)^{\alpha - j} }{2 \alpha} \left[ \left( \begin{array}{c} \alpha \\ j \end{array} \right) + \left( \begin{array}{c} \alpha-1 \\ j \end{array} \right) \right] \frac{(1 + x)^j }{j} ,
\label{eq:Cpos}
\end{align}
while for negative integers $\alpha \leq -1$ 
\begin{align}
C_{\alpha \leq -1} (x) &= \frac{(-1)^\alpha }{\alpha} \log \frac{1 + x}{x}  - \frac{(1 + 1/x)^{-\alpha} }{2 \alpha^2} + \sum_{j=1}^{-\alpha - 1} \frac{(-1)^{\alpha + j} }{2 \alpha j} \left[ \left( \begin{array}{c} -\alpha \\ j \end{array} \right) + \left( \begin{array}{c} -\alpha-1 \\ j \end{array} \right) \right] \left(\frac{1 + x }{x} \right)^j .
\end{align}
Note that each integral $C_\alpha(x)$ is defined only up to a constant.

With these analytic expressions for $C_\alpha(x)$ and $V_{nm}(x)$, $S_{mn}$ can be written as 
\begin{align}
S_{nm}(x) &= \left( V_{nm}(x) \, \log\left(\frac{\sqrt{x} }{1+x} \right) \right)  + \sum_{j = 0}^{m+2} \frac{1}{2} \left( \begin{array}{c} m + 2 \\ j \end{array} \right) C_{j + \frac{n - m}{2} }(x) ,
\label{eq:SNx}
\end{align}
which can in turn be used to evaluate the $k=m+2$ part of $U^{(\ell)}$.
The result:
\begin{align}
U^{(\ell)}_{nm}(v_2, [q_1, q_2]) &= 
\Bigg\{
 \sum_{k \neq m+2}^\ell   \!   \frac{ 2^{\ell-k}  }{k! (\ell - k)!  } \frac{ \Gamma\!\left( \frac{ k + \ell + 1 }{2} \right) }{ \Gamma\!\left( \frac{ k - \ell + 1 }{2} \right) }\left( 
 \left(\frac{v_2}{v_\star}\right)^{m+2-k}  \frac{B_{n,k}(x) }{m+2 - k} -  \frac{1}{2^{m+2 - k} }    \frac{B_{n, m+2}(x) }{m+2 - k} 
   \right)  
\nonumber\\&\hspace{-2em} +
\left[ 
\frac{2^{\ell - m- 2} \Gamma\!\left( \frac{3 +m + \ell}{2} \right) / \Gamma\!\left( \frac{3 +m - \ell}{2} \right) }{(m+2)! ( \ell - m - 2)!}  
\left( \log\frac{2 v_2 }{v_\star} \, B_{n, m+2}(x) + S_{nm} (x) \right) 
\right]_{\text{if } \ell \geq m+2}
\Bigg\} \Bigg|_{x = q_1^2/q_\star^2}^{q_2^2/q_\star^2} 
\end{align}

\subsection*{Main Result}
Now, $I_{nm}^{(\ell)}$ can be assembled from the explicit solutions for $T^{(\ell)}$ and $U^{(\ell)}$.
The general form is
\begin{align}
I_{nm}^{(\ell)}( [v_1, v_2], [q_2, q_3] ) &= U^{(\ell)}_{nm}( v_2, [q_1, q_2] )  +  T^{(\ell)}_{nm}( [v_1, v_2], [q_2, q_3] )  +  U^{(\ell)}_{nm}( v_2, [q_3, q_4] ) .
\label{Istar:UTU}
\end{align}
This admits all three types of boundary, Eqs.~(\ref{qrange:left}--\ref{qrange:right}), where $q_1$ is given either by $q_a$ or $\tilde q_{-}$; $q_4$ is given by $q_b$ or $\tilde q_+$; and $q_2$ and $q_3$ can be $q_-$ and $q_+$, respectively.
The first or second $U^{(\ell)}$ term disappears if $q_a > q_-$ or if $q_b < q_+$, respectively. 
Likewise, if the interval $[q_-, q_+]$ has no overlap with $[q_-, q_+]$, or if $v_1 < v_\star$, then there is no $T^{(\ell)}$ term: in these cases there is a single $U^{(\ell)}$ term, with the endpoints given by $\text{max}(q_a, \tilde q_-)$ and $\text{min}(q_b, \tilde q_+)$.  

Here are the values of $q_i$ to be used in each case. 
For the trivial cases:
\begin{align}
\text{if $v_2 < v_\star$ or $q_b < \tilde q_-$ or $q_a > \tilde q_+$:} \hspace{1em} q_1 \equiv q_2 \equiv q_3 \equiv q_4, 
&&\longrightarrow&& 
I_\star^{(\ell)} = 0.
\end{align}
For the nontrivial cases, the $q_1$ and $q_4$ endpoints can always be described as:
\begin{align}
q_1 = \text{max}(q_a, \tilde q_-) ,
&&
q_4 =   \text{min}(\tilde q_+, q_b) .
\end{align}
The definitions of $q_2$ and $q_3$ depend on the relationship between the intervals $[q_a, q_b]$ and $[q_-, q_+]$. 
If these are disjoint, then there is no $T^{(\ell)}$ region: that is,
\begin{align}
\text{if $q_b < q_-$ or $q_a > q_+$:} \hspace{1em} q_2 \equiv q_3,
&&\longrightarrow &&
I_{nm}^{(\ell)} =   U_{nm}^{(\ell)}(v_2, [q_1, q_4] ) .
\end{align}
Finally, the remaining cases can be summarized in a single line:
\begin{align}
\text{if $q_a < q_+$ and $q_b > q_-$:} 
&&
q_2 =   \text{max}(q_-, q_a)   ,
&&
q_3 = \text{min}(q_b,  q_+) ,
\end{align}
and $I_{nm}^{(\ell)}$ is given by \eqref{Istar:UTU}. 
Recall that $T^{(\ell)}$ and $U^{(\ell)}$ vanish when their $[q_i, q_j]$ arguments are identical: 
for example, if $q_a > q_-$, then $q_1 = q_2$ removes the first $U^{(\ell)}$ term in \eqref{Istar:UTU}.

\section{Interpolation and Wavelet Extrapolation} \label{sec:interpolation}

\subsection{Haar wavelets}

One corollary of Eqs.~(\ref{haar:f00}--\ref{haar:f11}) is that the parameters of a cubic interpolation can be extracted from the three wavelets $(0,0)$, $(1,0)$ and $(1,1)$:
\begin{align}
\Delta^1 f'_0 &= \frac{4}{3 A_\star \Delta} \left( - 7 \left\langle f \big| (0,0) \right\rangle + 4 \sqrt{2} \left\langle f \big| (1,0) \right\rangle + 4\sqrt{2} \left\langle f \big| (1,1) \right\rangle \right) 
\\
\Delta^2 f''_0 &= \frac{16\sqrt{2}}{A_\star \Delta} \left( \left\langle f \big| (1,0) \right\rangle - \left\langle f \big| (1,1) \right\rangle \right)
\\
\Delta^3 f^{(3)}_0 &= \frac{256}{A_\star \Delta} \left(   \left\langle f \big| (0,0) \right\rangle - \sqrt{2} \left\langle f \big| (1,0) \right\rangle - \sqrt{2} \left\langle f \big| (1,1) \right\rangle \right)  .
\end{align} 
Here $(0,0)$, $(1,0)$ and $(1,1)$ refer respectively to the $\ket{\lambda_\star \mu_\star}$, $\ket{\lambda_\star+1, 2\mu_\star}$ and $\ket{\lambda_\star + 1, 2\mu_\star + 1}$ wavelets,
and the derivatives $f^{(k)}_0$ refer to $f^{(k)}(x_2)$, with $x_2$ at the center of the $(0,0)$ wavelet. 
An interpolating function $f(x)$ also requires the value of $f(x_2)$: one can use the $f_0 = f(u_0)$ generated by the $\lambda \leq \lambda_\star$ wavelet expansion, but for the purposes of interpolation it is more precise to calculate $f_0 = f(x_2)$ by evaluating the function directly.

In the case of approximately constant $f^{(3)}(x)$, where $f(x)$ in $[x_a, x_b]$ is well approximated by a cubic polynomial, all of the higher-order wavelet coefficients can be obtained from $(0,0)$, $(1,0)$ and $(1,1)$. 
Take the $(\delta \lambda, \delta \mu)$ wavelet, i.e.~$\ket{\lambda, \mu}$ with $\lambda = \lambda_\star + \delta\lambda$ and $\mu = 2^{\lambda - \lambda_\star} \mu_\star +  \delta\mu$.
Direct integration of \eqref{fu:cubic} yields
\begin{align}
\left\langle f \big| (\delta\lambda,\delta\mu) \right\rangle &\simeq
- \frac{\sqrt{2}^{\delta\lambda} A_\star \Delta }{4 \cdot 2^{ \delta\lambda}} \bigg[
f_0' \left(\frac{\Delta}{2^{\delta\lambda}}\right) - \left( 2^{\delta\lambda} - 1  - 2 \, \delta \mu  \right) \frac{f_0''}{2} \left(\frac{\Delta}{2^{\delta\lambda}}\right)^2
\nonumber\\&\hspace{6em}
+ \left( \frac{ 7}{8}   + \frac{3}{4} \left( 4^{ \delta\lambda} - 2^{1 + \delta\lambda} \right) + 3\, \delta\mu \left( 2^{\delta\lambda} - 1 - \delta\mu \right)   \right) \frac{f_0^{(3)} }{6} \left(\frac{  \Delta}{2^{\delta\lambda}} \right)^3 ,
\bigg] 
\label{fH:extrapolated}
\end{align}
up to corrections of $\mathcal O(f^{(4)} \Delta^4 )$. 
Once the wavelet expansion of $f$ reaches the point where 
$f^{(k)} \Delta^k \lesssim \epsilon$ (for some error tolerance $\epsilon$) throughout the entire range $u \in [0, 1]$, the remaining wavelet coefficients can all be found from \eqref{fH:extrapolated}.

Note that the relative size of the  wavelet coefficients are determined by $f_0' \Delta 2^{-\delta\lambda}$, so the extrapolation is useful whenever
\begin{align}
\Delta^4 f_0^{(4)} < \epsilon < f_0' \frac{\Delta}{2 }.
\end{align}
That is, the values of the $\delta \lambda \leq 1$ coefficients can be used to infer the values for all $\delta \lambda \geq 2$, until $f_0' \Delta / 2^{\delta \lambda}$ and $\Delta^4 f^{(4)}$ become comparable in size.
Considering that the error in the cubic interpolation scales as $\Delta^4 f^{(4)}$, 
while the value of $\langle f | \lambda \mu \rangle$ scales as $f' \Delta$, 
we can use the extrapolation method to predict wavelet coefficients as far as:
\begin{align}
\delta \lambda \lesssim \log_2 \frac{\langle f_0' \rangle}{\Delta^3 \langle f_0^{(4)}\rangle } = 3 \lambda_\star + \log_2 \frac{\langle f'_0 \rangle }{\langle f_0^{(4)} \rangle} .
\end{align} 
Here $\langle f_0^{(p)} \rangle$ refers to the typical size of the $p$th derivative (dimensionless, with respect to $x \in [0,1]$) in this region. 
Using the cubic extrapolation, the information from the first $\lambda \leq \lambda_\star + 1$ wavelet generations can be used to predict the values of the following $3\lambda_\star -2$ generations. 
Comparing the number of coefficients generated this way to the original set obtained by integration, 
\begin{align}
\frac{ \text{$n_\text{coeffs}$ from cubic extrapolation} }{\text{$n_\text{coeffs}$ from direct integration }} 
&= \frac{2^{\lambda_\star + \delta \lambda + 1 } }{2^{\lambda_\star + 2} } \sim \frac{8^{ \lambda_\star }}{4}  ,
\label{extrapfraction}
\end{align}
it is clear that a relatively small number of inputs can generate a huge number of coefficients if $\lambda_\star \gtrsim \text{few}$. 

This scaling assumes that the derivatives $f^{(p)}$ are known exactly, in which case each new generation of wavelet coefficients can be used to increase the precision by a factor of $16$. 
When $f^{(p)}$ are extracted from the values of $\langle f | \lambda \mu \rangle$, the imperfect reconstruction of $f'$, $f''$ and $f^{(3)}$ introduces another small source of error.
As a result, the precision of the interpolating function near the edges of each bin increases by a factor of eight with each new generation, rather than 16, when $f^{(p)}$ is calculated from the wavelet coefficients.

\subsection{Spherical Wavelets} \label{sec:convergenceSph}

As in the previous section, let $\ket{ (0,0)}$ refer to a specific $\ket{\lambda_\star \mu_\star}$, where the cubic Taylor series is assumed to be a sufficiently precise approximation of $f(x)$ in $x_1 \leq x \leq x_3$. 
The only difference is that now $\ket{\lambda \mu}$ refers to a spherical wavelet, \eqref{def:sphericalHaar}, rather than a Haar wavelet.
Defining $\ket{(\delta\lambda, \delta \mu)}$ with respect to $\lambda_\star$, $\mu_\star$ as
\begin{align}
\ket{(\delta\lambda, \delta \mu) } &\equiv \ket{ \lambda_\star + \delta \lambda, 2^{\delta \lambda} \mu_\star + \delta\mu} ,
\end{align}
the projection $\langle f | (\delta \lambda, \delta \mu) \rangle$ can be taken directly from \eqref{sphH:cubic} with the following substitutions:
\begin{align}
\Delta &\rightarrow 2^{-\delta \lambda} \Delta_\star ,
&
\mu&\rightarrow 2^{\delta \lambda} \mu_\star  + \delta \mu ,
\end{align}
and with $F_p(\delta\lambda, \delta\mu)$ given as a function of $F_p^{(0)} \equiv F_p(\lambda_\star, \mu_\star)$ as:
\begin{align}
F_p(\delta\lambda, \delta\mu) &= \sum_{k \geq p} \frac{F_k^\star}{2^{\delta\lambda}} \frac{ k!}{ p! (k-p)!} \left( \frac{2\delta\mu +1 }{2^{\delta \lambda } } -1 \right)^{k-p} .
\label{Fpj}
\end{align}

\paragraph{Cubic Interpolation:}

Given $\langle f | 0 \rangle$, $\langle f | - \rangle$, and $\langle f | + \rangle$,
referring respectively to the $(0,0)$, $(1, 0)$, and $(1,1)$ wavelets, the values of $F_p \propto f^{(p)}_0$ for $p = 1,2,3$ can be extracted by inverting \eqref{sphH:cubic}. 
Once these derivatives are known, all higher order $\langle f | \lambda \mu \rangle$ coefficients are given by the cubic approximation of \eqref{sphH:cubic}.

Simplifying \eqref{Fpj} for the $\delta \lambda = 1$ case with $p \leq 3$, 
\begin{align}
F_1^\pm &= \tfrac{1}{2} F_1^\star \pm \tfrac{1}{2} F_2^\star + \tfrac{3}{8} F_3^\star,
&
F_2^{\pm } &= \tfrac{1}{2} F_2^\star \pm \tfrac{3}{4} F_3^\star,
&
F_3^\pm &= \tfrac{1}{2} F_3^\star
\end{align}
where the $\pm$ sign is positive for $\delta\mu=1$, negative for $\delta\mu = 0$.
In terms of $F_p$ and $H_p(\lambda, \mu)$, the three wavelet coefficients are given by
\begin{align}
\langle f | 0 \rangle &\simeq F_1^\star H_1^{(0)} + F_2^\star H_2^{(0)} + F_3^\star H_3^{(0)} ,
\\
\langle f | \pm \rangle &\simeq F_1^\star  C_1^{\pm}   + F_2^\star C_2^\pm  + F_3^\star C_3^\pm ,
\label{eq:fpmC}
\end{align}
with $0$, $+$ and $-$ referring to coefficients $(0,0)$, $(1,0)$ and $(1,1)$, respectively, and where
\begin{align}
C_1^\pm &\equiv \tfrac{1}{2} H_1(\pm), 
&
C_2^\pm &\equiv  \tfrac{1}{2} H_2(\pm) \pm \tfrac{1}{2}H_1(\pm) 
&
C_3^\pm &\equiv \tfrac{1}{2} H_3 (\pm) \pm \tfrac{3}{4} H_2(\pm) + \tfrac{3}{8} H_1(\pm)  .
\end{align}
Inverting this system of equations to find $F_p^{(0)}$ and $f_0^{(p)}$ is straightforward algebra. One way to represent the solution is:
\begin{align}
F_1^\star &\simeq d_1/D
&
F_2^\star &\simeq d_2/D,
&
F_3^\star &\simeq d_3/D, 
\label{solutionsF}
\\
d_1 &\equiv \left| \begin{array}{c c c} \langle f | 0 \rangle & H_2^0 & H_3^0 \\ \langle f | - \rangle & C_2^- & C_3^- \\ \langle f | + \rangle & C_2^+ & C_3^+ \end{array} \right| ,
&
d_2 &\equiv  \left| \begin{array}{c c c} H_1^0 & \langle f | 0 \rangle & H_3^0 \\ C_1^- & \langle f | - \rangle & C_3^- \\ C_1^+ & \langle f | + \rangle & C_3^+ \end{array} \right| ,
&
d_3 &\equiv  \left| \begin{array}{c c c} H_1^0 & H_2^0 &\langle f | 0 \rangle \\ C_1^- & C_2^- & \langle f | - \rangle \\ C_1^+ & C_2^+ & \langle f | + \rangle \end{array} \right| ,
\end{align}
with 
\begin{align}
D &\equiv \left| \begin{array}{c c c} H_1^0 & H_2^0 & H_3^0 \\ C_1^- & C_2^- & C_3^- \\ C_1^+ & C_2^+ & C_3^+ \end{array} \right| .
\label{cubic:D}
\end{align}
Given the \eqref{solutionsF} solutions to $F_p^\star$, recall that the derivatives themselves satisfy 
\begin{align}
f_0^{(p)} = 2^{p(\lambda_\star + 1) } F_p^\star /(p!).
\end{align}

Once $f^{(p)}_0$ are known, the descendent wavelets (with bases of support that overlap with the $\ket{\lambda_\star\mu_\star}$ wavelet) can all be calculated via 
\begin{align}
\langle f \ket{(\delta \lambda, \delta \mu)} &\simeq \sum_{p \geq 1} F_{p}(\delta \lambda, \delta \mu) H_p(\lambda, \mu),
\label{extrap:fH}
\end{align}
where $\lambda = \lambda_\star + \delta\lambda$, $\mu = 2^{\lambda_\star} \mu_\star + \delta\mu$. 
These solutions are approximate, with an accuracy controlled by the size of the missing $\Delta^4 f^{(4)}$ terms.
Considering that the wavelet coefficients scale as $\Delta^1 f^{(1)}$ at leading order, the precision of the wavelet extrapolation should improve by a power of~8 with each new generation of wavelet coefficients,
just like the Haar wavelets.

\paragraph{Generalization:}

Despite the focus on the cubic extrapolation, many of the results of this section leave $p$ generic, to make it easy to derive higher-order methods. 
For example, a seventh order wavelet extrapolation would involve derivatives out to $p \leq 7$, and it would include the values of the four $\delta \lambda = 2$ wavelets to constrain $F_{1\ldots 7}$.
Extending \eqref{eq:fpmC}, the linear system includes seven equations of the form 
\begin{align}
\langle f | \delta \lambda, \delta \mu \rangle &= \sum_{k=1}^7 \left( \sum_{p = 1}^k a_{k,p}(\delta\lambda, \delta \mu)\, H_p(\lambda, \mu)\right)  F_k^\star ,
\label{fdelta:seventh}
\\
a_{k,p} &\equiv \frac{1}{2^{\delta\lambda}} \frac{ k!}{ p! (k-p)!} \left( \frac{2\delta\mu +1 }{2^{\delta \lambda } } -1 \right)^{k-p} .
\end{align}
 The matrix appearing in \eqref{cubic:D} expands rightwards with $p \rightarrow 7$, and downwards to include the four new coefficients. 

Once $\Delta_\star$ is small enough that the Taylor series converges, subsequent increases in $\delta\lambda$ increase the precision of the wavelet expansion by factors of $2^7 = 128$. The computational price is that we must solve a $7\times 7$ system of equations for $F_k^\star$, rather than two neighboring $3\times 3$ systems.

For some applications, the simpler  $p=1$ linear extrapolation may be sufficiently precise. Here $f'_0$ is determined just by $\langle f | (0,0) \rangle$,
\begin{align}
\frac{\Delta \, f_0'}{2} = 
F_1 &\simeq \frac{ \langle f | \lambda_\star \mu_\star \rangle }{H_1(\lambda_\star, \mu_\star)} ,
\end{align}
on the assumption that all $f^{(p)}$ derivatives with $p \geq 2$ are irrelevant.

\paragraph{Interpolation:}

For the scattering rate calculation the coefficients $\langle f | \lambda \mu \rangle$ themselves are more important than the reconstruction of $f$. Even so, a wavelet-defined smooth interpolating function for $f(x)$ may be useful.   
Using \eqref{extrap:fH} to estimate the $\lambda \gg \lambda_\star$ wavelet coefficients produces a precise reconstruction of the function $f(x)$.
As in the Haar wavelet version, one must supply the value of $f_0 = f(x_2)$ at the center of the region, either from the inverse wavelet transformation or (more precisely) from $f(x)$ directly.

In many cases, e.g.~if $f(x)$ is a closed-form analytic function, it is easier to get the $f_0^{(p)}$ values directly from the derivatives $f^{(p)}(x)$, rather than from integrating $\langle f | \lambda \mu \rangle$. 
Bypassing the $\langle f | 0 \rangle$ and $\langle f | \pm \rangle$ integrals removes two generations of wavelets from the list of coefficients that must be obtained from integration,  reducing the difficulty of the calculation by a factor of 4, while also increasing the precision of the interpolation.
In this case, the relative precision improves by a factor of $2^4$ with every subsequent generation of wavelets, rather than $2^3$.

\section{Transformations From A Gaussian Basis} \label{sec:gaussian}

A normalized spherical 3d Gaussian has the form
\begin{align}
{g}_i(\vec u, \vec u_i, \sigma_i ) &=  \frac{e^{-|\vec u - \vec u_i |^2/2\sigma_i^2 } }{\sigma_i^3 \, (2\pi)^{3/2} },
&
\int\! d^3 u \, {g}_i(\vec u, \vec u_i, \sigma_i )  \equiv  1,
\end{align}
for a Gaussian with width parameterized by $\sigma_i$, centered at $\vec u_i$, as a function of $\vec u$.
 (Everything in this section applies equally well to 3d functions of velocity or momentum.)
The exponential of the dot product can be expanded in Legendre polynomials,
\begin{align}
e^{x \cos \theta} = \sum_{\ell=0}^\infty (2 \ell + 1)   \, i_\ell^{(1)}(x) \, P_\ell(\cos\theta) ,
&&
i_\ell^{(1)}(z) = i^{-\ell } j_\ell(iz) = \sqrt{ \frac{\pi}{2z} } I_{\ell + 1/2} (z) ,
\end{align}
where $i^{(1)}_{\ell}$ is the $\ell$th spherical modified Bessel function of the first kind.
For integer $\ell \geq 0$, $i_\ell^{(1)}(z)$ can be written in terms of rational functions involving $\sinh x$ and $\cosh x$, via~\cite{AandS}
\begin{align}
i_\ell(z) &= z^\ell \left[ \frac{d}{z \,dz} \right]^\ell \left( \frac{\sinh z}{z} \right) .
\label{def:i1z}
\end{align}
Note that for $z=0$, 
\begin{align}
i_\ell^{(1)}(z) \simeq z^\ell \left( \frac{ \sqrt{\pi} }{2^{1 + \ell} \Gamma(\frac{3}{2} + \ell) }  + \mathcal O(z^2) \right) ,
&&
i_0^{(1)}(0) = 1,
&&
i_{\ell \geq 1}^{(1)}(0) = 0,
\end{align}
while in the large $z$ limit $i_\ell^{(1)}(z)$ grows as $\exp(z)$. 
For integer $\ell$, $i_\ell^{(1)}$ can be written as rational functions of $z$ multiplying $\sinh z$ and $\cosh z$, 
\begin{align}
i_\ell^{(1)}(z) &= f_\ell(z) \, \sinh z + f_{-\ell -1}(z) \, \cosh z, 
\label{i1z:recursion}
\end{align}
for $f_\ell(z)$ given by the recursion relations~\cite{AandS}:
\begin{align}
f_0(z) = \frac{1}{z}, 
&&
f_1(z) = - \frac{1}{z^2}, 
&&
f_{\ell -1} - f_{\ell+1} = \frac{2 \ell + 1}{z} f_\ell .
\end{align}
The functions $z^{\ell + 1} f_\ell(z)$ and $z^{\ell + 1} f_{-(\ell + 1)}(z)$ are polynomials of order $\ell$ or $\ell-1$.

Using the orthogonality of the spherical harmonics,
\begin{align}
 g_i  &= \frac{4\pi  }{(2\pi)^{3/2} \sigma_i^3}  \exp\left( - \frac{u^2 + u_i^2 }{2\sigma_i^2} \right) \sum_{\ell = 0}^\infty \sum_{m = - \ell}^{\ell}  i_\ell^{(1)}\!\left( \frac{  u_i u }{ \sigma_i^2} \right) Y_{\ell m}(\hat u) Y_{\ell m } (\hat u_i) ,
\\
\langle  g_i(\vec u_i, \sigma_i) | \nlm \rangle &= \sqrt{ \frac{2}{\pi}} \frac{  Y_{\ell m } (\hat u_i) }{  u_0^3}   \int_0^\infty \frac{u^2 du}{  \sigma_i^3} r_n^{(\ell)}(u) \,e^{-(u^2+u_i^2)/2\sigma_i^2}  \,i_\ell^{(1)}\!\left( \frac{  u_i u }{ \sigma_i^2} \right) ,
\end{align}
where $ r_n^{(\ell)}(u)$ is the radial basis function, which can in principle depend on $\ell$.
So, $\langle  g_i | n \ell m \rangle$ is determined by a 1d integral rather than a 3d one:
\begin{align}
\langle  g_i(\vec u_i, \sigma_i) | \nlm \rangle &\equiv \frac{1}{u_0^3} Y_{\ell m } (\hat u_i)  \, \mathcal G_{n \ell} (u_i, \sigma_i),
\label{def:mathcalG}
\\
\mathcal G_{n \ell}(u_i, \sigma_i) 
&= \sqrt{ \frac{2}{\pi}} \int_0^\infty \frac{ u^2 du}{\sigma_i^3}  r_n^{(\ell)}(u) \,e^{-(u^2+u_i^2)/2\sigma_i^2}  \, i_\ell^{(1)}\! \left(   \frac{  u_i u }{\sigma_i^2 } \right) .
\end{align}
$\mathcal G_{n \ell}$ does not depend on the index $m$: so, not only is the $\mathcal G$ integral 1d rather than 3d, it also only needs to be evaluated once for every $(n \ell)$, rather than for every $(n \ell m)$. 
Following the pattern demonstrated elsewhere in this paper, one analytic simplification has led automatically to another one.

Using the \eqref{i1z:recursion} property of the $i_\ell^{(1)}$ functions, the \eqref{def:mathcalG} integrand can be simplified further. 
Introducing polynomials $P_\ell(z)$ and $Q_\ell(z)$, 
\begin{align}
P_\ell(z) &= z^{\ell + 1} f_\ell(z), 
&
Q_\ell(z) &= z^{\ell + 1} f_{-\ell - 1} (z), 
&
i_\ell^{(1)}(z) &= \frac{P_\ell(z)  \sinh z + Q_\ell(z) \cosh z }{z^{\ell+1} },
\end{align}
the integral $\mathcal G_{n \ell}$ is equivalently:
\begin{align}
\mathcal G_{n \ell} &= \sqrt{ \frac{2}{\pi}} \int_{0}^{\infty} \! \frac{x \,dx}{x^\ell} r_n^{\ell} (x \sigma_i)  \left[ \frac{ Q_\ell(x) + P_\ell(x) }{2} e^{- (x - x_i)^2 } + \frac{Q_\ell(x) - P_\ell(x) }{2} e^{- (x + x_i)^2 } \right],
\end{align}
defining $x =  u/(\sqrt{2}\sigma_i)$ and $x_i =   u_i /(\sqrt{2}\sigma_i)$. 
So, it is not actually necessary to refer to $I_{\ell + 1/2}(z)$ when evaluating $\mathcal G_{n \ell}$ numerically.

\subsection*{Functions in Gaussian Basis}

Functions $g(u)$ that are comprised of sums of gaussians of the form \eqref{gauss:norm} can be written in terms of $\mathcal G$ and $u_0$:
\begin{align}
\ket{g} &\equiv \sum_i c_i \, \ket{g_i(\vec u_i, \sigma_i) },
\\
\langle g | n \ell m \rangle &= 
\sum_{i} c_i \langle n \ell m | g_i(\vec u_i, \sigma_i) \rangle 
=\sum_i c_i \frac{Y_{\ell m} (\hat u_i)}{u_0^3}    \, \mathcal G_{n \ell} (u_i, \sigma_i).
\end{align}
So, the $\langle n \ell m | g_i(\vec u_i, \sigma_i) \rangle$ can be thought of as coefficients of a matrix that maps the Gaussian basis spanned by $\ket{ g_i }$ onto the orthogonal wavelet-harmonic basis spanned by $\ket{n \ell m}$.

\medskip

In Section~\ref{sec:extrapolation} I use a related analytic result to test the convergence of the radial function expansion.
The function $\tilde g_i$ can be projected onto an $(\ell m)$ spherical harmonic,
\begin{align}
\langle  g_i | \ell m \rangle &\equiv \int\! d\Omega\, Y_{\ell m}(\hat u) \, g_i(\vec u) 
= \frac{\sqrt{2} }{\sqrt{\pi} \sigma_i^3}\, Y_{\ell m}(\hat u_i) \,\exp\!\left( - \frac{u^2 + u_i^2 }{2\sigma_i^2} \right) \, i^{(1)}_{\ell} \!\left( \frac{  u_i u}{\sigma_i^2} \right) .
\label{gauss:lm}
\end{align}
This is an analytic function of the radial coordinate $u$, and the gaussian parameters $(\vec u_i, \sigma_i)$.
By comparing the second and third terms of
\begin{align}
\langle g | \ell m \rangle &= \sum_{n=0}^\infty \langle g | n \ell m\rangle \ket{n} 
= \sum_i c_i  \langle  g_i(\vec u_i, \sigma_i) | \ell m \rangle ,
\end{align}
for any function $g$ given as a sum of gaussians, 
\eqref{gauss:lm} can be used to check the local convergence for any basis of radial functions.

For the global convergence, the distributional energy $\mathcal E$ is a useful reference. For a gaussian function, integrated to $u \rightarrow \infty$, the total $\mathcal E$ is simply
\begin{align}
\mathcal E[g] &= \sum_{i, j } \frac{c_i c_j}{8\pi^3 \sigma_i^3 \sigma_j^3}  \int\! d^3 u \, \exp\left( - \frac{ \left| \vec u - \vec u_i\right|^2}{2\sigma_i^2} - \frac{ \left| \vec u - \vec u_j \right|^2}{2\sigma_j^2} \right)  ,
\\&=  \sum_{i, j } \frac{c_i c_j }{(2\pi)^{3/2} (\sigma_i^2+  \sigma_j^2)^{3/2}} \exp\left( \frac{u_{ij}^2}{2\sigma_{ij}^2}  - \frac{u_i^2}{2\sigma_i^2} - \frac{u_j^2}{2\sigma_j^2} \right),
\end{align}
where $\sigma_{ij}$ and $\vec u_{ij}$ are defined for each $(i,j)$ as
\begin{align}
\sigma_{ij}^2 &\equiv \frac{\sigma_i^2 \sigma_j^2}{\sigma_i^2+ \sigma_j^2} ,
&
\vec u_{ij} &\equiv \frac{\sigma_j^2 \vec u_i + \sigma_i^2 \vec u_j }{\sigma_i^2 + \sigma_j^2} .
\end{align}

\subsection*{Derivatives For Wavelet Extrapolation}

When the radial function $r_n(u)$ is a higher-order wavelet, with a base of support $\Delta u$ small enough that the integrand of \eqref{gi:nlm} is well approximated by a Taylor series, then $\langle g_i | n \ell m \rangle$ can be found using the extrapolation method of Section~\ref{sec:extrapolation}. 
The cubic method uses the first three derivatives of the radial function $\langle g | \ell m \rangle$ to approximate all of the high-$n$ $\langle g | n \ell m \rangle$ coefficients.
In the generic case, these derivatives are inferred from $\langle g | (\lambda_\star \mu_\star) \ell m \rangle$ and its two $\lambda_\star + 1$ descendent wavelets.
Here, given the analytic form for $\langle g | \ell m \rangle$, the derivatives can be calculated directly,
noting from \eqref{def:i1z} that
\begin{align}
\frac{d}{dz} i_\ell^{(1)}(z) &= i_{\ell + 1}^{(1)}(z) + \frac{\ell}{z} i_\ell^{(1)}(z) .
\end{align} 
Writing the results in terms of $x \equiv u/(\sqrt{2} \sigma_i)$, $x_i \equiv u_i / (\sqrt{2} \sigma_i)$:
\begin{align}
g_{i,\ell m}(u) &\equiv \langle g_i | \ell m \rangle = \mathcal Y_{i, \ell m}  \, e^{-x^2} i_\ell^{(1)}(2 x_i x) ,
&\mathcal Y_{i, \ell m}&\equiv \sqrt{ \frac{2}{\pi}} \frac{ Y_{\ell m}(\hat u_i) e^{-x_i^2} }{\sigma_i^3},
\end{align}
\begin{align}
\frac{d g_{i, \ell m} }{du} &= \frac{\mathcal Y_{i, \ell m} }{\sqrt{2} \sigma_i} e^{-x^2} \left\{ 2 x_i\, i_{\ell + 1}^{(1)}(2 x_i x) + \left(\frac{\ell}{x} - 2x \right)   i_\ell^{(1)}(2 x_i x) \right\},
\\
\frac{d^2 g_{i, \ell m} }{du^2} &=  \frac{\mathcal Y_{i, \ell m} }{2 \sigma_i^2} e^{-x^2}
\bigg\{ (2x_i)^2 i_{\ell+2}^{(1)}(2 x_i x) + \left[ \frac{2 \ell + 1 }{x} - 4x \right] (2 x_i ) i_{\ell + 1}^{(1)} (2 x_i x) 
\nonumber\\& \hspace{6em}
+ \left[  \frac{\ell (\ell - 1) }{x^2} - 2( 2 \ell + 1) + 4 x^2 \right] i_\ell^{(1)}(2 x_i x) 
\bigg\}
\\
\frac{d^3 g_{i, \ell m} }{du^3} &=  \frac{\mathcal Y_{i, \ell m} }{2^{3/2} \sigma_i^3} e^{-x^2}
\bigg\{ (2x_i)^3 i_{\ell + 3}^{(1)}(2x_i x) + 3 \left[ \frac{  \ell + 1}{x} - 2x \right] (2x_i)^2 i_{\ell + 2}^{(1)} (2 x_i x)  
\nonumber\\&\hspace{6em} + 3\left[ \frac{\ell^2}{x^2} - 4(\ell + 1) + 4 x^2 \right] (2 x_i) i_{\ell + 1}^{(1)} (2 x_i x) 
\nonumber\\&\hspace{6em} + \left[ \frac{\ell(\ell - 1) (\ell - 2) }{x^3} - \frac{6 \ell^2}{x} + 12 x (\ell + 1) - 8 x^3 \right] i_\ell^{(1)}(2 x_i x) 
\bigg\} .
\end{align}
This result can be used together  with \eqref{sphH:cubic} to approximate the values of $\langle g_i | n \ell m \rangle$ for wavelets $\ket{n}$ narrow enough that the cubic Taylor series for $g_{i, \ell m}$ is accurate.

\section{Timing Information} \label{sec:timing}

From \eqref{vectorsarethenewblack}, and the argument that matrix multiplication is easier than multidimensional numerical integration, it is clear that the wavelet-harmonic integration method should reduce the computation time by orders of magnitude for a direct detection analysis. 
In this appendix, I provide some specific results for the evaluation times of each part of the vector space calculation.
While substantial effort was made to simplify the analytic parts of the calculation (e.g.~deriving the gaussian basis $\mathcal G_{n \ell}$ method of Appendix~\ref{sec:gaussian}), 
very little effort was made to optimize the numeric implementation. 
Nothing was parallelized, even though the vector space calculation is massively parallelizable; the Python implementation \texttt{vsdm} uses relatively slow methods for combining vectors and arrays; and the method for evaluating spherical harmonics fails to take advantage of the iterative relations of the modified Legendre polynomials.
For these reasons (and perhaps others I have overlooked), it is likely that a couple more orders of magnitude of improvement could be coaxed out of the numeric implementation.

The evaluation times listed in this section come from running \texttt{vsdm 0.1.0} in a Jupyter notebook on a personal computer.\footnote{More recent versions of \texttt{vsdm} ($\geq 0.3.0$) use an improved iterative method to calculate $P_\ell^m$ and $Y_{\ell m}$; this is necessary for ensuring accuracy at large $\ell \gg 60$, and it is also much faster.}
For $g_\chi(\vec v)$, I use the four-gaussian velocity distribution \eqref{eq:model4}, evaluated with all $\ell \leq 36$ and all $|m| \leq \ell$, with $2^{8}$ radial basis functions defined on $| \vec{v} | \leq 960\,\text{km/s}$, for a total of 350,464 distinct $\ket{n\ell m}$ velocity basis functions. 
For the momentum form factor $f_s^2(\vec q)$, I compare two examples: the $\vec n = (1, 1, 2)$ and $\vec n = (3, 2, 1)$ excited states of the $(4 a_0, 7 a_0, 10 a_0)$ box, \eqref{fS2:box}, both including $2^{10}$ radial basis functions on $|\vec q| \leq 10 a_0^{-1}$. The wavelet-harmonic expansion of the simpler $\vec n = (1, 1, 2)$ form factor converges especially quickly, so the $(n \ell m )$ expansion is terminated at $\ell \leq 30$, while the $(3, 2, 1)$ expansion uses $\ell \leq 36$. For odd $\ell$, odd $m$, or negative $m$, $\langle n \ell m | f_s^2 \rangle = 0$ for both models,  so the two wavelet-harmonic expansions include a total of 114,688 and 194,560 nonzero coefficients, respectively. 
In order to isolate the impact of having more or fewer $\langle n \ell m | f_s^2 \rangle$ coefficients, I took the excitation energies for the two examples to be identical, $\Delta E = 4.03\, \text{eV}$, when calculating the scattering rate.

Each list of $\mathcal O(10^5)$ coefficients includes a substantial fraction of irrelevantly small values, and so each analysis is performed on a subset of relevant coefficients. 
With the $\langle n \ell m | n \ell m \rangle = 1$ normalization, and using the difference in the distributional norm $\Delta \mathcal E$ to measure the error, 
the optimal subsets are found by simply sorting the list of $\langle f | n \ell m \rangle$ coefficients and keeping only the $n \ell m$ with the largest values.

\subsection{Partial Scattering Rate Matrix}

After $g_\chi$ and $f_s^2$ have been projected onto their respective vector spaces, the only time-consuming step is to evaluate the partial rate matrices $K^{(\ell)}_{m m'}$.
From \eqref{rate:concise}, every $(\ell, m, m')$ coefficient is given by a double sum over indices $n_v < n_\text{max}^{(v)}$ and $n_q < n_\text{max}^{(q)}$, with each term in the sum calling the analytic function $\mathcal I^{(\ell)}_{n, n'}( m_\chi, F_\text{DM})$. 
When the sums over $n, n'$ are ``dense,'' where relatively few coefficients vanish identically, it is faster to evaluate the $\mathcal I^{(\ell)}_{n, n'}$ matrices first, and to reuse it for every $m,m'$ coefficient in $K^{(\ell)}$. 
Alternatively, if most values of $\langle g_\chi | n \ell m \rangle$ and $\langle n \ell m | f_s^2 \rangle$ vanish, it may be faster to perform a sparse sum, evaluating only those coefficients in $\mathcal I$ that are required. 
This depends on how many of the coefficients can be skipped, and this depends in turn on the precision goal.

So, there is a ``dense array'' method for finding $K^{(\ell)}$, where $\mathcal I^{(\ell)}$ is evaluated as a precursor to $K^{(\ell)}$, 
and a ``sparse array'' method, where $K^{(\ell)}$ is found by evaluating \eqref{def:mathcalK} for only the most important pairs of $\langle g_\chi | n \ell m \rangle$ and $\langle f_s^2 | n' \ell m' \rangle$ coefficients. 
To show how the evaluation time changes according to the precision, I constructed seven examples for each function $g_\chi$, $f_s^2(1, 1, 2)$, and $f_s^2(3, 2, 1)$, with precision goals of $\Delta \mathcal E / \mathcal E = 0.1, 0.03, \ldots, 0.003, 10^{-4}$. 
Table~\ref{table:timing} lists the total number of coefficients $N_v$ or $N_q$ required to meet these precision goals for each function, as well as the evaluation times for the $\mathcal I^{(\ell)}$ and $K^{(\ell)}$ matrices. 

The evaluation time for $K$ and $\mathcal I$ depends on how close $m_\chi$ is to the kinematic threshold, $m_\chi > 2 \Delta E / v_\text{max}^2$, so the analysis includes $m_\chi = 1, 10, 100\, \text{MeV}$.
For $m_\chi = 1\, \text{MeV}$, most elements of $\mathcal I^{(\ell)}$ vanish trivially: the velocity required to excite the system to the $\Delta E \simeq 4 \, \text{eV}$ final state is $v > 850\, \text{km/s}$, where $g_\chi$ has very little support. It is relatively fast to evaluate $\mathcal I^{(\ell)}$ for the handful of nonzero coefficients.
By $m_\chi \geq 10 \, \text{MeV}$, on the other hand, $T_\text{eval.$I$}$ largely loses its $m_\chi$ dependence, asymptoting to the larger value listed in Table~\ref{table:timing}. The lower end of each range corresponds to $m_\chi = 1\, \text{MeV}$, while the upper value usually applies to $m_\chi = \text{100} \, \text{MeV}$.

\begin{table}
\centering
\begin{tabular}{| l | c | cccl |  cccl |} \hline
$\Delta \mathcal E/\mathcal E$&	 $N_v$ 	
	& $N_q$	& $\ell_\text{max}$ 	& \hspace{.35em} $T_\text{eval.$I$}$ & $T_\text{eval.$K$}$  
	& $N_q^\text{alt}$	& $\ell_\text{max}^\text{alt}$ & \hspace{.35em} $T_\text{eval.$I$}^\text{alt}$ & $T_\text{eval.$K$}^\text{alt}$
\\ \hline 
\phantom{.0}10\%: sparse: & 1430  	
	&   16  	& 4	& --&  8\,ms 
	&	100	& 14	& --&  40\,ms 
\\
\phantom{.100\%:} dense: & 	  
		&&& 0.1--0.8\,s  &    18\,ms
	 	&&& 0.2--2.5\,s  &   23\,ms
\\ \hline
\phantom{.10}3\%: sparse: & 3062  	
	&   40  	& 8	& --&  40\,ms 
	&	206	& 16	& --&  0.3\,s 
\\
\phantom{.100\%:} dense: & 	  
		&&& 0.4--3.5\,s  &    45\,ms
	 	&&& 0.5--6.5\,s  &   54\,ms
\\ \hline
\phantom{.10}1\%: sparse: & 5229  	
	&   82  	& 10	& --&  0.15\,s 
	&	387	& 18	& --&  1.0\,s 
\\
\phantom{.100\%:} dense: & 	  
		&&& 1.8--18\,s  &   80\,ms
	 	&&& 2.3--30\,s  &   0.1\,s
\\ \hline
\phantom{0}0.3\%: sparse: & 8483  	
	&   153  	& 10	& --&  0.4\,s 
	&	718	& 20	& --&  3.0\,s 
\\
\phantom{.100\%:} dense: & 	  
		&&& 3.5--35\,s  &    0.15\,s
	 	&&& 4.6--70\,s  &   0.18\,s
\\ \hline
\phantom{0}0.1\%: sparse: & 12294  	
	&   274  	& 12	& --&  1.0\,s 
	&	1235	& 24	& --&  6.7\,s 
\\
\phantom{.100\%:} dense: & 	  
		&&& 8.4--90\,s  &    0.20\,s
	 	&&& 11--200\,s  &   0.26\,s
\\ \hline
0.03\%: sparse: & 17211  	
	&   493  	& 12	& --&  2.7\,s 
	&	2201	& 26	& --&  16\,s 
\\
\phantom{.100\%:} dense: & 	  
		&&& 15--180\,s  &    0.3\,s
	 	&&& 22--480\,s  &   0.4\,s
\\ \hline
0.01\%: sparse: & 22153  	
	&   771  	& 12	& --&  5.5\,s 
	&	3673	& 26	& --&  34\,s 
\\
\phantom{.100\%:} dense: & 	  
		&&& 18--240\,s  &    0.4\,s
	 	&&& 42--940\,s  &   0.5\,s
\\ \hline
\phantom{---} $\Delta \mathcal E \rightarrow 0$ & all &  all & 30	& 100--3200\,s &  12\,s  
	&  all & 36	& 140--5100\,s &  13\,s  
\\ \hline
\end{tabular}
\caption{The partial rate matrix $K^{(\ell)}$ was calculated for all $\ell \leq \ell_\text{max}$ in four examples, each using the $N_{v, q}$ most important coefficients to meet the precision goals $\Delta \mathcal E / \mathcal E = 10\%$, 3\%, \ldots, 0.01\%.  
Two models for $f_s^2(\vec q)$ are included: the first excited state, $\vec n = (1, 1, 2)$, and a higher excited state, $\vec n = (3, 2, 1)$, labeled here as ``alt.''
Each row provides $T_\text{eval.$K$}$ from the sparse and dense array versions of the calculation, with the full $\mathcal I^{(\ell)}$ matrix calculated only in the latter case.
The last example, $\Delta\mathcal E \rightarrow 0$, applies the dense array sum to the entire set (``all'') of $N_v = 350464$ and $N_q =  114688$ or $N_q^\text{alt} =  194560$  coefficients.
Where $T_\text{eval.}$ is provided as a range, the upper and lower limits correspond to the cases $m_\chi = 100\, \text{MeV}$ and $m_\chi = 1\, \text{MeV}$, respectively.
Two versions of $F_\text{DM} \propto 1/q^n$ were used, $n=0$ and $n=2$, but the difference in evaluation time is negligible.
$T_\text{eval.$K$}$ does not depend strongly on $m_\chi$. }
\label{table:timing}
\end{table}

From \eqref{timing:total}, the evaluation time for the rate calculation depends on the number of $g_\chi$ and $f_s^2$ models, $N_{g_\chi}$ and $N_{f_S}$; the number of DM particle models $(m_\chi, F_\text{DM}^2)$, $N_\text{DM}$; and the number of detector orientations, $N_{\mathcal R}$.  
\begin{align}
T_\text{total} &= N_{g_\chi} N_{f_S} N_\text{DM} \left( N_{\mathcal R} \cdot  T_\text{eval.R} 
+  T_\text{eval.$K$} \right) 
+ N_\text{DM} \cdot  T_\text{eval.$\mathcal I$} 
+ N_{\mathcal R} \cdot T_\text{eval.$G$} .
\end{align}
In the sparse version of the $K^{(\ell)}$ sum, one can drop the $T_\text{eval.$\mathcal I$}$ step, though $T_\text{eval.K}$ is often longer as a result. As long as  $N_{g_\chi } N_{f_S} T_\text{eval.K} < T_\text{eval.$\mathcal I$}$, the sparse sum saves time. 

In the ``dense array'' method, the calculation of $\mathcal I^{(\ell)}$ adds $T_\text{eval.$I$}$ to the evaluation time for each DM model, but streamlines the rest of the $K^{(\ell)}$ calculation.
The coefficients of $K_{m m'}^{(\ell)}$ are generated by multiplying vectors of $\langle g_\chi | n \ell m\rangle$ (with fixed $\ell,m$) and $\langle f_s^2 | n' \ell m'\rangle$ (of fixed $\ell, m'$) against the $\mathcal I^{(\ell)}_{n n'}$ matrix.
This procedure is marginally faster when most of these coefficients are set to zero, but $T_\text{eval.$K$}$  depends less strongly on $N_{v,q}$.
It does depend strongly on $\ell_\text{max}$: the number of operations to evaluate all $K^{(\ell)}$ matrices scales as $\ell_\text{max}^3 n^{(v)}_\text{max} n^{(q)}_\text{max} $. 

In contexts where extreme precision is required, the vectors $\ket{g_\chi}$ and $\ket{f_s^2}$ must be expanded to (much) larger values of $n_\text{max}$. 
The wavelet extrapolation methods of Section~\ref{sec:extrapolation} would extend $n_\text{max}$ to $2^k n_\text{max}$ for some power of $k$; 
alternatively, the $k$th order polynomial method suggested in Section~\ref{sec:polycap} would achieve equivalent accuracy with only $k n_\text{max}$ coefficients per $(\ell, m)$ harmonic. 
In either case, the evaluation times for a high-precision calculation can be estimated by scaling the $T_\text{eval}$ in Table~\ref{table:timing} according to:
\begin{align}
T_\text{eval.$K$} \propto \ell_\text{max}^3 n^{(v)}_\text{max} n^{(q)}_\text{max} ,
&&
T_\text{eval.$I$} \propto \ell_\text{max} n^{(v)}_\text{max} n^{(q)}_\text{max} .
\end{align}
Returning to the \eqref{counting:old} notation, an analysis with $N_{g_\chi}$ velocity distributions, $N_{f_S}$ detector models, and $N_\text{DM}$ dark matter particle models takes 
\begin{align}
T_\text{total}^K = N_\text{DM} \left(T_\text{eval.$I$}  +  T_\text{eval.$K$} \cdot N_{g_\chi} \cdot  N_{f_S}   \right)
\label{eq:timingK}
\end{align}
to evaluate all of the $K^{(\ell)}$ matrices at the specified level of precision. From Table~\ref{table:timing}, the dense array version is usually faster when $N_{g_\chi} N_{f_S} \gg 10^2$ for the ``alt'' model, $\vec n = (3, 2, 1)$. For $N_{g_\chi} N_{f_S} = 10^2$, the sparse array approach is faster for the  $\vec n = (1, 1, 2)$ examples at $\Delta\mathcal E /\mathcal E \gtrsim 0.3\%$.

After the partial rate matrices $K^{(\ell)}$ are calculated for every combination of DM model, $g_\chi$, and $f_s^2$, 
the only remaining step of the calculation is to evaluate the scattering rate itself, by multiplying $K^{(\ell)}$ by a Wigner $G^{(\ell)}$ matrix and taking the trace, as in \eqref{def:RKG}.

\subsection{Rotations}

When the scattering rate depends on the orientation of the detector, the analysis must be repeated for some list of $N_{\mathcal R}$ rotations $\mathcal R \in SO(3)$.
Evaluating the $\mathcal G^{(\ell)}(\mathcal R)$ matrices adds
\begin{align}
T_\text{total}^G &= N_{\mathcal R} \cdot T_\text{eval.$G$}
\end{align}
to the total evaluation time. Like $K^{(\ell)}$, the total number of coefficients scales as $\ell_\text{max}^3$; unlike $K$, the evaluation time for $G$ is independent of the number of models in the analysis.
Consequently, $T_\text{total}^G$ takes up a relatively small part of the evaluation time, unless $N_{\mathcal R}$ is especially large (e.g.~$N_{\mathcal R} \gtrsim 10^5$). 

The average evaluation time for $G^{(\ell)}(\mathcal R)$ is summarized in the table below for various $\ell_\text{max}$, using the \texttt{spherical} Python package to evaluate the Wigner $D^{(\ell)}$ matrix, and \eqref{Gmatrix} to find $G^{(\ell)}$: 
\begin{align}
\begin{tabular}{| c | c c c c c |} \hline
$\ell_\text{max}$: & $ 12$ & $16$	& $24$ & $ 36$  & 60 \\ \hline
$T_\text{eval.$G$}$: & 2.7\,ms	& 4.8\,ms & 13\,ms & 37\,ms & 160\,ms 
\\ \hline
\end{tabular}
\end{align}
As expected, $T_\text{eval.$G$} \propto \ell_\text{max}^3$. 
The $T_\text{eval.$G$}$ listed above include all even and odd values of $\ell \leq \ell_\text{max}$.  
In cases where $g_\chi$ or $f_s^2$ is invariant under a $\mathbbm Z_2$ central inversion symmetry, the odd values of $\ell$ can be dropped, reducing the evaluation time by a factor of 2.

At this stage, $K^{(\ell)}$ and $ G^{(\ell)}$ have been evaluated for every possibility in the analysis, and the only thing left to do is combine them. From \eqref{rate:concise}, the event rate in the detector target is:
\begin{align}
R(g_\chi, f_s^2, m_\chi, F_\text{DM}, \mathcal R) &= \frac{k_0}{T_\text{exp}} \sum_{\ell = 0}^{\ell_\text{max}} \sum_{m,m' = - \ell}^{\ell} G^{(\ell)}_{mm'}(\mathcal R)   \cdot  K^{(\ell)}_{mm'}(g_\chi, f_s^2, m_\chi, F_\text{DM}).
\end{align}
This trace must be evaluated once for every point in the analysis, i.e.~$N_{\mathcal R} N_\text{DM} N_{g_\chi} N_{f_S}$ many times, 
adding
\begin{align}
T_\text{total}^\text{Tr} &= N_{\mathcal R} N_\text{DM} N_{g_\chi} N_{f_S} \cdot T_\text{eval.R}
\end{align}
to the total evaluation time. Each $\ell$ term is essentially a dot product of two $(2\ell + 1)^2$ dimensional vectors, so  $T_\text{eval.R}$ is measured in microseconds: 
\begin{align}
\begin{tabular}{| c | c c c c c  |} \hline
$\ell_\text{max}$: & $ 12$ & $16$	& $24$ & $ 36$  & 60 \\ \hline
$T_\text{eval.R}$: & 3.5\,$\mu$s & 5.3\,$\mu$s & 12\,$\mu$s & 41\,$\mu$s & 100\,$\mu$s 
\\ \hline
only even $\ell$: & 2.1\,$\mu$s & 3.0\,$\mu$s & 5.7\,$\mu$s & 15\,$\mu$s & 47\,$\mu$s 
\\ \hline
\end{tabular}
\end{align}
These averages were calculated using the same set of $1000$ randomly selected rotations $\mathcal R$ for various DM models.
The first row  includes all $\ell \leq \ell_\text{max}$ in the sum, while the second row only includes the even values of $\ell$. 

If the vector space versions of $\ket{g_\chi}$ and $\ket{f_s^2}$ are not known at the start of the calculation, then the calculation of $\langle f_s^2 | n \ell m \rangle$ and $\langle g_\chi | n \ell m \rangle$ adds
\begin{align}
T_\text{total}^\text{proj} &= N_{g_\chi} T_\text{proj.$V$} + N_{f_s} T_\text{proj.$Q$} 
\end{align}
to the total evaluation time, where $T_\text{proj.$V,Q$}$ is the time to project one model of $g_\chi$ or $f_s^2$ onto the $V$ or $Q$ vector space.
Unless $N_\text{DM}$ is extremely large, this is usually the most time-consuming portion of the calculation.
In the Section~\ref{sec:demo} examples, $T_\text{proj.$Q$} \sim 30\,\text{hours}$, while the $\mathcal G_{n \ell}$ gaussian basis trick permits $T_\text{proj.$V$} <  15\,\text{min}$ when neglecting even $\ell$. 

In conclusion, a complete analysis (starting without pre-evaluated $\ket{g_\chi}$ and $\ket{f_s^2}$) can be completed in a time:
\begin{align}
T_\text{total} &= T_\text{total}^K + T_\text{total}^G + T_\text{total}^\text{Tr} + T_\text{total}^\text{proj}.
\end{align}
This should be compared to the current standard, which requires
\begin{align}
T_\text{total}^\text{\,slow} &= N_{\mathcal R} N_\text{DM} N_{g_\chi} N_{f_s} \times T(\text{multidimensional numeric integration}) .
\end{align}
In previous analyses~\cite{Blanco:2019lrf,Blanco:2021hlm} this type of integral might require tens of minutes to achieve comparable accuracy.

\subsection{Comparison: Direct Integration}

To compare the wavelet-harmonic integration with the standard approach of Section~\ref{sec:standard}, I calculate the rate $R$ directly from \eqref{rate:discrete}, using the $\delta$ function to reduce the 6d integral into five dimensions: $\vec q$, $v_x$ and $v_y$. For simplicity I used Cartesian coordinates within a rectangular  box, bounded by $ | q_i | \leq 10\, \alpha m_e$ and $|v_i| \leq 960\, \text{km/s}$ for $i = x,y,z$.

Aside from the fact that $g_\chi(\vec v)$ is not isotropic in any rest frame (so the \eqref{eq:gXiso} method cannot be used to replace $g_\chi(\vec v)$ with $g(\vec q)$), 
the analytic expression for $f_s^2(\vec q)$ makes the Section~\ref{sec:demo} demonstration one of the simplest test cases. 
Even so, numeric integration is rather expensive. 
Using the same Monte Carlo integrator (\texttt{vegas} in Python) that was used to find $\ket{g_\chi}$ and $\ket{f_s^2}$, 
the integration time as a function of the precision goal is approximately:
\begin{align}
\begin{tabular}{|  c |  c c c  |}  \hline
&	1\%	&	0.3\%	&	0.1\%	 
\\ \hline
$T_\text{int.}$: & 40\,s	& 300\,s	& 600\,s	 
\\ \hline
\end{tabular}
\end{align}
In cases where $f_s^2(\vec q)$ must be found from some other numerical method, it may be evaluated on a grid of $\vec q$ and saved. Integrating \eqref{rate:discrete} then requires some interpolation of the $f_s^2(\vec q)$ grid, so the evaluation time may be noticeably slower (while also requiring much more computer memory).

\bibliography{dmvec_refs}

\end{document}